\documentclass[final,5p,times,twocolumn]{elsarticle}

\usepackage{medima}
\usepackage{framed,multirow}

\usepackage{graphicx}
\usepackage{amssymb}

\usepackage{lineno}

\usepackage{xcolor}
\usepackage{amsmath}
\usepackage{amssymb}
\usepackage[space]{grffile}
\usepackage{fixmath}
\usepackage{multirow}
\usepackage{array}
\usepackage{paralist}
\usepackage{color,soul}
\usepackage{hyperref}

\colorlet{usercolorname}{red!0}
\sethlcolor{usercolorname}

\setstcolor{red}

\definecolor{fuchsia}{HTML}{ca2c92}
\definecolor{gold}{HTML}{FFD700}
\definecolor{orchid}{HTML}{DA70D6}
\definecolor{cyan}{HTML}{00EEEE}
\definecolor{springgreen}{HTML}{00FF7F}
\definecolor{c1}{HTML}{41b6c4}
\definecolor{c2}{HTML}{225ea8}
\definecolor{c3}{HTML}{31a354}
\definecolor{c4}{HTML}{a1d99b}
\definecolor{c5}{HTML}{e5f5e0}

\definecolor{white}{HTML}{F5F9F7}
\definecolor{cm1}{HTML}{fef0d9}
\definecolor{cm2}{HTML}{fdcc8a}
\definecolor{cm3}{HTML}{fc8d59}
\definecolor{cm4}{HTML}{d7301f}

\definecolor{cm5}{HTML}{FABFD1}
\definecolor{cm6}{HTML}{FC718E}

\usepackage{algorithm}
\usepackage[noend]{algorithmic}

\usepackage[font=footnotesize,labelfont=bf]{caption}
\usepackage[font=footnotesize,labelfont=bf]{subcaption}

\usepackage{afterpage}
\newcommand{\model}{DR$\vert$GRADUATE}
\newcommand{\kaggle}{Kaggle DR detection}
\newcommand{\qwk}{\kappa}

\usepackage{xcolor}

\newenvironment{packed_item}{
\begin{itemize}
  \setlength{\itemsep}{1pt}
  \setlength{\parskip}{0.25pt}
  \setlength{\parsep}{0.25pt}
}{\end{itemize}}

\usepackage{pifont}
\newcommand{\cmark}{\ding{51}}%
\newcommand{\xmark}{\ding{55}}%

\usepackage{footmisc}

\newcounter{daggerfootnote}
\newcommand*{\daggerfootnote}[1]{%
    \setcounter{daggerfootnote}{\value{footnote}}%
    \renewcommand*{\thefootnote}{\fnsymbol{footnote}}%
    \footnote[2]{#1}%
    \setcounter{footnote}{\value{daggerfootnote}}%
    \renewcommand*{\thefootnote}{\arabic{footnote}}%
    }
\makeatletter
\newcommand\footnoteref[1]{\protected@xdef\@thefnmark{\ref{#1}}\@footnotemark}
\makeatother

\journal{Medical Image Analysis}

\begin{document}

\verso{Teresa Ara\'{u}jo \textit{et~al.}}

\begin{frontmatter}


\title{\model{}: uncertainty-aware deep learning-based diabetic retinopathy grading in eye fundus images}

\tnotetext[mytitlenote]{$\copyright$ 2019. Licensed under the Creative Commons CC-BY-NC-ND 4.0 license \url{http://creativecommons.org/licenses/by-nc-nd/4.0/.}}

\address[inesc]{INESC TEC - Institute for Systems and Computer Engineering, Technology and Science, 4200-465 Porto, Portugal}
\address[feup]{Faculty of Engineering of University of Porto, 4200-465 Porto, Portugal}
\address[hbr]{Department of Ophthalmology of Hospital de Braga, Braga, Portugal}
\address[hsj]{Department of Ophthalmology of Centro Hospitalar S\~{a}o Jo\~{a}o, Porto, Portugal}
\address[fmup]{Department of Surgery and Physiology of Faculty of Medicine of University of Porto, Porto, Portugal }

\author[inesc,feup]{Teresa Ara\'{u}jo}
\ead{tfaraujo@inesctec.pt}
\cortext[correspondingauthor]{Corresponding author}

\author[inesc,feup]{Guilherme Aresta}
\ead{guilherme.m.aresta@inesctec.pt}

\author[hbr]{Lu\'{i}s Mendon\c{c}a}

\author[hsj,fmup]{Susana Penas}

\author[hsj]{Carolina Maia}

\author[hsj,fmup]{\^{A}ngela Carneiro}

\author[inesc,feup]{Ana Maria Mendon\c{c}a}
\ead{amendon@fe.up.pt}

\author[inesc,feup]{Aur\'{e}lio Campilho}
\ead{campilho@fe.up.pt}


\begin{abstract}

Diabetic retinopathy (DR) grading is crucial in determining the adequate treatment and follow up of patients, but the screening process can be tiresome and prone to errors. 
Deep learning approaches have shown promising performance as computer-aided diagnosis (CAD) systems, but their black-box behaviour hinders clinical application.
We propose \model{}, a novel deep learning-based DR grading CAD system that supports its decision by providing a medically interpretable explanation and an estimation of how uncertain that prediction is, allowing the ophthalmologist to measure how much that decision should be trusted. We designed \model{} taking into account the ordinal nature of the DR grading problem. A novel Gaussian-sampling approach built upon a Multiple Instance Learning framework allow \model{} to infer an image grade associated with an explanation map and a prediction uncertainty while being trained only with image-wise labels.
\model{} was trained on the \kaggle{} training set and evaluated across multiple datasets. In DR grading, a quadratic-weighted Cohen's kappa ($\qwk{}$) between 0.71 and 0.84 was achieved in five different datasets. We show that high $\qwk{}$ values occur for images with low prediction uncertainty, thus indicating that this uncertainty is a valid measure of the predictions' quality. Further, bad quality images are generally associated with higher uncertainties, showing that images not suitable for diagnosis indeed lead to less trustworthy predictions. Additionally, tests on unfamiliar medical image data types suggest that \model{} allows outlier detection. The attention maps generally highlight regions of interest for diagnosis. 
These results show the great potential of \model{} as a second-opinion system in DR severity grading.

\end{abstract}


\end{frontmatter}



\section{Introduction}
\label{sec:intro}

Diabetic retinopathy (DR) is a complication of diabetes and is one of the leading causes of blindness worldwide~\citep{DRGuidelines2012,Wild2004} and the number of diabetic patients is expected to grow from 346 million in 2012 to 552 million people in 2030~\citep{Schully2012,DRGuidelines2012}.
The majority of visual loss cases can be prevented with early detection and adequate treatment~\citep{Garg2009,Wu2013}. However, one third of the diabetic patients suffer from DR without having symptoms of visual problems, leading to the progression of the disease without treatment~\citep{Acharya2012}. Consequently, regular check-ups via screening programs are essential to achieve early diagnosis of DR. 

\begin{figure}[t]
 \begin{center}
     \includegraphics[width=0.48\textwidth]{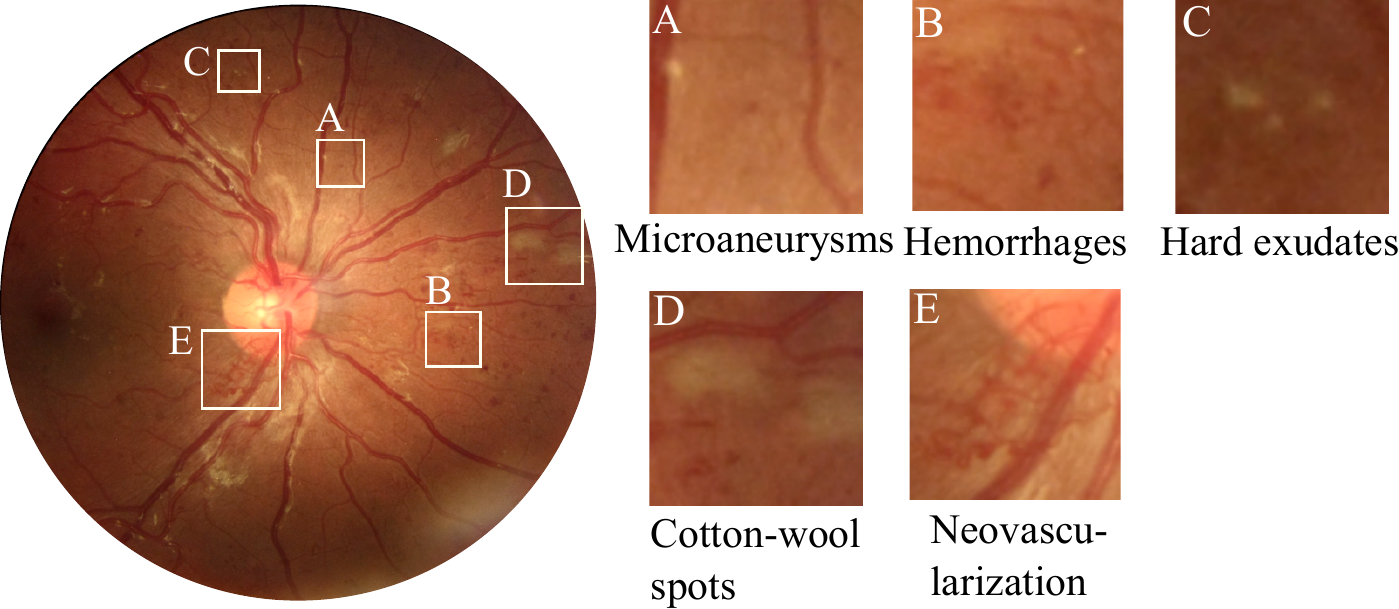} 
  \caption{Examples of typical retinal lesions characteristic of diabetic retinopathy. 
  }
    \label{fig:eye_lesions}
     \end{center}
 \end{figure}

DR screening programs have been developed across the world, namely in the United Kingdom\footnote{\url{https://www.gov.uk/topic/population-screening-programmes/diabetic-eye}}, Ireland\footnote{\url{https://www.diabeticretinascreen.ie}} Netherlands\footnote{\url{http://www.eyecheckklant.nl}}~\citep{Abramoff2005}, France\footnote{\url{reseau-ophdiat.aphp.fr}}~\citep{Massin2008}, United States\footnote{\url{https://www.eyepacs.com}}~\citep{Cuadros2009}, Canada~\citep{Nguyen2009} and Portugal~\citep{DutraMedeiros2015}.
In these programs, DR detection and DR severity grading are performed by trained specialists by visual analysis of color eye fundus images. These images are captured by fundus photography, allowing for non-invasive visual assessment of several retinal structures, as depicted in Fig.~\ref{fig:eye_lesions}:
\begin{inparaenum}[1)]
\item the fovea, a reddish area without vessels located at the center of the macula;
\item the optic disc (OD) which is a round structure close to the macula that corresponds to the exit of the axons of ganglion cells that constitute the optic nerve and the door for the retinal vessels that supply the retina, and 
\item blood vessels, that radiate from the OD center. 
\end{inparaenum}
During screening, specialists search for abnormalities in these structures and in the retinal tissue and classify the severity of the disease according to the type and quantity of the findings.
Despite the overall success of these programs, the diagnosis process is still prone to errors due to the large number of patients, poor image acquisition and variety of lesions.


DR is generally classified according to the absence or presence of new abnormal vessels as non-proliferative DR (NPDR) or proliferative DR (PDR), respectively. NPDR can be further graded as mild, moderate and severe~\citep{DRGuidelines2012}. These stages can be ordered according to their risk of progression~\citep{Arevalo2013}. \hl{Namely,} Table~\ref{tab:dr_scale} \hl{represents the International Clinical DR Scale}~\citep{Wilkinson2003}, \hl{adapted to consider only one image for diagnosis}~\citep{Porwal2020} \hl{by removing the quadrant concept, since in the case of a single fundus image analysis quadrant assessment is limited due to the under-representation of some of the quadrants. Thus, in practice, the exact counting of lesions in each quadrant is not performed by the specialists, but rather the estimation of what happens in the full eye based on their experience and medical knowledge.}

\hl{Mild NPDR is characterized by the presence of microaneurysms (MAs), which are the first sign of DR. 
Hemorrhages (HEMs) start to appear with the evolution of the disease in the deeper layers of the retina.  
The disease progression results in an increase of MAs, dot-hemorrhages (red lesions), retinal thickening, hard exudates (EXs), and the appearance of intraretinal microvascular abnormalities (IrMAs), venous beading and cotton-wool spots (CWSs), also known as soft exudates, in the innermost retina.  
Finally, PDR results from abnormal vessel formation, i.e., neovessels (NVs). 
Other signs of PDR are the vitreous/pre-retinal hemorrhages (PHEMs) and pre-retinal fibrosis (PFIB)}~{\citep{Hall2011}}.
Fig.~\ref{fig:eye_lesions} shows examples of some of these lesions.  





\begin{table}[t]
\caption{\st{International d}Diabetic retinopathy (DR) severity scale. NPDR: Non-proliferative DR; PDR: proliferative DR. Non-referable DR: R0 and R1; Referable DR: R2, R3 and R4. \label{tab:dr_scale}} 

\begin{tabular}{|ll|l|}
\hline
\multicolumn{2}{|l|}{Grade}                                    & Description                                                                                                                                                                                             \\ \hline
R0 - & No DR                                                    & No visible sign of abnormalities                                                                                                                                                                        \\ \hline
R1 - & \begin{tabular}[c]{@{}l@{}}Mild \\ NPDR\end{tabular}     & Presence of microaneurysms only                                                                                                                                                                         \\ \hline
R2 - & \begin{tabular}[c]{@{}l@{}}Moderate \\ NPDR\end{tabular} & \begin{tabular}[c]{@{}l@{}}More than just microaneurysms but \\ less than severe NPDR\end{tabular}                                                                                                      \\ \hline
R3 - & \begin{tabular}[c]{@{}l@{}}Severe \\ NPDR\end{tabular}   & \begin{tabular}[c]{@{}l@{}}Any of the following:\\     \textgreater 20 intraretinal hemorrhages\\     Venous beading\\     Intraretinal microvascular abnormalities \\     no signs of PDR\end{tabular} \\ \hline
R4 - & PDR                                                      & \begin{tabular}[c]{@{}l@{}}Either or both of the following:\\     Neovascularization\\     Vitreous/pre-retinal hemorrhage\end{tabular}                                                                 \\ \hline
\end{tabular}

\end{table}

Computer-aided diagnosis (CAD) systems can improve the DR screening pipeline both reducing the burden~\citep{Scotland2007} and by proving a second objective opinion to the ophthalmologists, reducing subjectivity in the diagnosis~\citep{Gargeya2017,Abramoff2016, Gulshan2016, Quellec2017}. Namely, deep learning has recently allowed for these systems to achieve near-human performance in DR detection, \textit{i.e.} detection of at least one DR-related lesion.
On the other hand, DR grading, \textit{i.e.} staging of the pathology according to its severity, is a more complex problem since it requires the identification and integration of different abnormalities. 
In fact, specialists tend to disagree on the grade of complex cases~\citep{Krause2018} and thus the external opinion of CAD systems may contribute to further improve the success of DR grading. However, due to this higher complexity and the inherent black box behaviour of deep learning systems, it may be difficult for specialists to understand and accept the system's decision. With this in mind, we propose a novel DR grading CAD system that not only supports its decision by providing a medically interpretable explanation but also estimates how uncertain that prediction is. By doing so, the ophthalmologist not only has a reasoning behind the system's decision, but also a measure of how much that decision should be trusted, making the DR diagnosis process less prone to errors.

\subsection{State-of-the-art}
\label{subsec:state_art}

Deep learning has become the preferred approach, over field knowledge-based feature engineering approaches, for DR grading since it allows to better exploit the large number of data available and to better deal with the labeling noise resulting from the complexity of the task.
For instance, \cite{Krause2018} fine-tuned an Inception-v4 with  1,600,000 images of size 779$\times$779 pixels to avoid the elimination of small lesions such as MAs. Also, to account for data imbalance, the authors apply a cascade of thresholds on the output probabilities of the model for each DR grade to obtain the final image classification. These thresholds are sequentially applied in a severity descending order to ensure that a minimum grade-wise sensitivity is achieved.
Instead, \cite{DelaTorre2018} considered DR grading as an ordinal classification problem and thus aimed at reducing the distance between predicted and true labels of an image. For that, they proposed a quadratic-weighted-Kappa-based loss function, which allowed to improve the performance of a DR grading network comparing to training with the cross-entropy loss. 
The authors tested different input images sizes (squares with side of 128, 256, 384 and 512 pixels), and the highest resolution yielded the best performance.
The network, which has 11.3 million parameters for the 512$\times$512 case, was trained in a public dataset (\kaggle{}). To overcome the high unbalancing of the dataset, the data is artificially balanced using data augmentation, oversampling the least represented classes. 

Despite the high classification performance of these systems, their black box behavior hinders the application on screening settings. Indeed, providing explanations for DR grading is challenging due to the complexity of the task, and thus research is mainly focused on the binary detection and referral tasks.
For instance, \cite{Quellec2017} analyzed several explanation methods for the referable DR decisions of their CNNs'. Namely, the authors compare deconvolution~\citep{Zeiler2014}, sensitivity analysis~\citep{Simonyan2014} and layer-wise relevance propagation~\cite{Bach2015}. Based on~\citep{Simonyan2014}, the authors also proposed the creation of a pixel-wise explanation map that is jointly optimized with the CNN prediction during training.
\cite{Costa2019} proposed the attention map generation via a pre-trained network coupled with a Multiple Instance Learning (MIL)-based approach for DR detection. Namely, the network is cropped at early layers to reduce the model's receptive field, and the classification is performed by a 1$\times$1 convolution and global max-pooling, following the assumption that the presence of one unhealthy patch is enough to label the entire image. Because of this, the output of this convolution is a map where each element indicates the malignancy score of a corresponding patch of the input image, thus explaining why the image label was predicted.
Instead, \cite{Gonzalez-gonzalo2018} has applied an iterative inpainting approach that, at each step, allows to identify progressively less discriminative pixels of the image for DR classification. This method involves the computation of the saliency map at each step, \textit{i.e.}, the derivative of the classification output with respect to the input image using guided back-propagation. The authors use a pre-trained VGG-16 on the Kaggle dataset (512$\times$512 pixel images). Similarly to other grading approaches, the problem is treated as being ordinal and thus the model is optimized using a mean-square error loss. To avoid overfitting due to the high class imbalance, the classes are balanced taking into account the referal/non-referal sample distribution.

Besides explaining the decision, uncertainty estimation of a model's prediction is of interest as it helps to avoid consequences of the blind use of the model response~\citep{Kendall2017}. This is particularly true on medical settings, where misdiagnosis can have severe consequences for the patients. With this in mind, \cite{Leibig2017} analysed the uncertainty information from deep neural nets for DR detection and referable DR detection. 
The authors tested the dropout-based Bayesian uncertainty estimation, comparing it with alternative techniques, such as directly analysing the network's softmax output. In the dropout-based approach, dropout is switched on during inference. Performing multiple predictions on the same input image can be interpreted as having an ensemble of highly similar models, and thus disagreement between the inferences is indicative of the global model uncertainty~\citep{Gal2016}. The authors state that the Bayesian treatment worked better than the softmax output as a proxi for uncertainty and also show that uncertainty-aware decision referral can improve the diagnosis.

\par~\par


\subsection{Contributions}

Many research work has already addressed DR grading, explainability and uncertainty estimation separately. Instead, in this study we propose \model{}, a system that is capable of simultaneously providing an explanation and a measure of uncertainty together with a DR severity grade label. 
Our main contributions are the following:

\begin{packed_item}
    \item estimation of a DR grade with an associated uncertainty, thus allowing to better establish which cases require further inspection by specialists, which is of high interest in computer-aided diagnosis systems; 
    \item explanation of the algorithm decision, easily and directly interpretable by the retinal specialists, in the form of an attention map, which will leverage its use by medical professionals since it partially mitigates the black-box behaviour associated with deep nets;
    \item extensively validated deep learning-based method for DR severity grading of eye fundus images in the internationally recognized DR levels. 

\end{packed_item}

\section{\model{} for DR grading}

\begin{figure*}[tb]
    \centering
    \includegraphics[width=\textwidth]{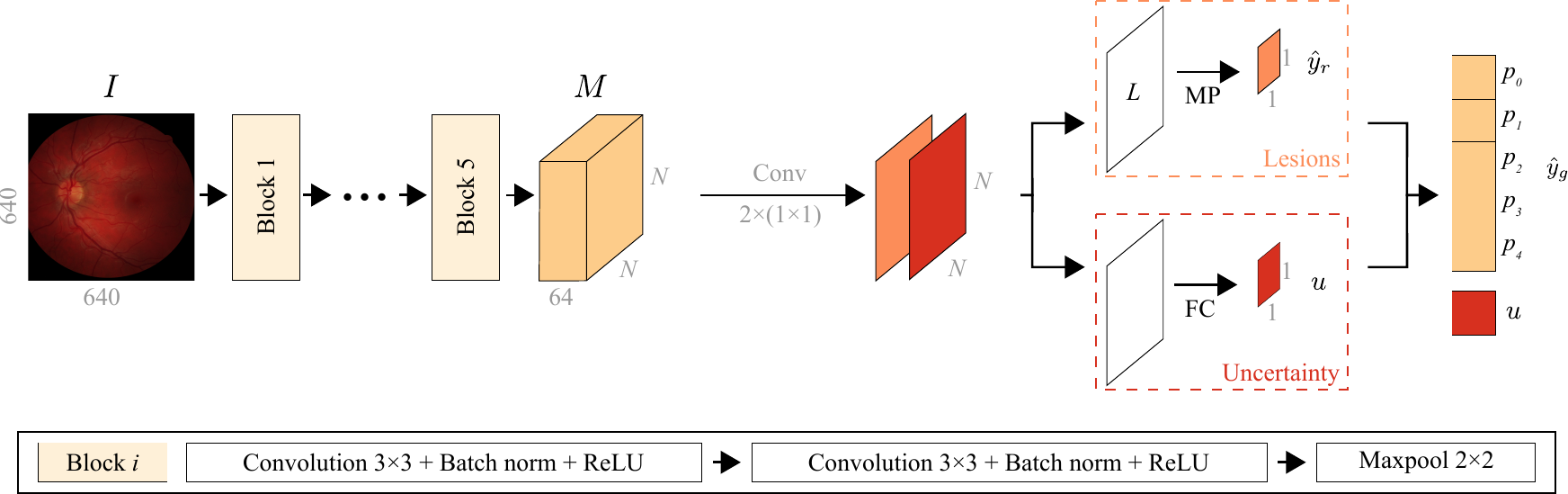}
    \caption{\model{}'s architecture. The network is composed of several convolutional-batch normalization blocks interleaved with max-pooling layers. 
    For each input image $I$, \model{} predicts a diabetic retinopathy (DR) grade $\hat{y}_g$ and a grade uncertainty $u$. $M$ is the output of the last layer of \model{}'s backbone after assessing $I$.
    The lesion map $L$ is computed by performing a $1\times 1 \times 1$ convolution with a linear activation over $M$. $L$ is a map where the value of each element $L_{\,i\,j}$ indicates the presence of a lesion of grade $\lfloor L_{\,i\,j} \rceil$\textsuperscript{\ref{foot:floor_ceil}} on a patch of size dependent on the model's receptive field in the input. Following the Multiple Instance Learning approach, the grade of an image is initially estimated as $\hat{y}_r=max(L)$. The output probabilities ($p_c, c\in\{0...4\}$) are then computed from $\hat{y}_r$ based on a Gaussian distribution centered on $\hat{y}_r$ with variance equal to the uncertainty $u$, with $u$ learned during training.}
    \label{fig:architecture}
\end{figure*}

The backbone of \model{}, depicted in Fig.~\ref{fig:architecture}, is composed of several convolutional-batch normalization blocks interleaved with max-pooling layers that has already shown potential for DR grading tasks~\citep{DelaTorre2017}. The top of the network is composed of a novel component that allows for the inference of the image's grade $\hat{y}_g$ together with the corresponding explainability $\mathcal{E}$ and the explicit uncertainty $u$ of the decision, without using labels other than the DR grade, such as lesion-wise annotations.

Let \model{} be a classification model that predicts $\hat{y}_g=argmax\left(p\right)$, where $p_c$ is the class-wise probability and $\sum_{c}^{C} p_c=1$, with $C$ denoting the number of considered classes. The following assumptions regarding DR grading are made:
    [\hl{\texttt{assumption~1}}] grading is a discrete ordinal classification problem with an image label $y\in \mathcal{G}=\{0,1,2,3,4\}$, where $\mathcal{G}$ is the set of possible DR grades; in this scenario, we want to optimize \model{} by minimizing the distance between the target and the prediction, $min(|y-\hat{y}_g|)$, and
    [\hl{\texttt{assumption~2}}] there is always a different lesion type per grade on the universe of possible DR-related lesions, $DR_{lesions}$; \hl{thus, if one associates a numerical value to each lesion from $DR_{lesions}$, accordingly to the DR stage in which they appear,} \st{and thus} $\hat{y}_g$ can be predicted by following the standard MIL assumption, \textit{i.e.} $max\left(DR_{lesions}\right)\mapsto\hat{y}_g$.

DR grading can be seen as a regression problem and consequently the predicted image grade $\hat{y}_g$ can be inferred from a continuous output $\hat{y}_r\in\mathbb{R}_+$. This also allows for the implicit Multiple Instance Learning (MIL) of the problem, as we can to retrieve $\hat{y}_r$ from a predicted lesion map $L$, $\hat{y}_r=max(L)$. Furthermore, the uncertainty $u$ could be estimated by $u\propto|\lfloor\hat{y}_r\rceil-\hat{y}_r|$\daggerfootnote{\label{foot:floor_ceil}$\lfloor\, \rceil$ denotes rounding to the nearest integer, $\lfloor \,\rfloor$ rounding down and $\lceil\,\rceil$ rounding up.}. However, this uncertainty does not properly hold in scenarios where the regression cut-off thresholds are adjusted to better fit the data distribution and malignancy priority, as in many Kaggle competition solutions\footnote{https://storage.googleapis.com/kaggle-forum-message-attachments/88655/2795/competitionreport.pdf}. Also, in terms of clinical applicability, it is of higher interest to provide a class-wise probability to clinicians since it is easier to interpret. Because of this, [\hl{\texttt{assumption~3}}] \model{} assumes that the prediction and its uncertainty can be modeled by a Gaussian curve centered near the most probable class. Namely, for each image, \model{} predicts a variance $\sigma^2$, which allows for $\hat{y}_r\in\mathbb{R}_+ \mapsto \hat{y}_g=argmax(p)$. With this, \model{} is capable of providing an uncertainty $\sigma^2$ and a class-wise probability that still attempts at minimizing the distance between the prediction and the ground truth. The next sections describe \model{} with more detail.


\subsection{Predicting a grade with uncertainty}
\label{subsec:uncertainty}

For each input image $I$, \model{} predicts a DR grade $\hat{y}_g$ and a grade uncertainty $u$. Let $M$ be the output of the last layer of \model{}'s backbone after assessing an input image $I$. $M$ has size $N\times N \times F$, where $N$ and $F$ are, respectively, the side and number of features. The lesion map $L$ is computed by performing a $1\times 1 \times 1$ convolution with a linear activation over $M$, so that $L \in \mathbb{R}$ has size $N\times N \times 1$. $L$ is a map where the value of each element $L_{\,i\,j}$ indicates the presence of a lesion of grade $\lfloor L_{\,i\,j} \rceil$\textsuperscript{\ref{foot:floor_ceil}} on a patch of size dependent on the model's receptive field in the input.
Following the MIL approach, the grade of an image is initially estimated as $\hat{y}_r=max(L)$~\citep{Costa2019}.

\begin{figure*}[tb]
    \centering
    \begin{subfigure}{0.27\textwidth}
    \includegraphics[width=\textwidth]{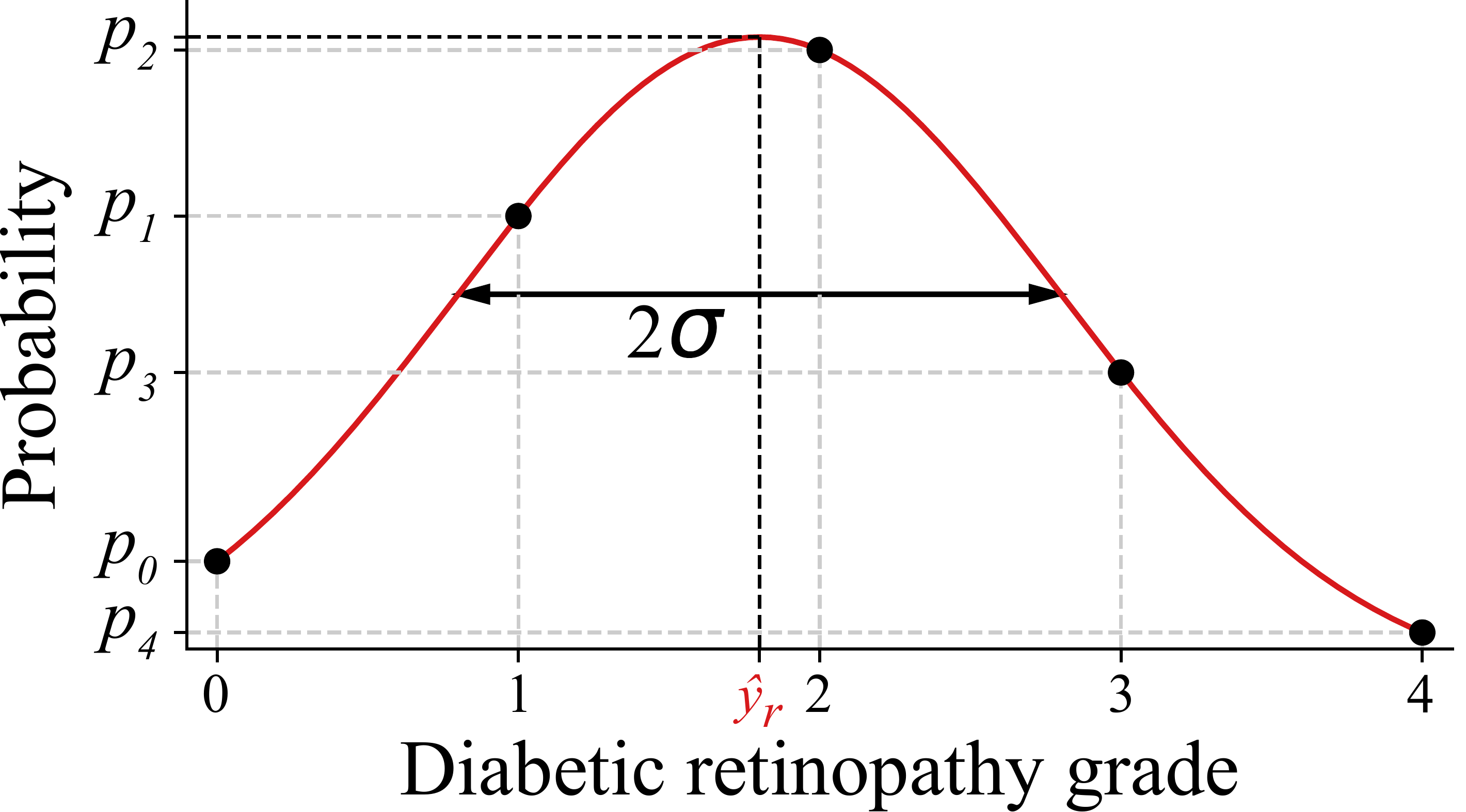}
    \caption{Probability Gaussian distribution with mean $\hat{y}_r$ and standard deviation $\sigma$ (Eq.~\ref{eq:prob_sigma}).  \label{fig:unc_prob}}
    \end{subfigure}
    \hfill
    \begin{subfigure}{0.703390\textwidth}
    \includegraphics[width=\textwidth]{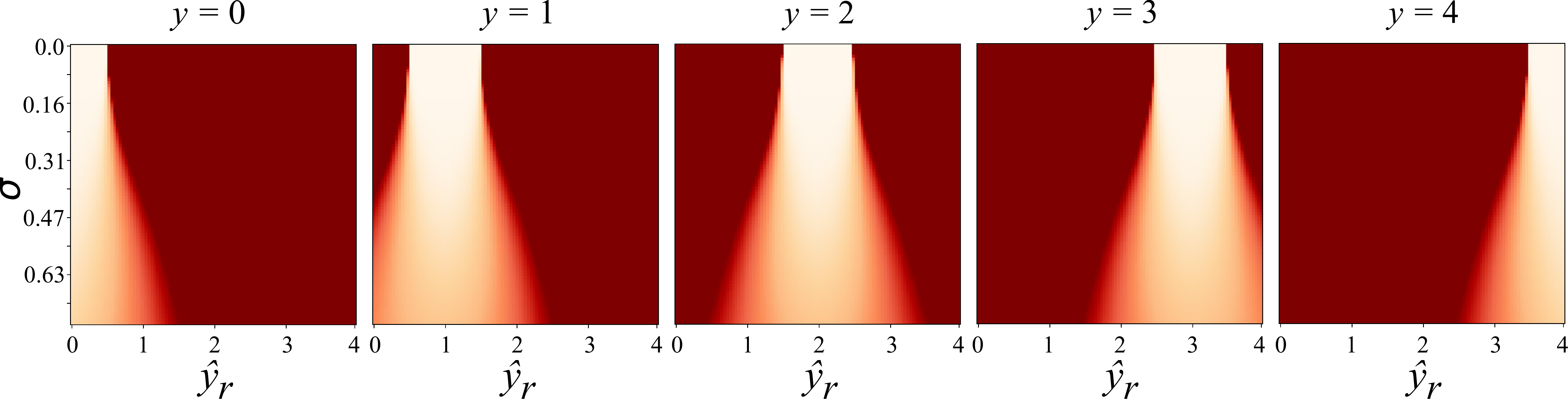}
    \caption{$\mathcal{L}$ loss values (Eq.~\ref{eq:our_loss}, with $\alpha=0.7$) as function of $\hat{y}_r$ and $\sigma$, for $y\in \mathcal{G}$. Loss colorbar: $0$~\protect\includegraphics[height=.75em,width=5em]{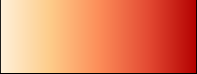}~$2$ . \label{fig:unc_loss_exs}}
    \end{subfigure}
    \caption{Uncertainty estimation (Eq.~\ref{eq:prob_sigma}) and loss representation (Eq.~\ref{eq:our_loss}).  
    \label{fig:unc_prob_loss}}
\end{figure*}

Having into account \hl{\texttt{assumptions 1} and \texttt{ 3}}, $\hat{y}_r \mapsto \hat{y}_g$ must ensure that
$ p_{\lfloor \hat{y}_r  \rfloor}>p_{\lfloor \hat{y}_r  \rfloor-1}>p_{\lfloor \hat{y}_r  \rfloor-2}> \ldots
$
, and 
$ p_{\lceil \hat{y}_r  \rceil}>p_{\lceil \hat{y}_r  \rceil+1}>p_{\lceil \hat{y}_r  \rceil+2}> \ldots
$ ,
\textit{i.e.}, that the classes near to $\hat{y}_r$ have higher probability than the others. For each image, \model{} must thus predict a generalized Bernoulli distribution biased around the classes closer to $\hat{y}_r$. This bias is introduced in the model by means of a Gaussian distribution $\mathcal{N}$ centered on $\hat{y}_r$ and thus the image-wise grade probability $p$ is:

\begin{align}
    p' &= \left\{\frac{1}{\sqrt{2 \pi}  \sigma^2} \text{exp}\left(-0.5\frac{[i-\hat{y}_r]^2}{\sigma^2}\right)
    ,\,  \forall c\in\mathcal{G} \right\} \nonumber \\
    p &= \frac{p'}{\sum_{c \in \mathcal{G}} p'_c}
    \label{eq:prob_sigma}
\end{align}

\noindent where $\sigma^2$ is the variance of $\mathcal{N}$ and $p \in [0\,, 1]$. Fig.~\ref{fig:unc_prob} illustrates the sampling of the probability values $p$ from the Gaussian probability distribution. Knowing that the uncertainty of the prediction can be measured by the value of its entropy~\citep{Friston2010}, \textit{i.e.} $u$ is directly related to $S = -\sum_c p_c log(p_c) $, it follows that since $S$ depends on $\sigma^2$, then $u$ depends on $\sigma^2$. Thus by estimating $\sigma^2$ it is possible to infer the uncertainty associated with each predicted grade, $\hat{y}_g = argmax(p)$.

In \model{}, $\sigma^2$ is inferred by using a fully-connected layer on top of $M$. Since both $\hat{y}_r$ and $\sigma^2$ depend on $M$, both grading and the associated uncertainty affect the backbone of the model, easing its convergence. Indeed, \model{} is trained end-to-end minimizing the loss function:

\begin{equation}
    \mathcal{L} = -\alpha\sum_{c \in \mathcal{G}}[b_c log(p_c)] + (1-\alpha)\sigma^2
    \label{eq:our_loss}
\end{equation}

\noindent where $b_c$ is the hot-encoded (i.e., a binary representation of) $i$ and $\alpha\in[0,1]$ is a weighting factor. The first term, the categorical cross-entropy, leads \model{} to predict $\hat{y}_r$ and thus $\hat{y}_g$ as close to the truth as possible. The second term regularizes $\sigma^2$ to avoid excessive Gaussian spreads and thus the stagnation of the learning on the local minimum of near equiprobable predictions. 

The global loss $\mathcal{L}$ and the structure of \model{} implicitly introduce an inter-dependency between $\hat{y}_r$ and $\sigma$ that allows to deal with images of different complexity. 
For less complex images, both $\sigma$ and the preliminary class prediction error, $|\lfloor\hat{y}_r\rceil-\hat{y}_r|$, should be low. However, for difficult images, high values of $\sigma^2$ can compensate for classification errors since the higher spread of the Gaussian tends to increase the predicted probability of the correct class, lowering the cross-entropy loss. 
Fig.~\ref{fig:unc_loss_exs} shows the evolution of $\mathcal{L}$  (Eq.~\ref{eq:our_loss}) as function of $\hat{y}_r$ and $\sigma$, for different targets $y$, $y\in \mathcal{G}$. These plots show that when $\hat{y}_r$ is close to $y$ the $\mathcal{L}$ loss reaches its minimum. Further, if one freezes $\hat{y}_r$ at a value close to the ground truth, the $\sigma$ that leads to a lower $\mathcal{L}$ loss is smaller than if $\hat{y}_r$ is fixed at a value far from the ground truth.

\subsection{Explainability of the DR grade}

The goal of this study does not include a pixel-level prediction of lesions. Even human-performed lesion segmentation shows high variability, due to the extremely small size of some lesions and also to differences in the annotation style. Indeed, some ophthalmologists prefer to only annotate clusters of lesions, whereas others are more detailed and prefer to give lesion-wise markings. 
Since DR image grading is performed based on the presence of the typical DR lesions, pinpointing relevant lesions should be a valuable tool for doctors to assess the algorithm's prediction without the need for a refined segmentation.

\begin{figure}
    \centering
    \includegraphics[width=0.45\textwidth]{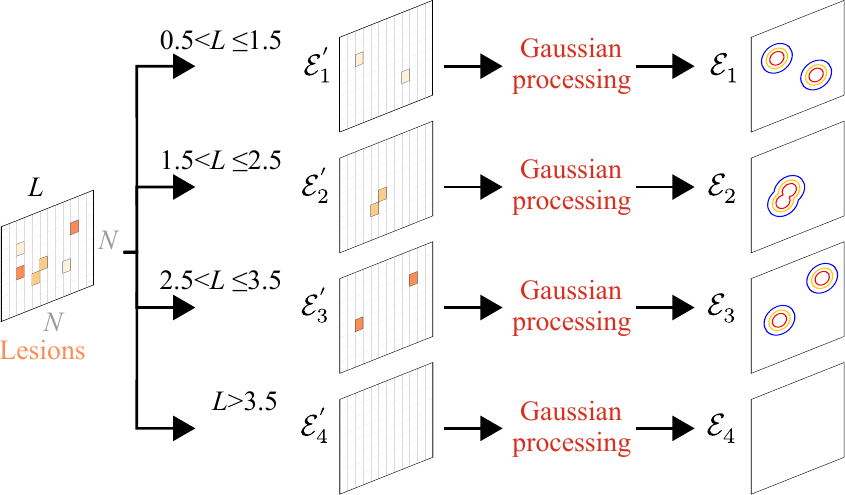}
    \caption{Attention maps construction. Gaussian processing is described in Eq.~\ref{eq:gaussian_maps}.}
    \label{fig:lesion_maps}
\end{figure}

\model{} provides a grade-wise explanation map, as illustrated in Fig.~\ref{fig:lesion_maps}. For each grade $c$ the lesion presence encoded in $L$ is translated to an explainability map $\mathcal{E}_c$ of the same size of $I$ via:

\begin{align}
  \mathcal{E}'_c(s\cdot i, s\cdot j) & = 
\begin{cases}
1 & \text{if } c-0.5 \leq L_{\,i\,j} < c+0.5 \nonumber \\
0 & \text{otherwise}
\end{cases} \\
\mathcal{E}_c &=  \mathcal{E}'_c \circledast \mathcal{N}_{2\,, \sigma'}   
\label{eq:gaussian_maps}
\end{align}

\noindent where $\mathcal{E}'_c$ is a matrix of the same size of $\mathcal{E}_c$, $c\in \mathcal{G}\setminus\{0\}$ is one of the possible pathological grades, $i,j$ are each row and column position, respectively, in the $L$ map of  side $N$ (${i,j\in  \mathbb{N}~\vert~i,j<N}$),  and $\mathcal{N}_{2\,, \sigma'}$ is a convolutional kernel from a two-dimensional multivariate normal distribution with standard deviation $\sigma'$, with $\sigma'$ related to the receptive field of the network, and $s$ is the stride of \model{}.
The convolution with the Gaussian kernel allows to give higher priority to central pixels of the receptive field, which have a higher contribution to the value of each $L_{\,i\,j}$.

\subsection{Dealing with class imbalance}

DR grading is an imbalanced problem skewed towards the healthy cases, similarly to many other medical imaging problems. Data imbalance introduces convergence issues during model training, demanding for class balancing strategies. Oversampling has shown to be an overall good choice for dealing with class imbalance~\citep{Buda2018}. However, excessively reducing class imbalance may lead to an unnecessarily high trade-off between the model's sensibility and specificity, \textit{i.e.} increase the misclassifications of healthy images, since the most represented classes will now tend to be relatively over-classified. With this in mind, \model{} is trained using an iterative batch balancing scheme that has already shown success for DR grading\footnote{\url{https://www.kaggle.com/blobs/download/forum-message-attachment-files/2797/report.pdf}}:

\begin{equation}
{w_c}_t = r^{t-1}{w_c}_o + (1-r^{t-1}){w_c}_f
\label{eq:class_balance}
\end{equation}

\noindent where ${w_c}_t$ is the expected ratio of images of grade $c$ on the batch at epoch $t$, ${w_c}_0$ is the representativeness of class $c$ on the dataset,
${w_c}_f$ is the ratio of class $c$ on the batch after $f$ training epochs and $r$ is the decay rate that controls the update of ${w_c}_t$. With this, in the beginning of the training the network is confronted with a completely balanced set and does the initial learning based on this, and then progressively starts to adjust (or fine-tune) to a more real data distribution, but still not as skewed as the original training set distribution. The oversampling of the least represented classes is performed via online horizontal and vertical flips, rotations, brightness adjustments and contrast normalization.

\section{Experiments}

Intra- and inter-datasets experiments were performed for evaluating \model{} in terms of:
\begin{inparaenum}[1)]
\item DR grading performance,
\item uncertainty estimation, and
\item explainability. 
\end{inparaenum}
The DR grading performance was quantitatively evaluated on several DR-labeled eye fundus image datasets, and class-wise performance was analysed.
The relation between the uncertainty and the algorithm's performance was assessed. Further, the uncertainty estimation on good and bad quality images was compared, using image quality-annotated datasets, in order to assess whether bad quality images would be associated with higher uncertainties. Finally, uncertainty estimation on unfamiliar data types was also analysed, aiming at inspecting whether the estimated uncertainties could allow to detect unknown image types (i.e., outliers). Different medical images were considered: colon (colonoscopy), skin, cataracts surgery and breast biopsy microscopy images.
For the explainability, the attention maps produced by \model{} were qualitatively evaluated through visual inspection, as well as quantitatively, by comparison with lesion annotations performed by an ophthalmologist.

\subsection{Datasets}

\subsubsection{DR grading datasets}

\begin{figure}[tb]
    \centering
    \includegraphics[width=0.45\textwidth]{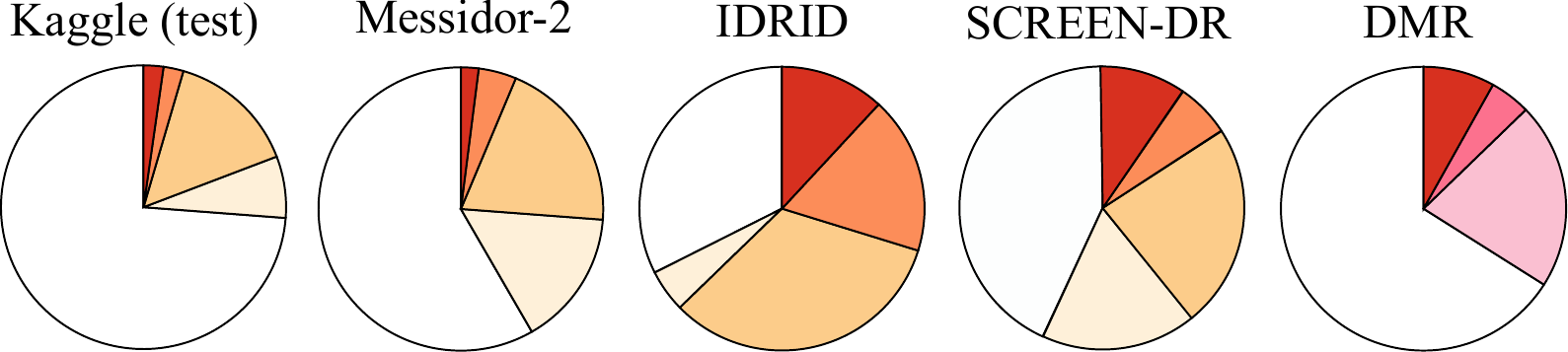}
    \caption{Class distribution of the DR grading datasets used for evaluation. {$\square$}~R0/no DR, {\color{cm1}$\blacksquare$}~R1, {\color{cm2}$\blacksquare$}~R2, {\color{cm3}$\blacksquare$}~R3, {\color{cm4}$\blacksquare$}~R4/PDR, {\color{cm5}$\blacksquare$}~SDR, {\color{cm6}$\blacksquare$}~PPDR.
    \label{fig:datasets_classes}}
\end{figure}

The \kaggle{} training dataset was used for training \model{}, and several public and private datasets have been used for evaluation. Fig.~\ref{fig:datasets_classes} shows the class distribution of these datasets.

\textit{\kaggle{}:} This dataset was published by Kaggle\footnote{\url{https://www.kaggle.com/c/diabetic-retinopathy-detection}}, comprising a large number of high resolution images (approximately 35\,000 in the training set, 55\,000 in the test set). Images were acquired under a large variety of conditions, using different types of cameras.
Images are labeled by a single clinician with the respective DR grade, out of 4 severity levels: 1 - mild, 2 - moderate, 3 - severe, and 4 - proliferate DR. Level 0 stands for the absence of DR.
The organizers state that this dataset is similar to any real-world dataset, in which noise is present in both images and labels. Consequently, images may contain artifacts, be out of focus and/or with inadequate exposition, aiming at promoting the development of robust and reliable algorithms that behave well in the presence of noise and across a variety of images.

\textit{Messidor-2:} The Messidor-2 public dataset~\citep{Abramoff2016}\footnote{\st{https://medicine.uiowa.edu/eye/abramoff}\hl{http://www.adcis.net/en/third-party/messidor2/}} consists of color eye fundus images, one fovea-centered image per eye, of 874 subjects with diabetes, for a total of 1748 images.  
Images were acquired using a color video 3CCD camera on a Topcon TRC NW6 non-mydriatic fundus camera with a 45$^\circ$ FOV, at 1440$\times$960, 2240$\times$1488, or 2304$\times$1536 pixels (px). 
The ground truth~\citep{Krause2018}\footnote{\url{https://www.kaggle.com/google-brain/messidor2-dr-grades}} corresponds to an adjudicated consensus of three specialists. A total of 1744 images were used, as the remaining were adjudicated as ungradable.



\textit{IDRID:} The Indian Diabetic Retinopathy Image Dataset (IDRiD) dataset~\citep{Porwal2018} is composed of 413 images of 4288$\times$2848 px, 50$^{\circ}$ FOV, acquired using a Kowa VX-10 alpha digital fundus camera. Images were classified accordingly to the International Clinical Diabetic Retinopathy Scale.

\textit{DMR:} The dataset announced in~\cite{Takahashi2017} (DMR) is composed of 9939 posterior pole color fundus images (2720$\times$2720 px) from 2740 diabetic patients. Images were captured using a NIDEK AFC-30 fundus camera, and have a 45$^{\circ}$ FOV. 
One to four images may exist from a given patient. Images were graded by modified Davis grading, described in Table \ref{tab:davis_scale}. The international scale grades (Table~\ref{tab:dr_scale}) were converted to the modified Davis scale (Table~\ref{tab:davis_scale}) using the following assumption: NDR~=~R0, SDR~=~R1 $\cup$~R2, PPDR~=~R3 and PDR~=~R4.

\begin{table}[tb]
\small{
\caption{Modified Davis grading used in~\cite{Takahashi2017} for the DMR dataset annotation.}
\label{tab:davis_scale}
\begin{tabular}{|l|l|}
\hline
Grade     & Description                                                                 \\ \hline
No DR     & No signs of diabetic retinopathy                                                                  \\ \hline
SDR  & \begin{tabular}[c]{@{}l@{}}  Microaneurysm, retinal hemorrhage, hard exudate, \\ retinal edema, and more than 3 small soft exudates  \end{tabular}     \\ \hline
PPDR & \begin{tabular}[c]{@{}l@{}}Soft exudate, varicose veins, intraretinal microvascular \\abnormality, and non-perfusion area over one disc area  \end{tabular}   \\ \hline
PDR  & \begin{tabular}[c]{@{}l@{}} Neovascularization, pre-retinal hemorrhage, vitreous \\ hemorrhage, fibrovascular proliferative membrane, and \\ tractional retinal detachment \end{tabular}                         \\ \hline
\end{tabular}
}
\end{table}



\begin{figure}[tb]
\centering
\begin{subfigure}{0.23\textwidth}
\includegraphics[width=\textwidth]{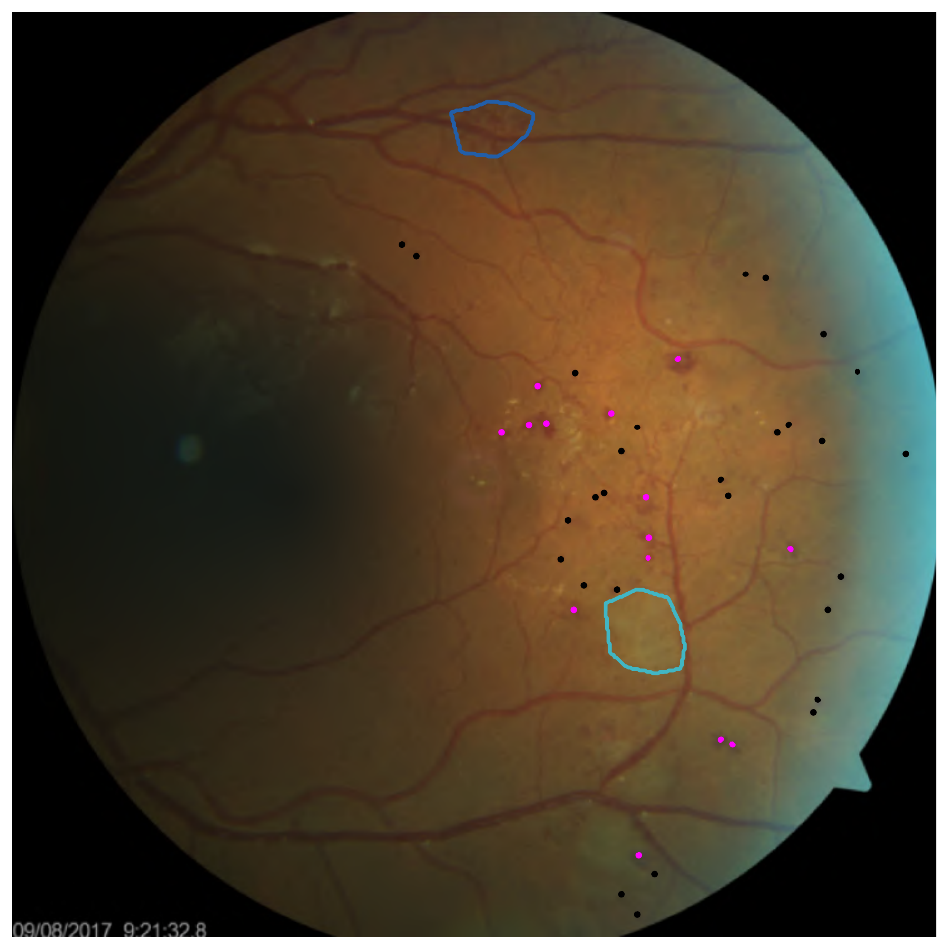} 
\end{subfigure}
\hfill
\begin{subfigure}{0.23\textwidth}
\includegraphics[width=\textwidth]{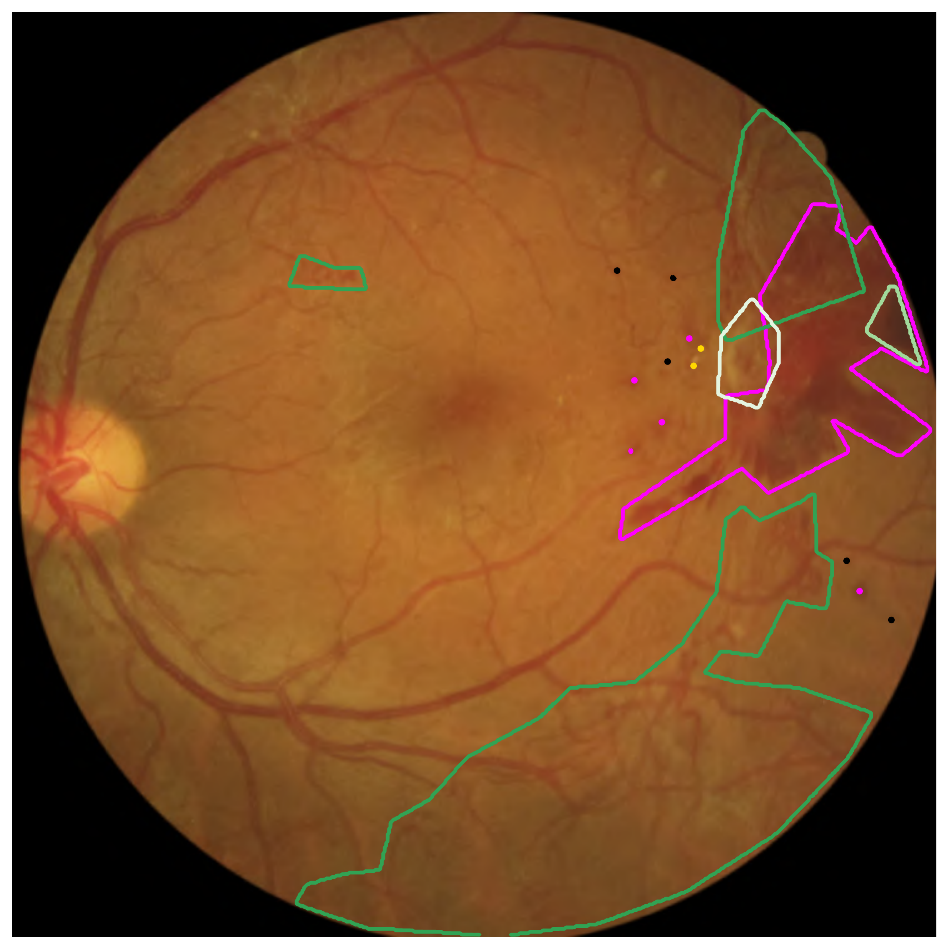}
\end{subfigure}
\caption{Examples of ophthalmologist's annotations on the SCREEN-DR dataset. {\color{black}$\blacksquare$}~MA, {\color{fuchsia}$\blacksquare$}~HEM, {\color{gold}$\blacksquare$}~EX, {\color{c1}$\blacksquare$}~CWS, {\color{c2}$\blacksquare$}~IrMA, {\color{c3}$\blacksquare$}~NV, {\color{c4}$\blacksquare$}~PHEM, {\color{c5}$\blacksquare$}~PFIB.
\label{fig:annot_screendr}
}
\end{figure}

\textit{SCREEN-DR:} This private dataset consists of retinal images from 
a portuguese DR screening program, managed by the Portuguese North Health Administration (ARSN). A subset of 966 images was selected (SCREEN-DR dataset). These were acquired using Canon CR-2 AF nonmydriatic retinal cameras and have different resolutions (with width and height ranging approximately from 1500 px to 2600 px).
Images from both left and right eyes were included, which may be centered at the macula, at the optic disc or elsewhere.
Images were graded in 5 levels by a retinal specialist.
For 348 of these images, pixel-wise annotations of the lesions MAs, HEMs, CWSs, IRMAs, EXs, NVs, PHEMs and PFIB  are available.
\hl{MAs appear as reddish small and circular dots}~\citep{Zhang2010}, \hl{and may occur alone or in groups}~\citep{Mahendran2015}. 
\hl{HEMs are bright red spots and, contrarily to MAs, their shape and appearance has substantial variability}~\citep{SilKar2017}.
\hl{Exudates present a large variety of shapes, sizes and contrast}~\citep{Harangi2013}. \hl{EXs have a more yellowish and bright appearance and sharp edges, whereas CWSs are more pale yellow/white without well defined borders}~\citep{Prentasic2016}.
\hl{NVs are thin and weak, exhibiting irregular appearance such as tortuous patterns and loops}~\citep{Gupta2017}. 
Different tools were available for annotating the images, such as dots, circles, polygons and free-hand tool, as exemplified in Fig.~\ref{fig:annot_screendr}.  
The annotated lesions were grouped in the different grades to allow a direct correspondence with the predicted attention map for each grade. 
The R1 ground truth map includes only the annotated MAs; R2 contains only HEMs; R3 contains HEMs, CWSs and IRMAs; R4 includes NVs, PHEMs and PFIB. EXs were also annotated but not included in the grade maps since they are not directly evaluated in the grading process.

\subsubsection{Quality assessment datasets}
\label{subsec:datasets_qual}

Three different eye fundus images datasets with image quality information were used on uncertainty-related experiments. 

\textit{DR1:}
This dataset~\citep{Pires2012} is constituted by 5776 images with resolution 640$\times$480 px, captured using a Topcon TRC-50X mydriatic camera (45$^\circ$ FOV). From these, 1300 are labeled as being of good quality (do not contain blur and are centered on the macula) and 1392 have poor quality (blur).

\textit{DRIMDB:} The Diabetic Retinopathy Image Database (DRIMDB)~\citep{Sevik2014} is composed of 216  images divided into three classes: good (125), poor (69), and outlier (22). The images were captured by a Canon CF-60UVi fundus camera (FOV of 60$^\circ$), and have a resolution of 570$\times$760 px. Images were annotated by an expert, and good-quality images are suitable for use in medical diagnosis by an ophthalmologist.

\textit{HRF:} The High-Resolution Fundus (HRF) database~\citep{Kohler2013}\footnote{\url{https://www5.cs.fau.de/research/data/fundus-images/}} contains 18 pairs of images from the same eye from different subjects, captured using a Canon CR-1 fundus camera (FOV of 45$^\circ$). In each pair there is a poor and a good quality image. 
These poor quality images have decreased sharpness, which can be local or global, due, for instance, to a defocused camera. Other quality features, such as image contrast or illumination conditions, are not contemplated.

\subsection{Evaluation}

\subsubsection{DR grading}
The DR grading performance of \model{} is evaluated using the Cohen's quadratic weighted Kappa ($\qwk{}$), used for measuring inter-rating agreement between raters in ordinal multi-class problems. This metric 
penalizes discrepancies between ratings, in a manner that depends quadratically on the distance between the prediction and the ground-truth, as follows:
\begin{equation}
\qwk{} = 1 - \frac{\sum_i^C \sum_j^C w_{i,j}O_{i,j}}{\sum_i^C \sum_j^C w_{i,j}E_{i,j}}
\label{eq:qwk}
\end{equation}
\noindent where $C$ is the number of classes, $w$ is the weight matrix, $O$ is the observed matrix and $E$ the expected matrix. $w_{i,j}$ is the weight penalization for $i,j$ element, which is given by: $\frac{(i-j)^2}{(C-1)^2}$. $O_{i,j}$ is the number of samples from the $j^{th}$ class (ground truth) which are classified by the model as being from the $i^{th}$ class. $E_{i,j}$ is the outer product between the prediction and ground truth classification histogram vectors, normalized such that $\sum_{i,j}{E_{i,j}} = \sum_{i,j}{O_{i,j}}$.
$\qwk{}$ values range from $-1$ (complete disagreement) to 1 (perfect agreement). 

However, $\qwk{}$ alone does not completely reflect the performance of a DR grading algorithm. When dealing with a very unbalanced problem, as DR grading (Fig.~\ref{fig:datasets_classes}), $\qwk{}$ is dominated by the most represented classes. Using confusion matrices allows to perform grade-wise evaluation and thus understand better the algorithm's behaviour.
We computed normalized versions of the confusion matrices by dividing each value of the original matrix by the sum of the values of each row. 

\paragraph{MIL assumption validation}
\hl{Preliminary experiments were performed to test whether the MIL assumption hampered the DR grading performance of the network. This was done by training a network in which the MP layer of} \model{} (Fig.~\ref{fig:architecture}) \hl{is replaced by a fully connected (FC) layer, keeping the remaining network as it was. No significant performance difference was found between the non-MIL and MIL models, with the first tending to present better R0 detection but a higher confusion among the disease cases. Because of this, no further experiments for this model were performed.} 

\subsubsection{Uncertainty estimation}

Experiments were performed in order to assess how the prediction performance relates with the uncertainty estimation of those predictions. In order to test whether low uncertainty images are indeed associated with a higher performance, we defined a set of thresholds on the uncertainty values and, for each threshold, select the subset of images with estimated uncertainty below that threshold. We then compute the $\qwk{}$ of our model predictions on these subsets.  
Since very low thresholds may lead to the inclusion of a number of images which do not truly represent the dataset, and the $\qwk{}$ computation for these few samples would not be reliable, the uncertainty thresholds start at 0.15 to ensure an acceptable image subset size. \hl{This threshold was determined considering the dataset with the smaller number of images (SCREEN-DR), for which an uncertainty threshold of 0.15 lead to the inclusion of ~5\% of the dataset, corresponding to 48 images.} \hl{Furthermore, cumulative histograms of the class-wise number of images as function of the uncertainty were produced. This allows to exclude the possibility of variations in} $\qwk{}$ \hl{being influenced by the different proportion of images per grade instead of being related with the uncertainty.}

\hl{Finally, matrices of the estimated uncertainty were computed by averaging the} \model's \hl{produced uncertainties for all images from each position of the confusion matrix (predicted grade \textit{vs} ground truth grade).}

\paragraph{Sensitivity analysis}
\hl{A sensitivity analysis was performed to assess the image quality's influence in the uncertainty estimation. Namely, the influence of image blur was studied by smoothing the input image with Gaussian filters of increasing standard deviation values and inspecting the effect on} \model's \hl{uncertainty estimation. Likewise, contrast was studied since lower contrasts are also characteristic of bad quality images. For that, the image's contrast was decreased via linear histogram stretching, by setting the image's maximum intensity to a progressively lower value, and evaluating the influence on the prediction's uncertainty.}

\subsubsection{Explainability}
\label{sec:exp_att_maps}

With the purpose of evaluating the explanation map, a predicted object $e_c$ corresponds to each connected component from the grade-wise explanation $\mathcal{E}_c$ after applying a fixed probability threshold. The explanation was quantitatively evaluated as follows:
\begin{inparaenum}[1)]
\item percentage of all $e_c$ that overlap with the ground truth map of the corresponding grade ($O_{obj\_g}$);
\item percentage of all $e_c$ that overlap with the ground truth map of the corresponding grade or lower ($O_{obj}$);
\item percentage of correctly predicted images for which the maximum activation object overlaps with the ground truth map of the corresponding grade ($O_{max}$); 
\item percentage of correctly predicted images for which at least one $e_c$ overlaps with the ground truth map of the corresponding grade ($O_{class}$);
\item percentage of ground truth objects that overlap with the predicted map of the corresponding grade or higher ($O_{gt}$);
\item percentage of all $e_c$ that overlap with the ground truth map of any grade. ($O_{any}$).
\end{inparaenum}
Note that the evaluations 2) and 5) are justified by the max-pooling assumption made in our pipeline: it is acceptable to assume that if a given region is detected as being from grade 2, for instance, it may also contain signs from a lower grade which are not highlighted since only the higher grade prevails after the max-pooling.

\subsubsection{Statistical tests}
\label{subsec:stats}

Statistical tests were performed to assess significant differences between the uncertainties inferred for different datasets. Since these data samples did not follow a normal distribution, which is an assumption on most parametric statistical hypothesis tests, the non-parametric Kruskal-Wallis H Test was used~\citep{Kruskal2012}. 
Further analysing the \textit{p-value} may not be enough when one is dealing with sufficiently large samples, since a statistical test will indicate a significant difference in most of the cases~\citep{sullivan2012}, \textit{i.e.} very large sample sizes may lead to the detection of differences that are quite small and possibly non relevant. It is thus important to report the \textit{p-value} results but also the effect size, which measures the magnitude of the difference between groups. 
The effect size is defined as $ES = \mu_1 - \mu_2$, where $\mu_1$ and $\mu_2$ are the averages of the two groups. 
The Cohen's $d$ was used as an effect size measure, and can be defined as follows:
\begin{equation}
    d = ES/S
\end{equation}
where $S$, the pooled standard deviation, is given by:
\begin{equation}
    S = \sqrt{\frac{(n_1-1)\times s_1^2+(n_2-1)\times s_2^2}{n_1+n_2-2}}
\end{equation}
where $n_1$ and $n_2$ are the sample sizes for the two groups and $s_1$ and $s_2$ are the standard deviations for the two groups.

The interpretation of $d$, considering the thresholds proposed by Cohen and revised by~\cite{Sawilowsky2009}, can be established as follows:
\begin{inparaenum}[1)]
    \item $d=0.1$: very small
    \item $d=0.2$: small
    \item $d=0.5$: medium
    \item $d=0.8$: large
    \item $d=1.3$: very large
    \item $d=2$: huge.
\end{inparaenum}

Having into account that our datasets are commonly on the order of more than 1000 samples, the effect size is reported together with the significance level. Since the significant difference will most likely be achieved in these datasets, $d$ provides a measure of the magnitude of these differences.

\subsection{Parameter setting}

The input size to the network (Fig.~\ref{fig:architecture}) is 640$\times640\times3$, and the side of the output feature map, $N$, is 20.
The receptive field, $RF$ of the network is of 156, and the stride, $s$, is of 32. 
Regarding the attention maps generation, the $\sigma'$ of $N_{2,\sigma'}$ from Eq.\ref{eq:gaussian_maps} was set to $RF$/6, since it allows to fit the majority of the Gaussian distribution in the kernel window.
The weight $\alpha$ in the loss (Eq.\ref{eq:our_loss}) was set to 0.7, in order to preserve the greater importance of the log-loss part but still impose some level of regularization. 
Regarding the class balancing, we chose $r$ = 0.99 and $w_f= (0.5,2,2,3,3)$ ($f=300$) in Eq.~\ref{eq:class_balance}, aiming at not overfitting the original dataset distribution and to increase the generalization capability of the network.


\subsection{Training details}

Images were cropped around the FOV and resized to the input size of the network, 640$\times$ 640 pixels. 
The model was trained for 240 epochs with the \kaggle{} training set, using a batch size of 30, the Adam optimizer, and a learning rate of 3$\times10^{-4}$. 
Experiments were performed on an Intel Core i7-5960X, 32Gb RAM, $2\times$GTX1080 desktop with Python 3.5, Keras 2.2 and TensorFlow 1.8. 

\section{Results}

\subsection{DR grading}


\begin{table*}[tb]
\centering
\caption{Quadratic weighted kappa for \model's{} DR grading in four different datasets. $\triangleleft$ 7025 images from the \kaggle{} training dataset, * 7000 images from the \kaggle{} training dataset, $\star$ 4 classes, $\diamond$ 496 out of the 9443 images. \label{tab:qwk}}
\begin{tabular}{ll|l|l|l|l|l}
& \begin{tabular}[c]{@{}l@{}}Kaggle test \\ \end{tabular} & \begin{tabular}[c]{@{}l@{}}Kaggle \\  \end{tabular} & Messidor-2 & IDRID & SCREEN-DR & DMR$\star$ \\ \hline

\multicolumn{1}{l|}{\model{} (proposed)}                                     & 0.74                                                   & 0.75$\triangleleft$                                   & 0.71     & 0.84  & 0.74      & 0.78  \\

\multicolumn{1}{l|}{\cite{DelaTorre2018}}        & 0.74                                                                                         & -                                                             & -        & -     & -     & -      \\

\multicolumn{1}{l|}{\cite{Gonzalez-gonzalo2018}} & -                                                      & \begin{tabular}[c]{@{}l@{}}0.72* \end{tabular}                                                             & -        & -     & -        & -   \\

\multicolumn{1}{l|}{\cite{Takahashi2017}}        & -                                                                                                & -                                                             & -        & -     & -     &  0.79$\diamond$  

\end{tabular}

\end{table*}



\begin{figure*}[t]
\centering
\begin{subfigure}{0.185\textwidth}
\includegraphics[width=\textwidth]{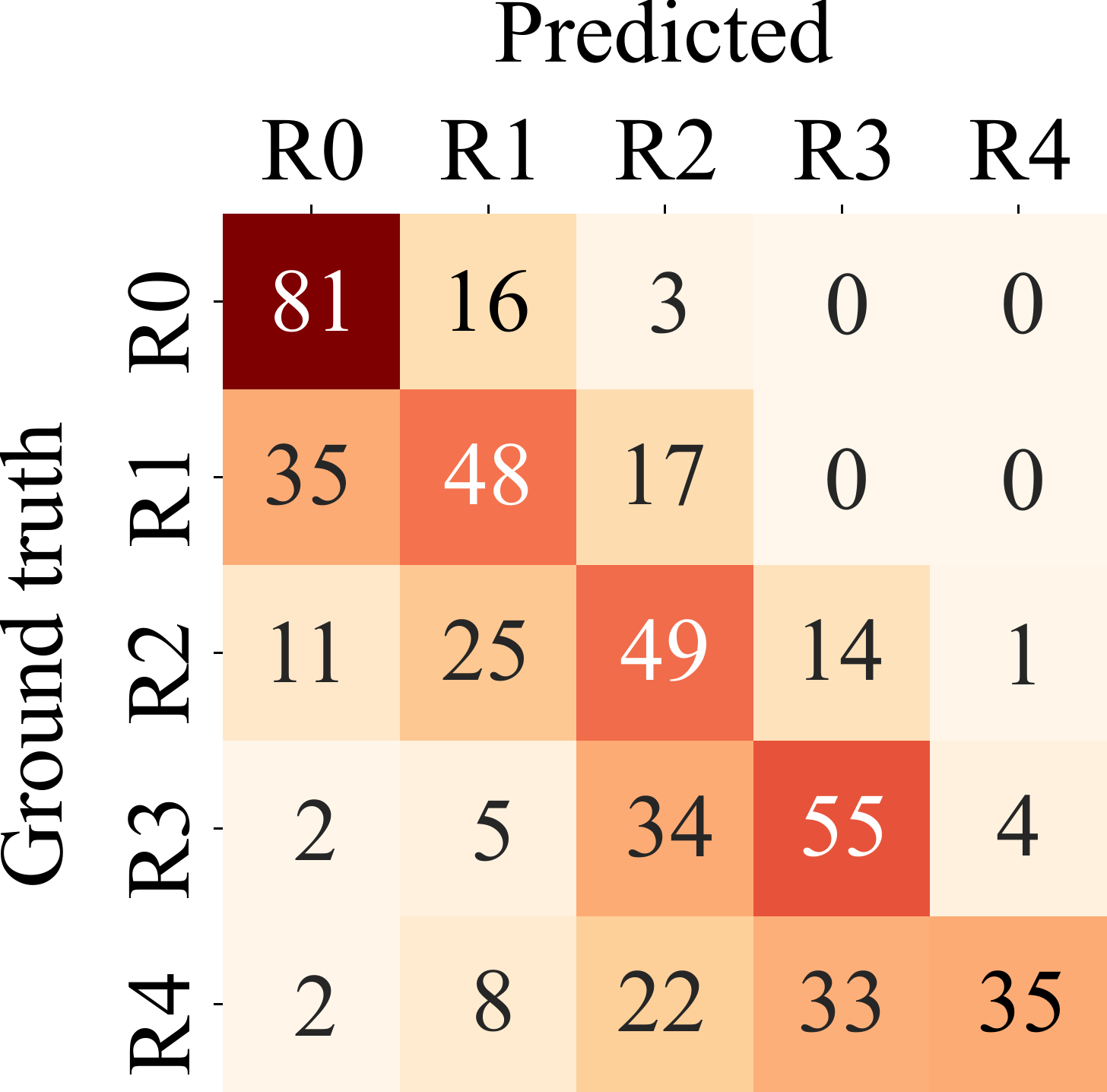}
\caption{\kaggle{} test \label{fig:conf_Kaggle_test}}
\end{subfigure}
\hfill
\begin{subfigure}{0.185\textwidth}
\includegraphics[width=\textwidth]{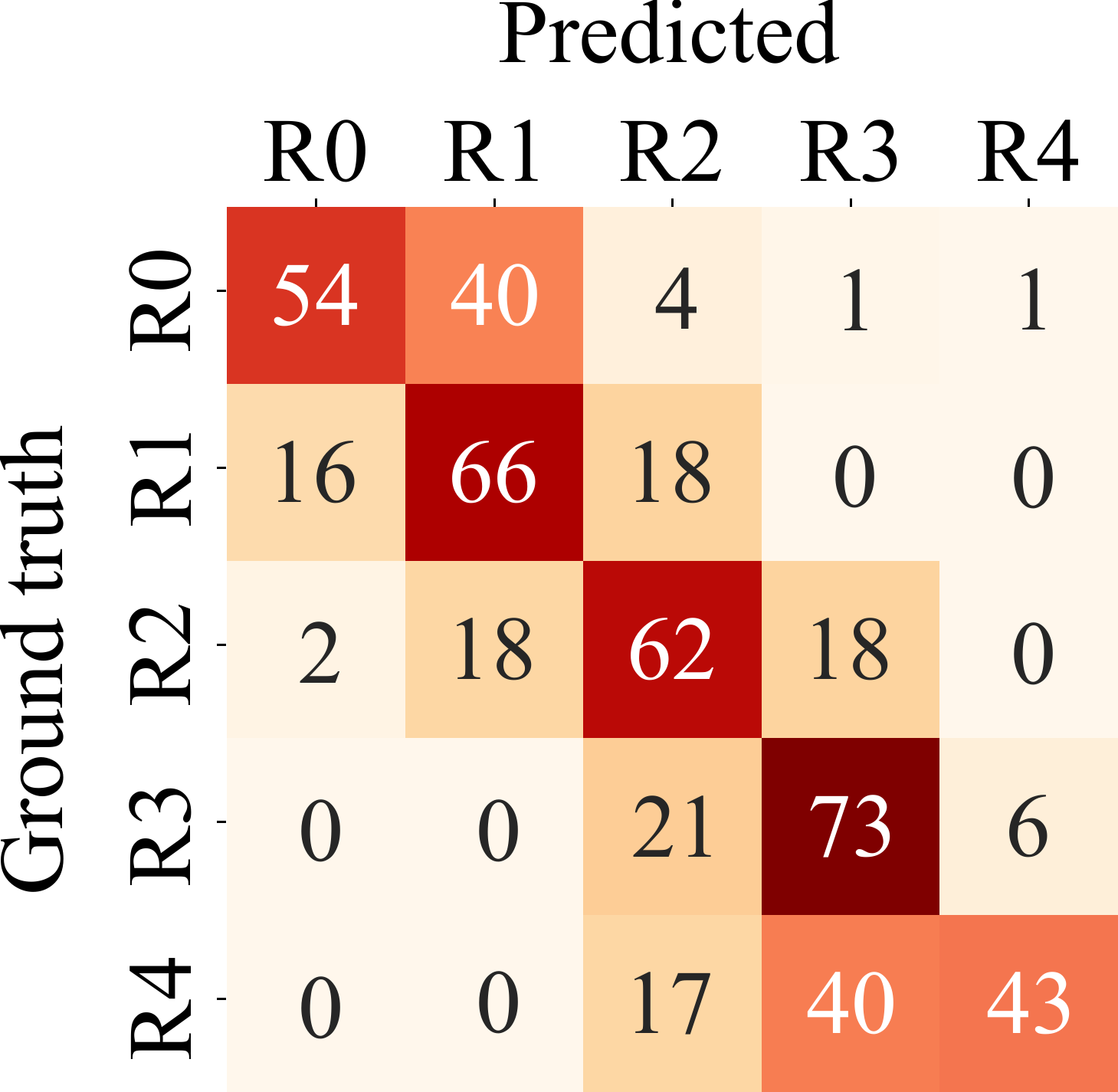}
\caption{Messidor-2 dataset \label{fig:conf_messidor}}
\end{subfigure}
\hfill
\begin{subfigure}{0.185\textwidth}
\includegraphics[width=\textwidth]{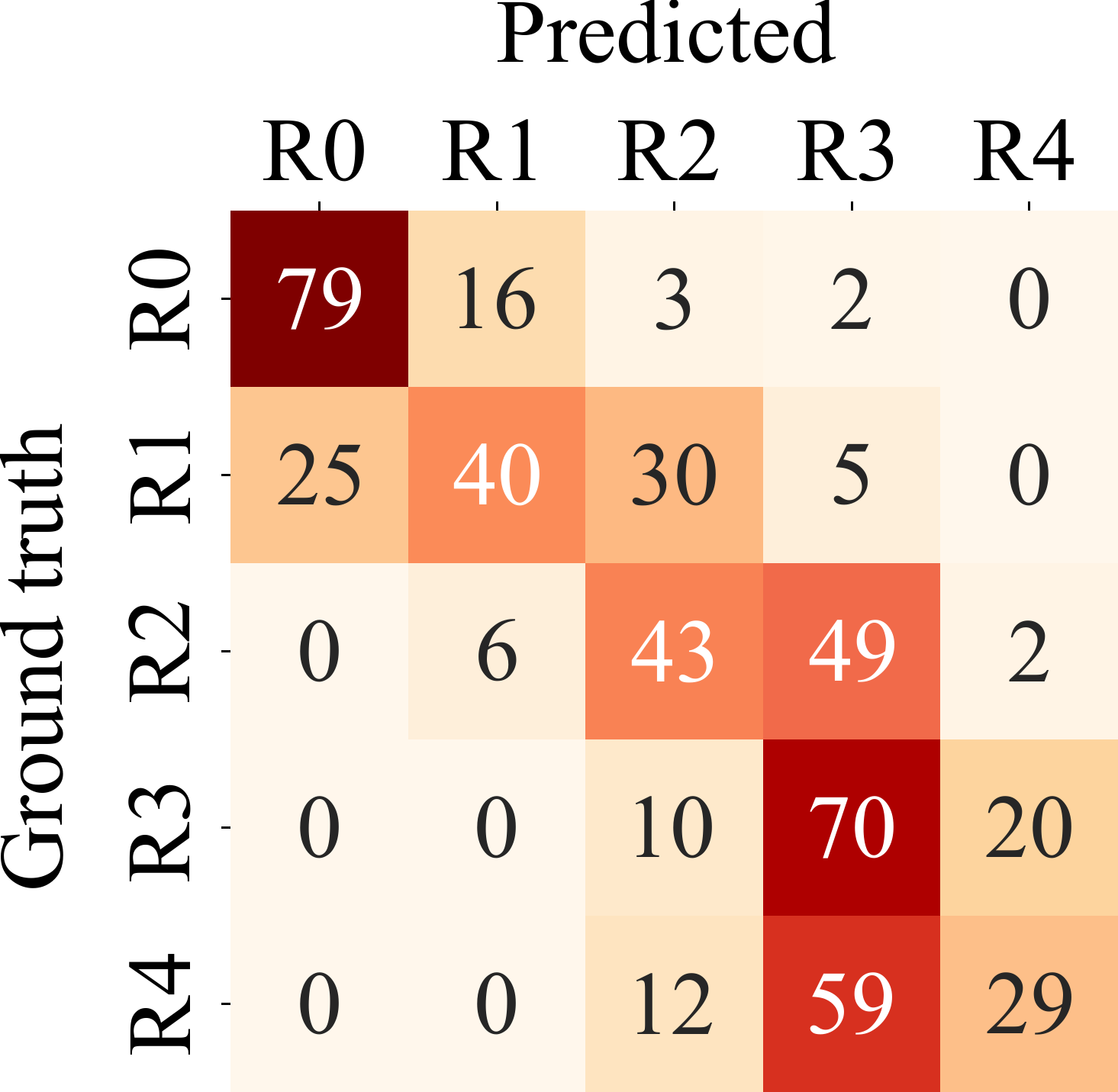}
\caption{IDRID dataset \label{fig:conf_idrid}}
\end{subfigure}
\hfill
\begin{subfigure}{0.185\textwidth}
\includegraphics[width=\textwidth]{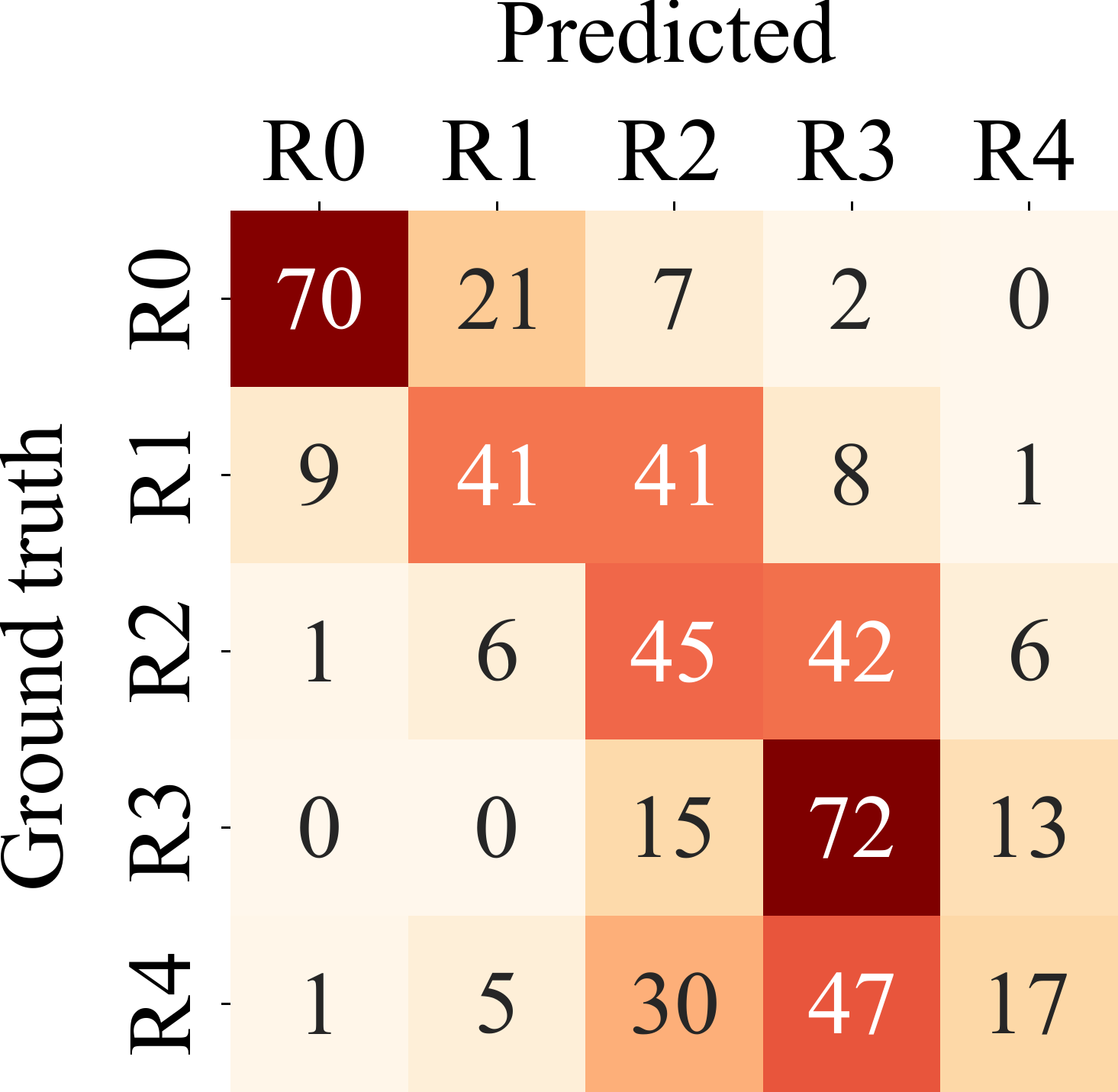}
\caption{SCREEN-DR dataset \label{fig:conf_screen}}
\end{subfigure}
\hfill
\begin{subfigure}{0.185\textwidth}
\includegraphics[width=\textwidth]{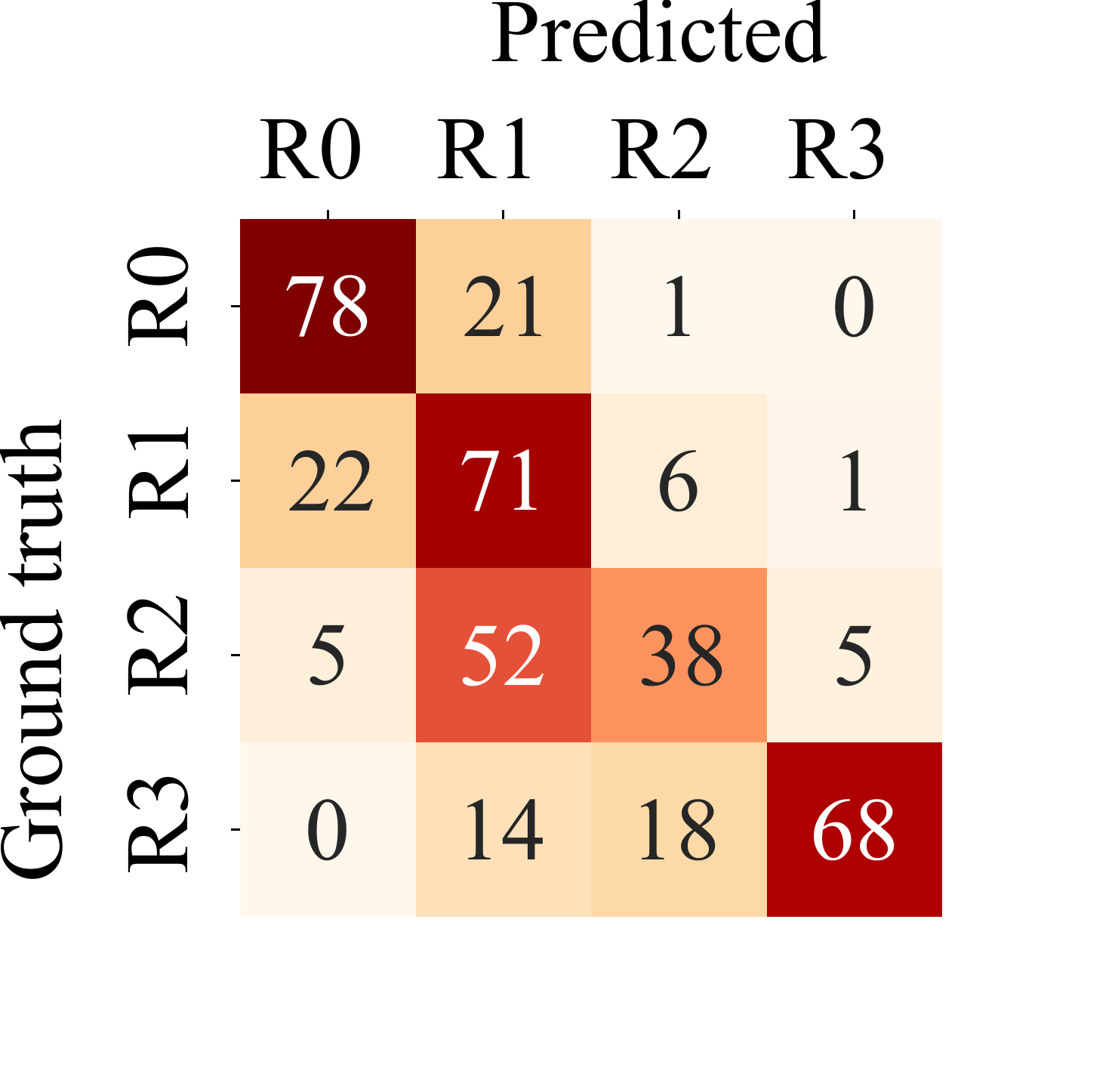}
\caption{DMR dataset \label{fig:conf_dmr}}
\end{subfigure}
\caption{Confusion matrixes of \model{}'s predictions \textit{vs} the ground truth DR grades for different test datasets. Matrixes were normalized by dividing each value of the original confusion matrix by the sum of the values from the corresponding row (total number of images from that grade in the dataset), \hl{and the shown values correspond to percentages}.\label{fig:conf_matrixes}}
\end{figure*}

\model{} achieves similar $\qwk{}$ to other state-of-the-art methods, as depicted in Table \ref{tab:qwk}. 
Note that the majority of the state-of-the-art methods did not test on different datasets, and thus an assessment of their generalization capabilities is not possible. \model{} was evaluated in different test datasets and the consistency of the results suggests that the system has properly learned discriminant features for DR grading. 
The grade-wise normalized confusion matrices \hl{(in percentage)} are shown in Fig.~\ref{fig:conf_matrixes}. In general, all the matrices have a diagonal tendency, \textit{i.e.}, the predictions are close to the ground truth. R4 is the most misclassified grade. For instance, for both the IDRID and SCREEN-DR datasets (Fig.~\ref{fig:conf_idrid}~and~\ref{fig:conf_screen}), the majority of the R4 images is confounded with R3. From the misclassifications reported in Fig.~\ref{fig:conf_matrixes}, the percentage of over-diagnosis is between 24$\%$ (\kaggle{} test set) and 55\% (SCREEN-DR) for all datasets, which is higher than the 21$\%$ expected for human observers~\citep{Krause2018}.
Consequently, the percentage of under-diagnosis is between 45\% (SCREEN-DR) and 76\% (\kaggle{} test set), being always lower than the 79\% reported for human observers.


\subsection{Uncertainty estimation}

\begin{figure}[htb]
\centering
\begin{subfigure}{0.5\textwidth}
\includegraphics[width=\textwidth]{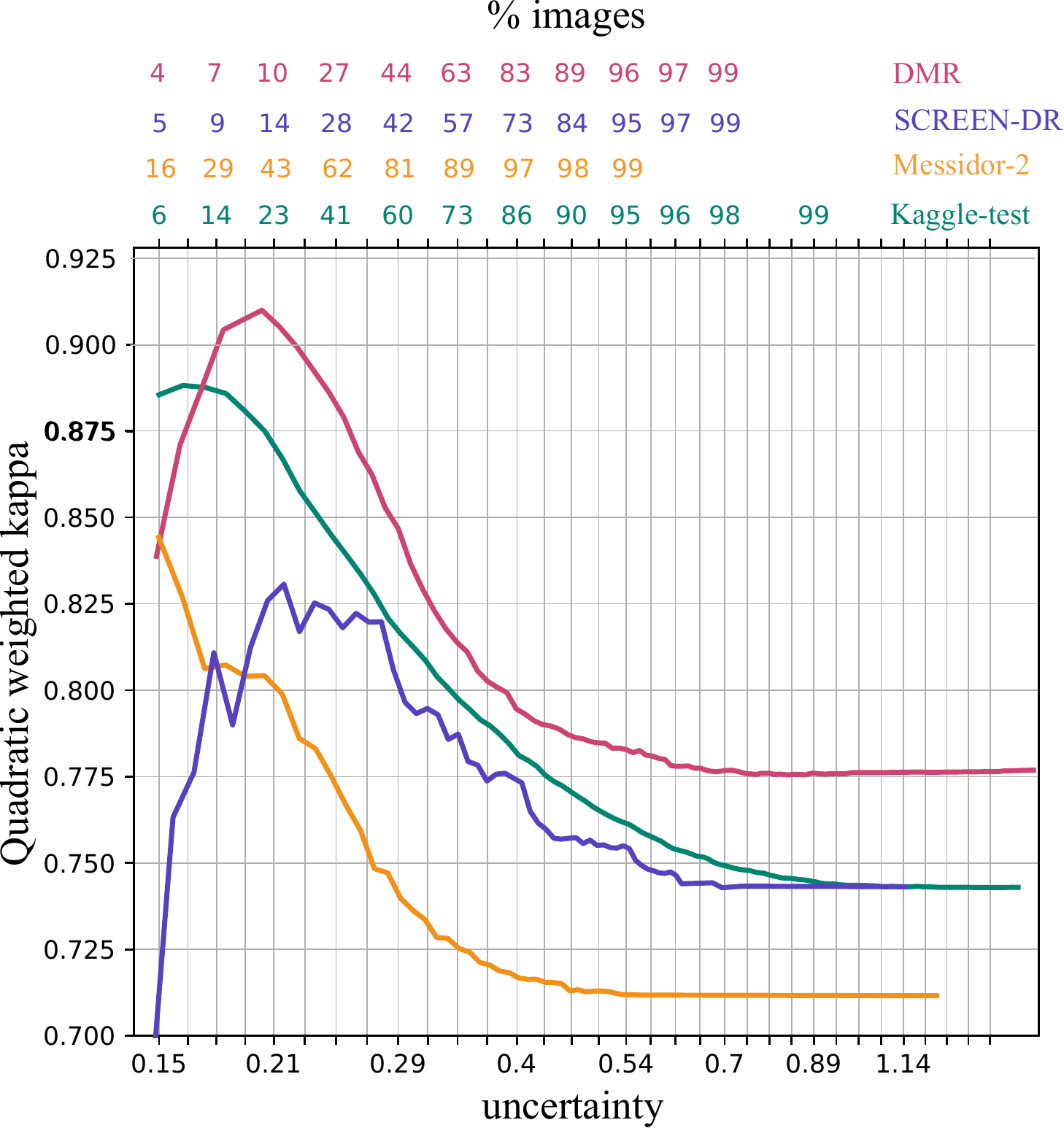}
\caption{Quadratic weighted Kappa \textit{vs} uncertainty ($xx$-axis in logarithmic scale) for four different datasets (DMR: 4 classes).\label{fig:unc_kappa_datasets} }
\end{subfigure}
\begin{subfigure}{0.28\textwidth}
\includegraphics[width=\textwidth]{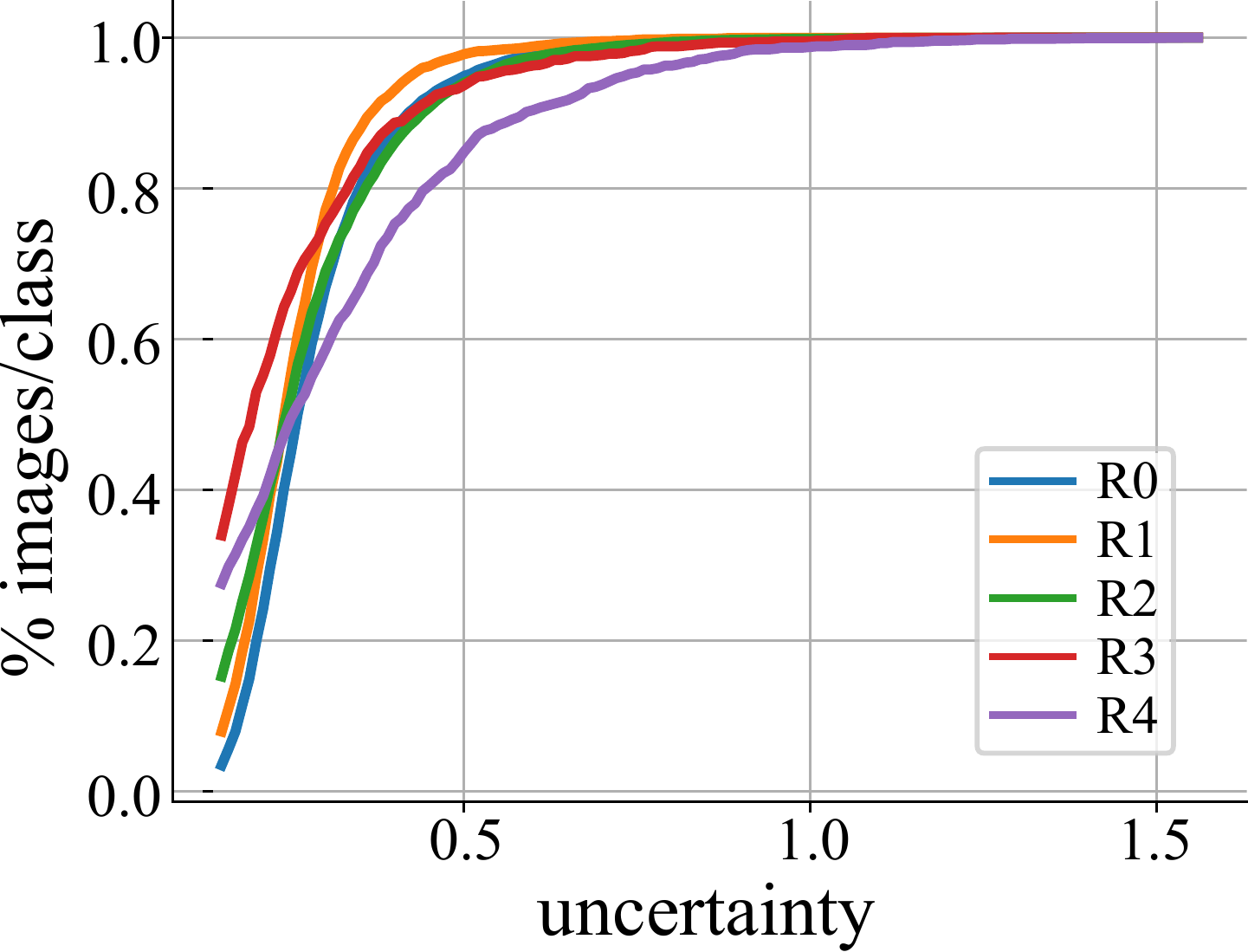}
\caption{Cumulative percentage of images from each grade included per uncertainty threshold, for the \kaggle{} test set. \label{fig:unc_kappa_grade} }
\end{subfigure}
\hfill
\begin{subfigure}{0.19\textwidth}
\includegraphics[width=\textwidth]{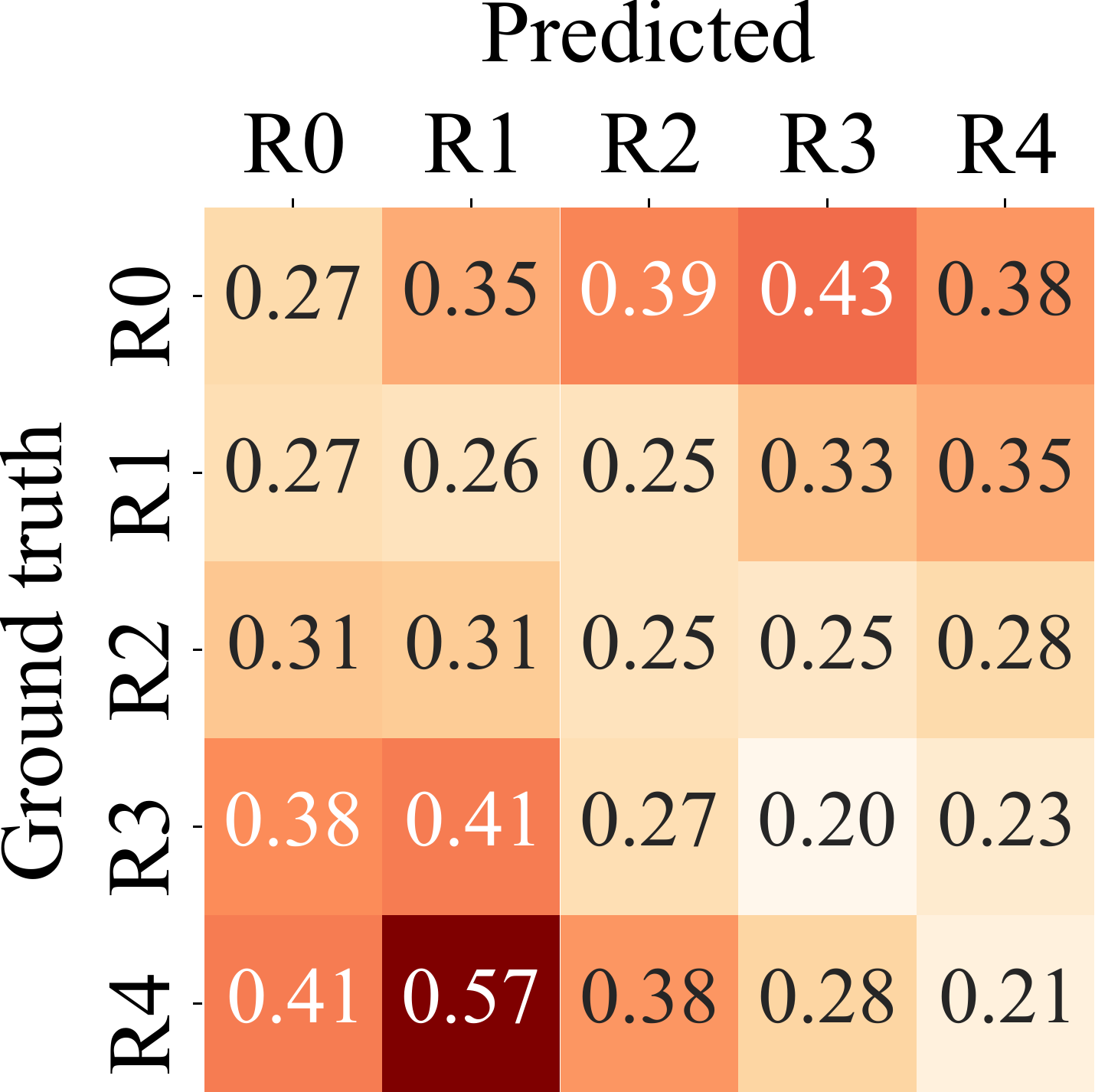}
\vspace*{0.2mm}
\caption{Average predicted uncertainty matrix for the \kaggle{} test set. \label{fig:unc_matrix}}
\end{subfigure}
\caption{Uncertainty of the predictions in relation with the \model{}'s performance. \label{fig:unc_kappa}}
\end{figure}

\begin{figure*}[tb]
\centering
\begin{subfigure}{0.16\textwidth}
\includegraphics[width=\textwidth]{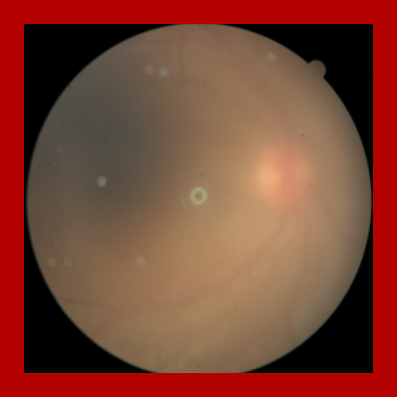}
\caption{$\hat{y}_g=1, y=2$.\label{fig:unc_14_p1}}
\end{subfigure}
\hfill
\begin{subfigure}{0.16\textwidth}
\includegraphics[width=\textwidth]{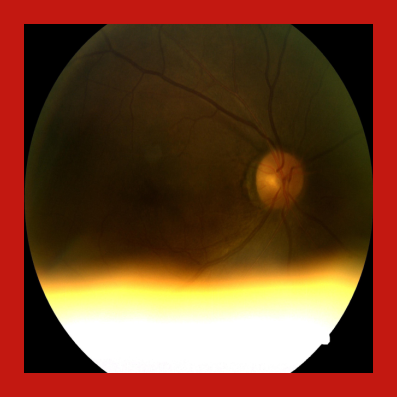}
\caption{$\hat{y}_g=0, y=0$. \label{fig:unc_13_p0}}
\end{subfigure}
\hfill
\begin{subfigure}{0.16\textwidth}
\includegraphics[width=\textwidth]{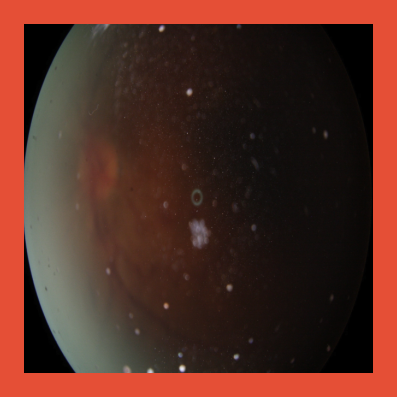}
\caption{$\hat{y}_g=2, y=3$. \label{fig:unc_1_p2}}
\end{subfigure}
\hfill
\begin{subfigure}{0.16\textwidth}
\includegraphics[width=\textwidth]{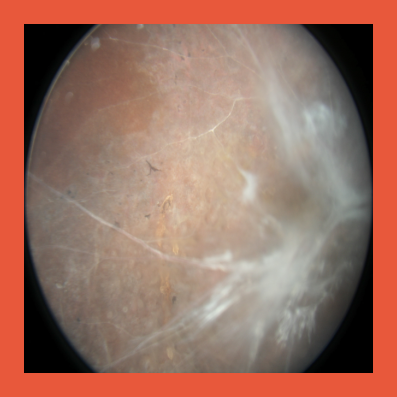}
\caption{$\hat{y}_g=4, y=4$. \label{fig:unc_1_p4}}
\end{subfigure}
\hfill
\begin{subfigure}{0.16\textwidth}
\includegraphics[width=\textwidth]{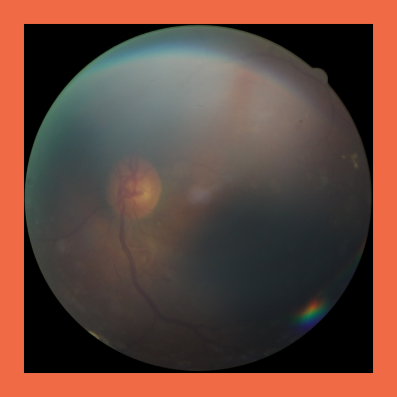}
\caption{$\hat{y}_g=2, y=4$. \label{fig:unc_09_p2}} 
\end{subfigure}
\hfill
\begin{subfigure}{0.16\textwidth}
\includegraphics[width=\textwidth]{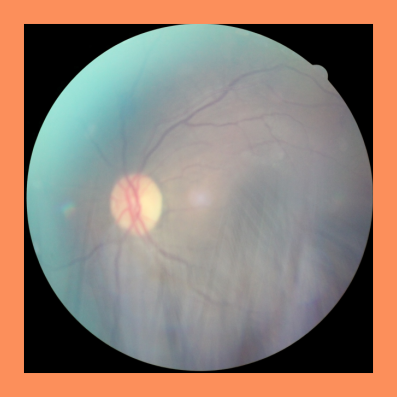}
\caption{$\hat{y}_g=0, y=2$. \label{fig:unc_07_p0}}
\end{subfigure}

\begin{subfigure}{0.16\textwidth}
\includegraphics[width=\textwidth]{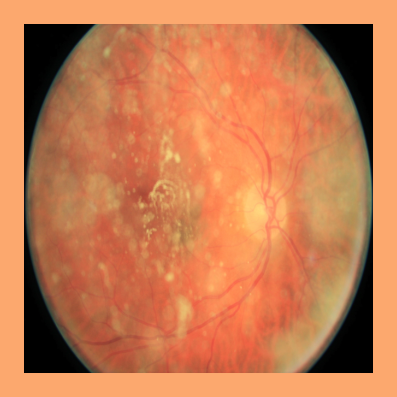} 
\caption{$\hat{y}_g=4, y=0$.\label{fig:unc_058_p4}} 
\end{subfigure}
\hfill
\begin{subfigure}{0.16\textwidth}
\includegraphics[width=\textwidth]{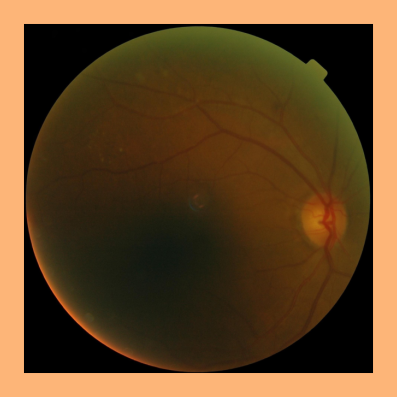}
\caption{$\hat{y}_g=2, y=2$.\label{fig:unc_05_p2}} 
\end{subfigure}
\hfill
\begin{subfigure}{0.16\textwidth}
\includegraphics[width=\textwidth]{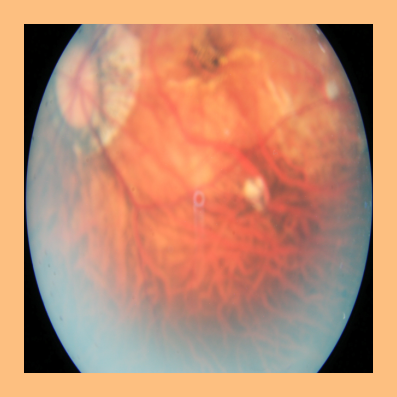}
\caption{$\hat{y}_g=3, y=0$. \label{fig:unc_045_p3}}
\end{subfigure}
\hfill
\begin{subfigure}{0.16\textwidth}
\includegraphics[width=\textwidth]{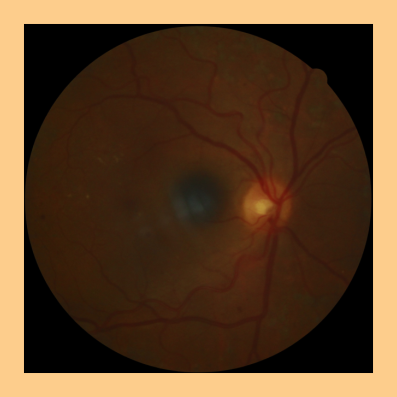}
\caption{$\hat{y}_g=3, y=4$. \label{fig:unc_037_p3}}
\end{subfigure}
\hfill
\begin{subfigure}{0.16\textwidth}
\includegraphics[width=\textwidth]{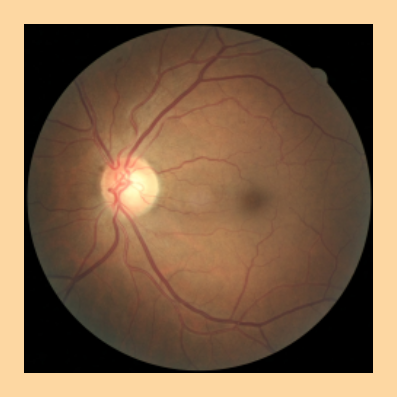}
\caption{$\hat{y}_g=1, y=1$. \label{fig:unc_029_p1}}
\end{subfigure}
\hfill
\begin{subfigure}{0.16\textwidth}
\includegraphics[width=\textwidth]{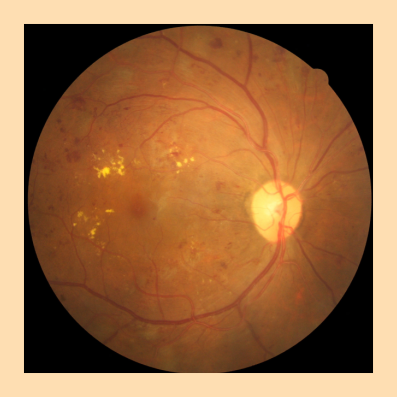}
\caption{$\hat{y}_g=3, y=3$. \label{fig:unc_022_p3}}
\end{subfigure}

\begin{subfigure}{0.16\textwidth}
\includegraphics[width=\textwidth]{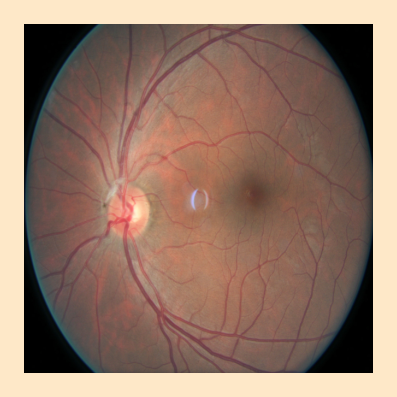}
\caption{$\hat{y}_g=2, y=2$. \label{fig:unc_012_p2}}
\end{subfigure}
\hfill
\begin{subfigure}{0.16\textwidth}
\includegraphics[width=\textwidth]{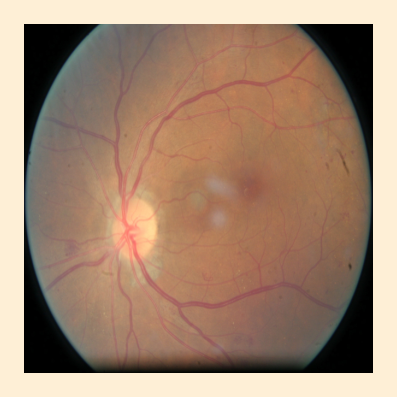}
\caption{$\hat{y}_g=0, y=0$. \label{fig:unc005_p0}}
\end{subfigure}
\hfill
\begin{subfigure}{0.16\textwidth}
\includegraphics[width=\textwidth]{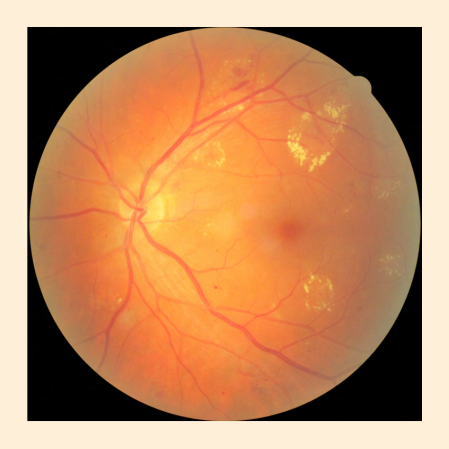}
\caption{$\hat{y}_g=2, y=2$. \label{fig:unc005_p2}}
\end{subfigure}
\hfill
\begin{subfigure}{0.16\textwidth}
\includegraphics[width=\textwidth]{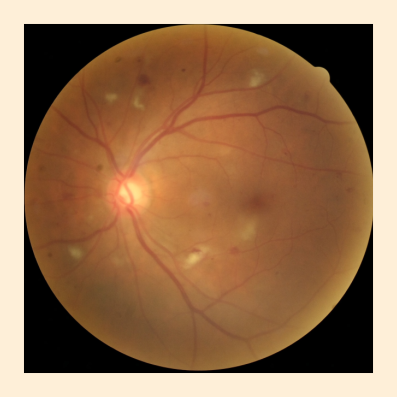}
\caption{$\hat{y}_g=3, y=2$. \label{fig:unc_005_p3}} 
\end{subfigure}
\hfill
\begin{subfigure}{0.16\textwidth}
\includegraphics[width=\textwidth]{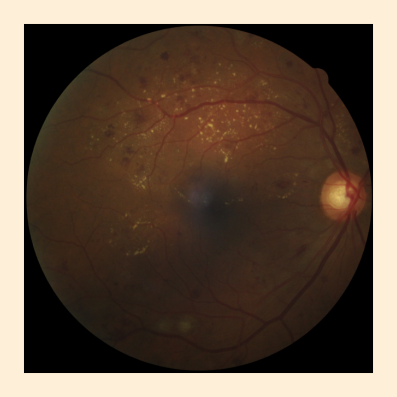}
\caption{$\hat{y}_g=3, y=3$. \label{fig:unc_005_p3_}} 
\end{subfigure}
\hfill
\begin{subfigure}{0.16\textwidth}
\includegraphics[width=\textwidth]{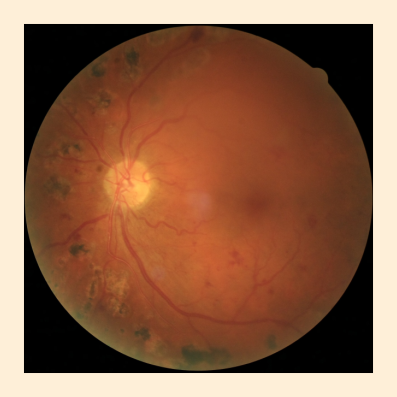}
\caption{$\hat{y}_g=4, y=4$. \label{fig:unc_005_p4}}
\end{subfigure}

\caption{Uncertainty-associated predictions for images from the \kaggle{} test set. The color of the frame represents the uncertainty, $\hat{y}_g$ is the prediction and $y$ the ground truth grade. Images were cropped around the field-of-view and resized to a square image. Uncertainty colorbar: $0.05$~\protect\includegraphics[height=.75em,width=5em]{figures/uncertainty_gradient_bar.pdf}~$1.406$ \label{fig:unc}}
\end{figure*}

\begin{figure*}

\begin{subfigure}{0.4\textwidth} 
\includegraphics[width=\textwidth]{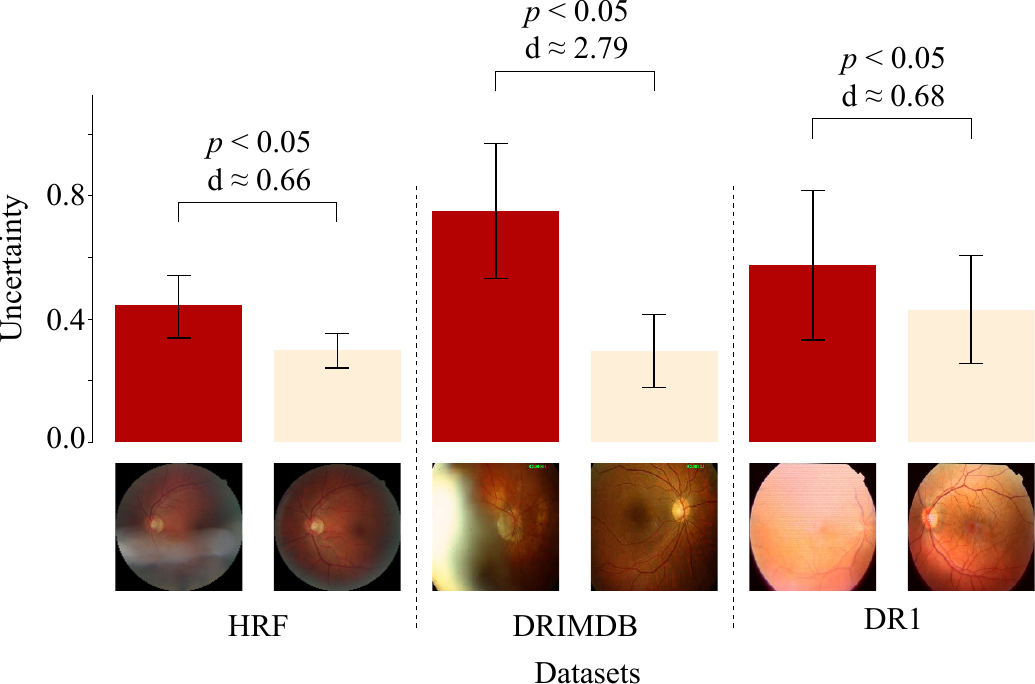}
\caption{Uncertainties predicted by \model{} for good and bad quality images from public datasets. $p$-values obtained through the Kruskal-Wallis H Test and Cohen's $d$ between each good and bad sets from a given dataset are presented. \label{fig:bad_good_qual}} 
\end{subfigure}
\hfill
\begin{subfigure}{0.586\textwidth} 
\includegraphics[width=\textwidth]{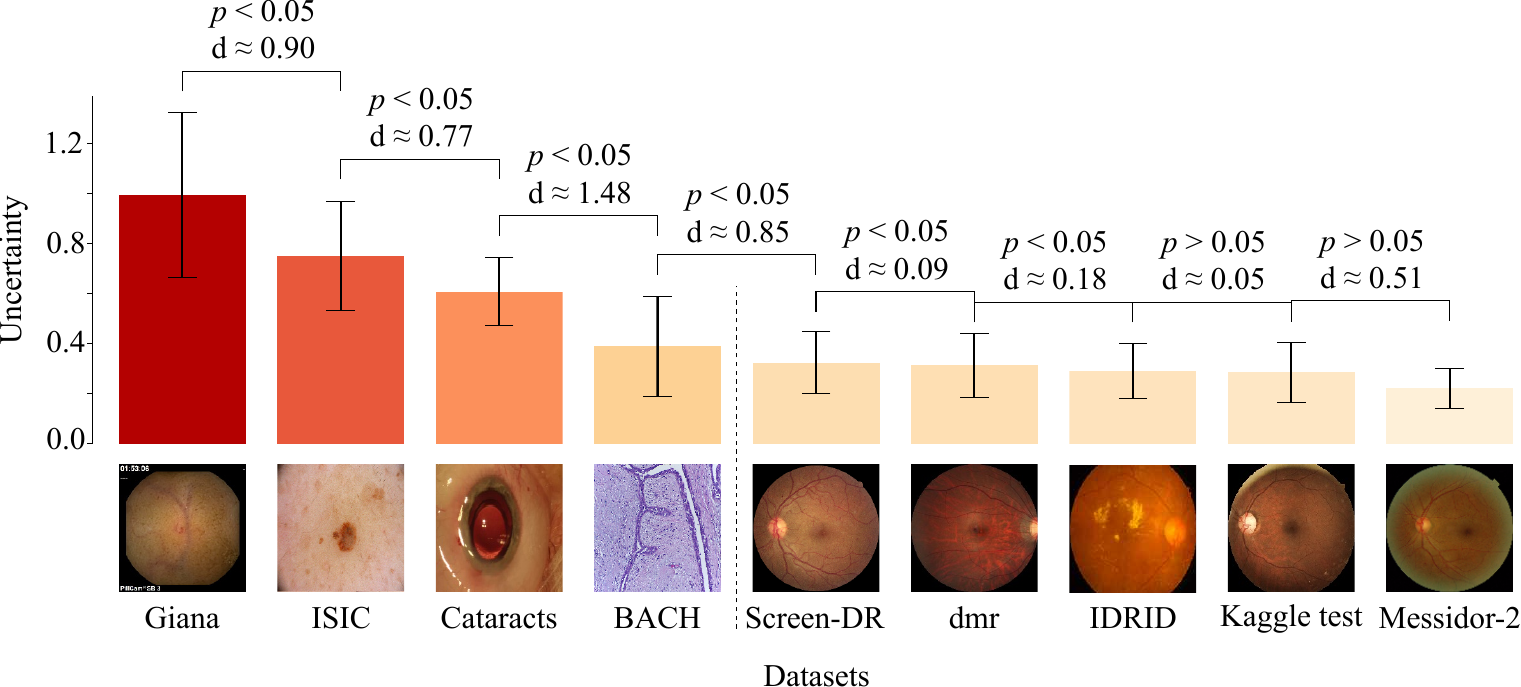}
\caption{Uncertainties predicted by \model{} for unfamiliar data types from different medical image datasets and for the studied eye fundus images datasets. $p$-values obtained through the Kruskal-Wallis H Test and Cohen's $d$ between each dataset and its right neighbour on the plot are presented.\label{fig:unfamiliar_data}} 
\end{subfigure}

\caption{Uncertainty prediction for different types of tasks.}
\label{fig:my_label}
\end{figure*}


The evolution of $\qwk{}$ with the uncertainty level threshold for four different datasets
is shown in Fig.~\ref{fig:unc_kappa_datasets}. For the \kaggle{} test set, a $\qwk{}$ of approximately 0.82 is achieved for the 50\% lowest uncertainty images. Likewise, considering the 75$\%$ lowest uncertainty predictions, $\qwk{}$ is of approximately 0.8. 
The average uncertainty matrix is shown in Fig.~\ref{fig:unc_matrix}, Fig.~\ref{fig:unc_kappa_grade} depicts the grade-wise normalized cumulative histogram as function of the inferred uncertainty. 
Finally, Fig.~\ref{fig:unc} shows examples of \kaggle{} test set images along with \model{}'s uncertainties of the predictions.

The uncertainty estimation results for three quality-labeled datasets (section~\ref{subsec:datasets_qual}) are shown in Fig.~\ref{fig:bad_good_qual}. For all datasets, the inferred uncertainty is statistically higher for the bad quality subsets ($p$-$value<0.05$), and the respective Cohen's $d$ is huge for the DRIMDB and medium/large for the other datasets.

Fig.~\ref{fig:unfamiliar_data} shows the uncertainty behavior of \model{} for datasets of unfamiliar data types along with the eye fundus image datasets used for evaluation.
Between the unfamiliar data types the statistical hypothesis tests indicate that, for each dataset, the results are statistically different ($p$-$value~<~0.05$) from the neighbouring dataset, and the Cohen's $d$ values suggest a large to very large magnitude of the differences. 
From the BACH to the SCREEN-DR dataset a statistical difference is found, with Cohen's $d$ indicating a medium magnitude of the differences.
Regarding the eye fundus images, some datasets show a statistically different average uncertainty. However, in these cases, Cohen's $d$ values indicate a small to very small magnitude of the differences.

\subsubsection{Sensitivity analysis}

\begin{figure}[tb]
\centering
\begin{subfigure}{0.11\textwidth}
\includegraphics[width=\textwidth]{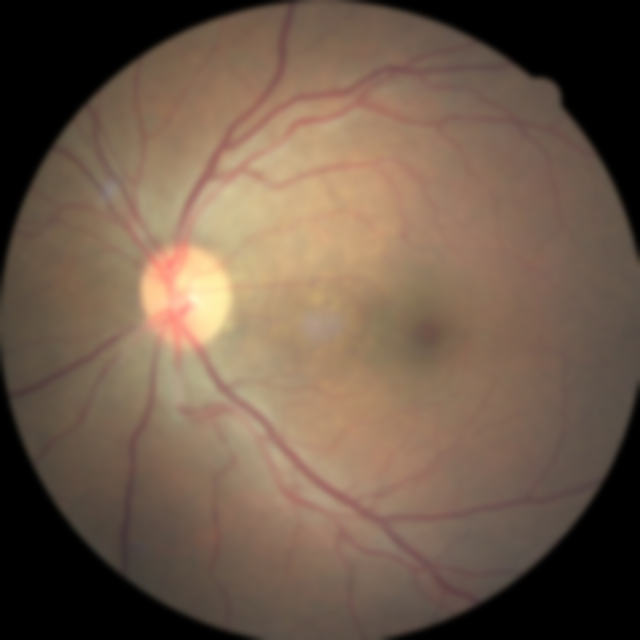}
\caption{$std = 3$}
\end{subfigure}
\hfill
\begin{subfigure}{0.11\textwidth}
\includegraphics[width=\textwidth]{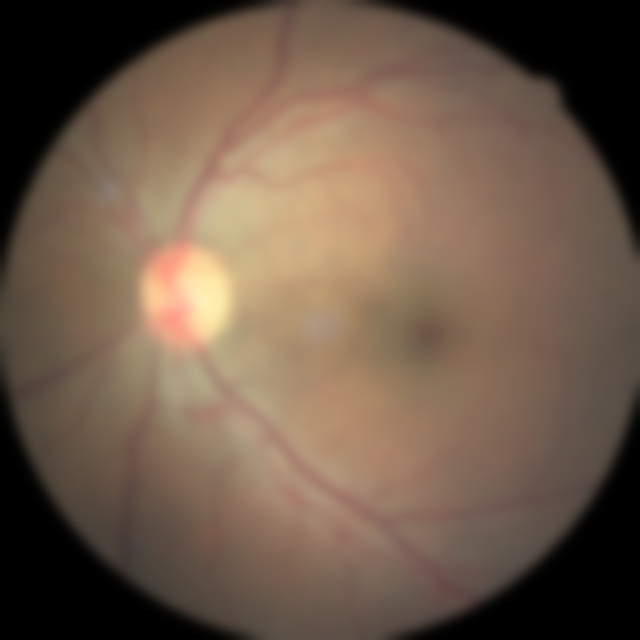}
\caption{$std = 6$}
\end{subfigure}
\hfill
\begin{subfigure}{0.11\textwidth}
\includegraphics[width=\textwidth]{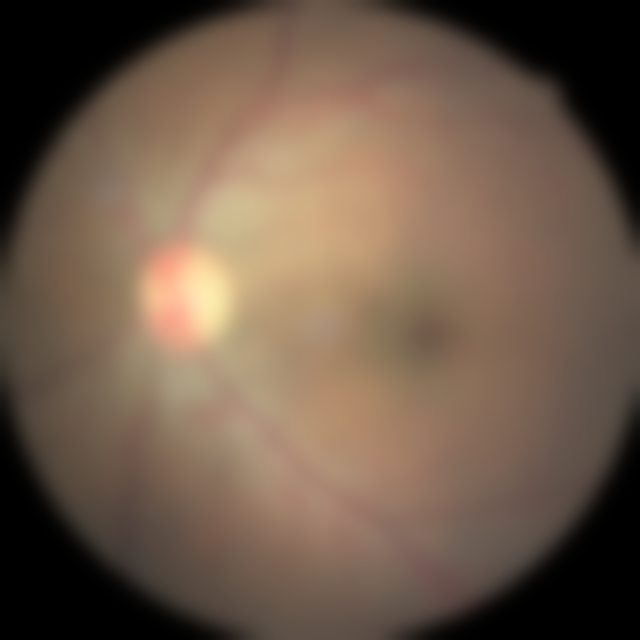}
\caption{$std = 10$}
\end{subfigure}
\hfill
\begin{subfigure}{0.11\textwidth}
\includegraphics[width=\textwidth]{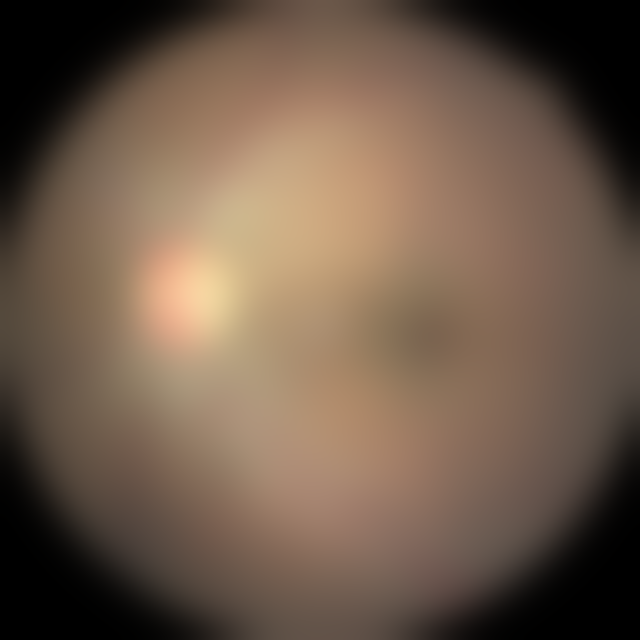}
\caption{$std = 20$}
\end{subfigure}
\caption{\hl{Examples of convolution with Gaussian filters of increasing standard deviation ($std$)}. \label{fig:sens_gauss}}
\end{figure}

\begin{figure}[tb]
\centering
\includegraphics[width=0.3\textwidth]{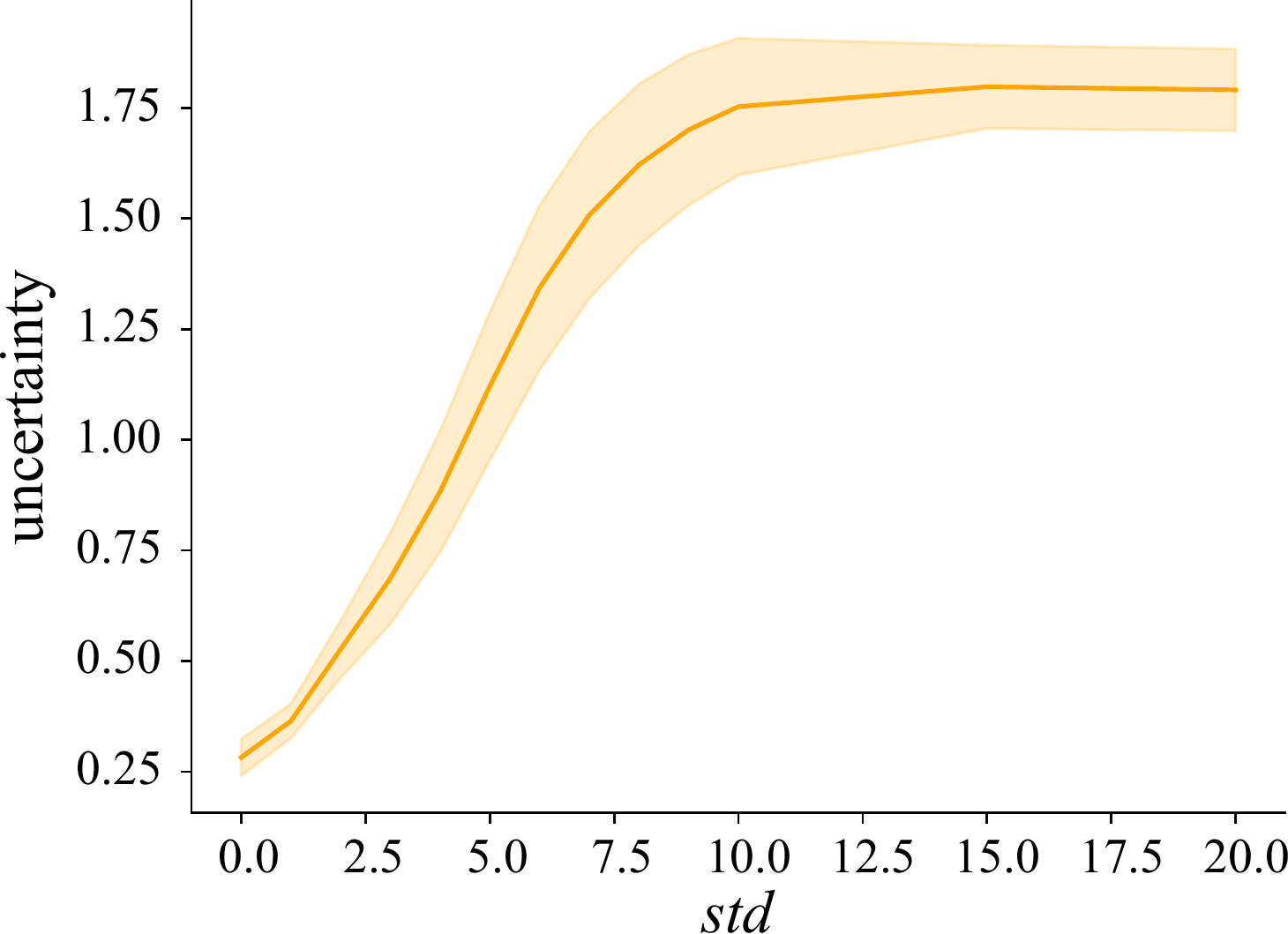} 
\caption{\hl{Uncertainty estimation of} \model{} \hl{for images resulting from convolution with Gaussian filters of increasing standard deviation} ($std$). \hl{The uncertainty values are the average of the uncertainties for 5 datasets} (\kaggle{} \hl{test set, Messidor-2, IDRID, SCREEN-DR and DMR) considering a 95$\%$ confidence interval.}
\label{fig:sens_plot}}
\end{figure}

 Fig.~\ref{fig:sens_gauss} \hl{shows examples of images blurred with the Gaussian filters of increasing standard deviation ($std$).} Fig.~\ref{fig:sens_plot} \hl{depicts the results obtained for this test in 5 different datasets} (\kaggle{} \hl{test set, Messidor-2, IDRID, SCREEN-DR and DMR), where the curve represents the average of the uncertainty values for each dataset, and the shaded area corresponds to a confidence interval of 95$\%$.} \hl{The sensitivity analysis via contrast adjustment did not suggest any relation between image contrast and uncertainty estimation.}

\subsection{Explainability}

\begin{figure*}[tb]
\centering
\begin{subfigure}{0.48\textwidth}
\includegraphics[width=\textwidth]{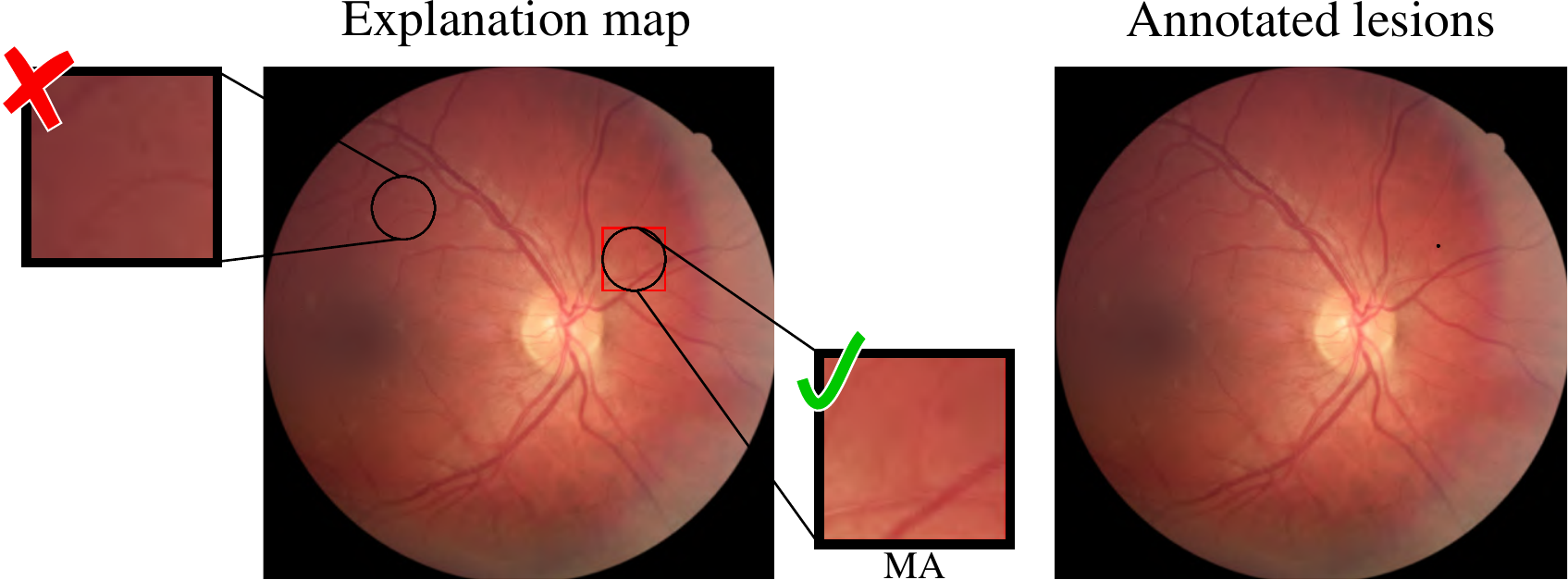}
\caption{$\hat{y}_g=1, y=1$.\label{fig:gt1_pred1}}
\end{subfigure}
\hfill
\begin{subfigure}{0.48\textwidth}
\includegraphics[width=\textwidth]{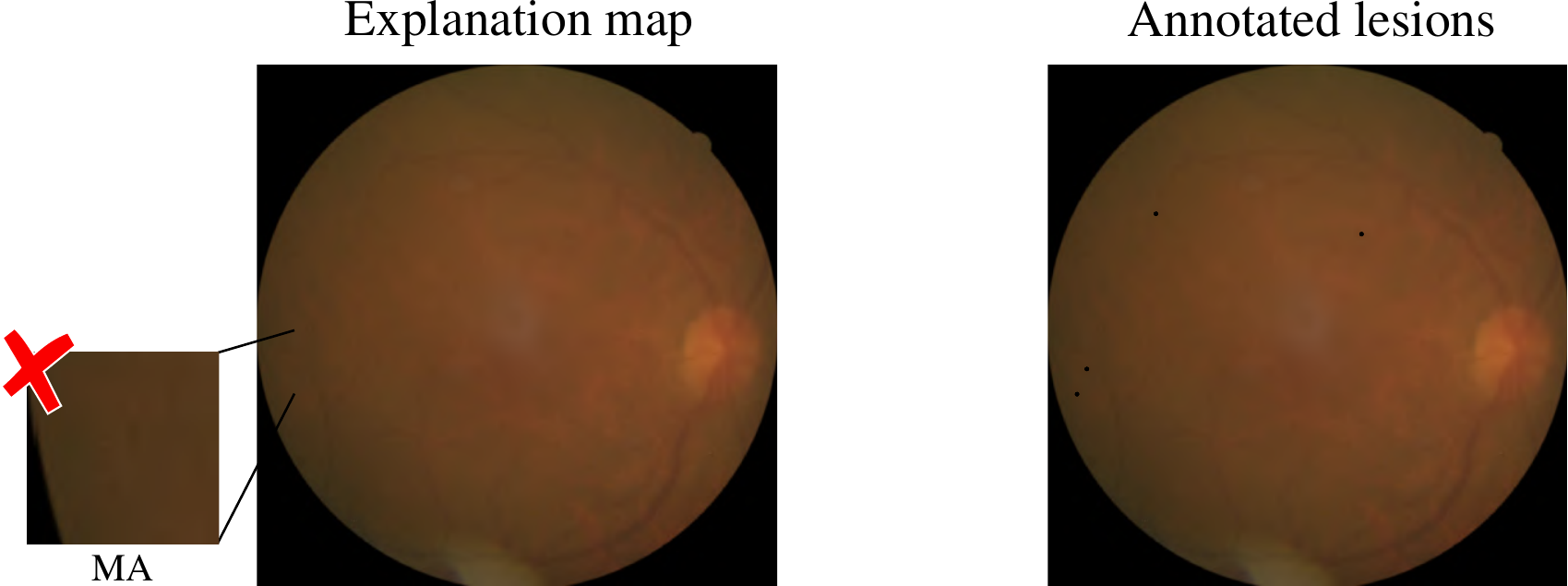} 
\caption{$\hat{y}_g=0, y=1$. \label{fig:gt1_pred0_badqual}}
\end{subfigure}

\begin{subfigure}{0.48\textwidth}
\includegraphics[width=\textwidth]{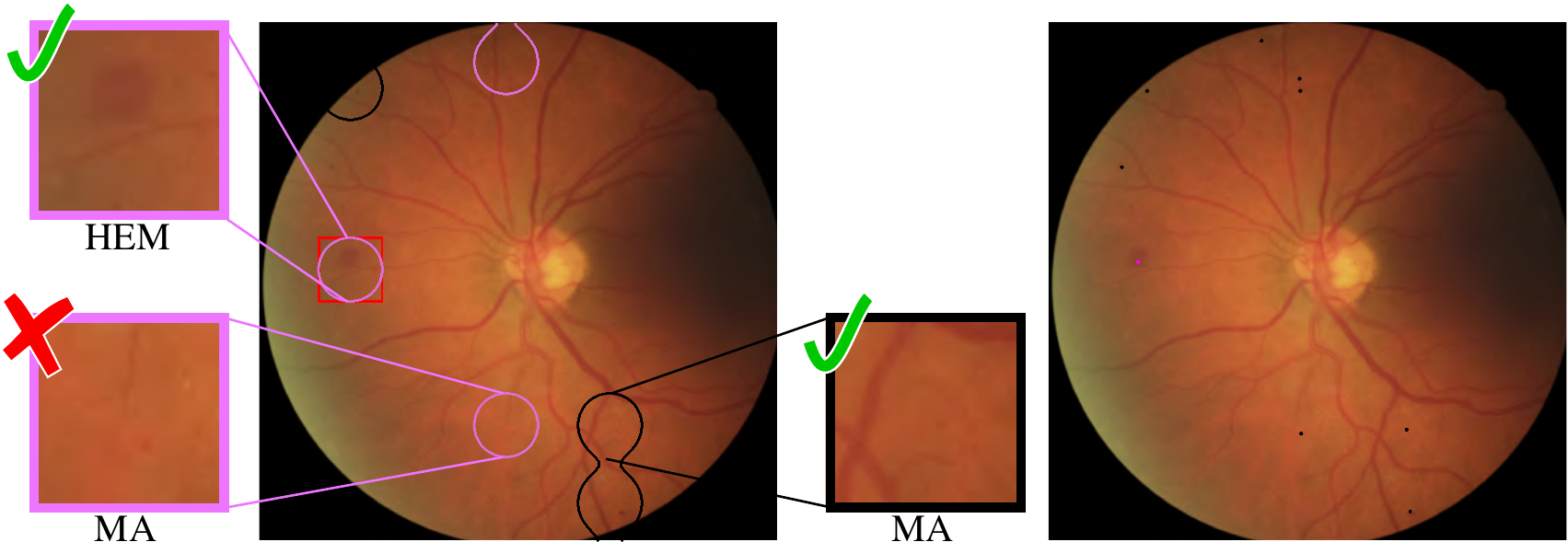}
\caption{$\hat{y}_g=2, y=2$. \label{fig:gt2_pred2}}
\end{subfigure}
\hfill
\begin{subfigure}{0.48\textwidth}
\includegraphics[width=\textwidth]{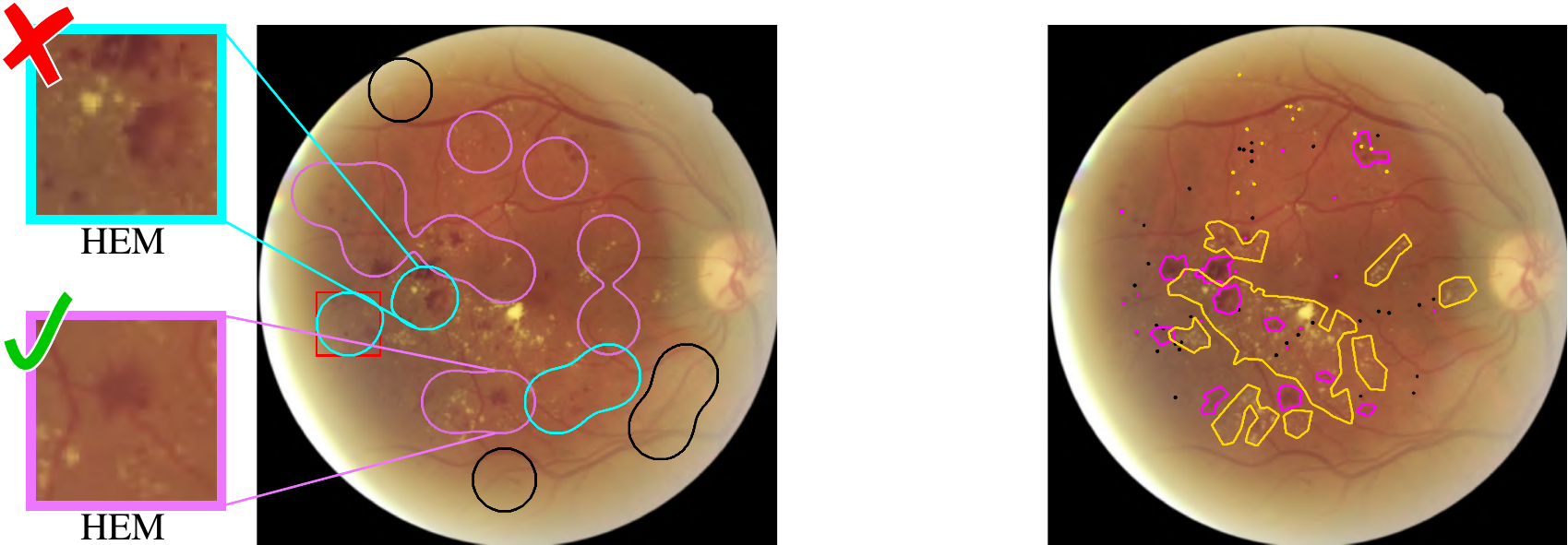}
\caption{$\hat{y}_g=3, y=2$. \label{fig:gt2_pred3_hems}}
\end{subfigure}

\begin{subfigure}{0.48\textwidth}
\includegraphics[width=\textwidth]{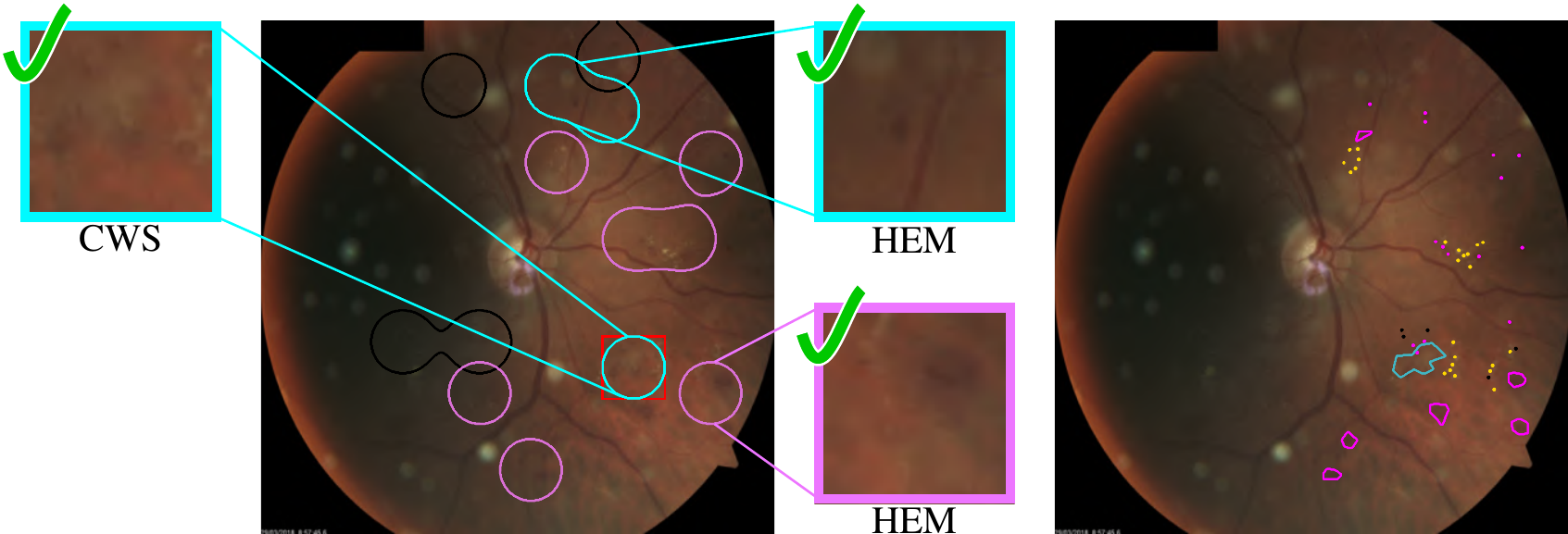} 
\caption{$\hat{y}_g=3, y=3$. \label{fig:gt3_pred3_cws}}
\end{subfigure}
\hfill
\begin{subfigure}{0.48\textwidth}
\includegraphics[width=\textwidth]{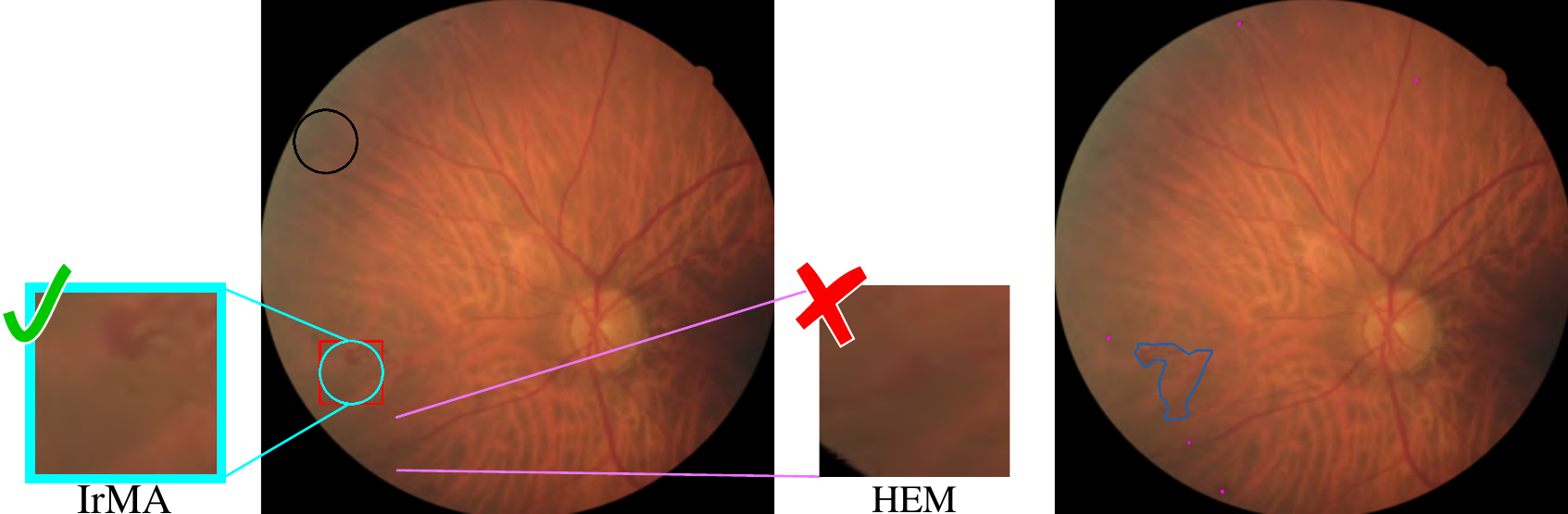}
\caption{$\hat{y}_g=3, y=3$. \label{fig:gt3_pred3_irma}}
\end{subfigure}

\begin{subfigure}{0.48\textwidth}
\includegraphics[width=\textwidth]{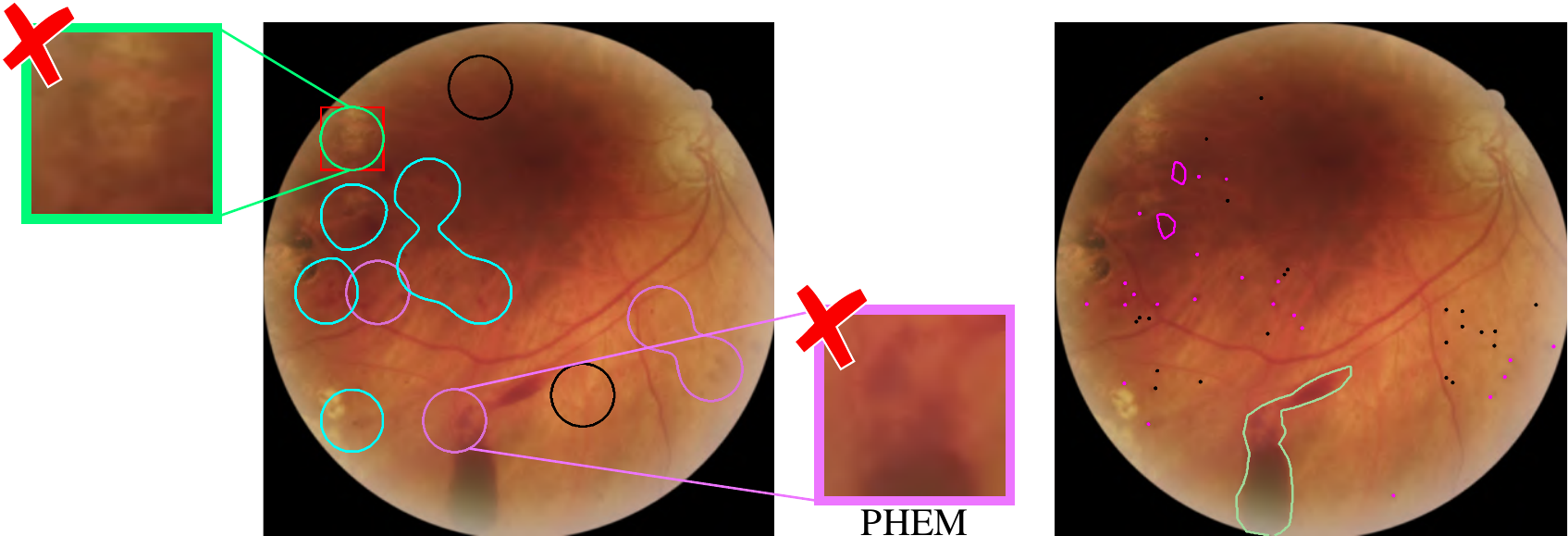} 
\caption{$\hat{y}_g=4, y=4$. \label{fig:gt4_pre4_badmap}}
\end{subfigure}
\hfill
\begin{subfigure}{0.48\textwidth}
\includegraphics[width=\textwidth]{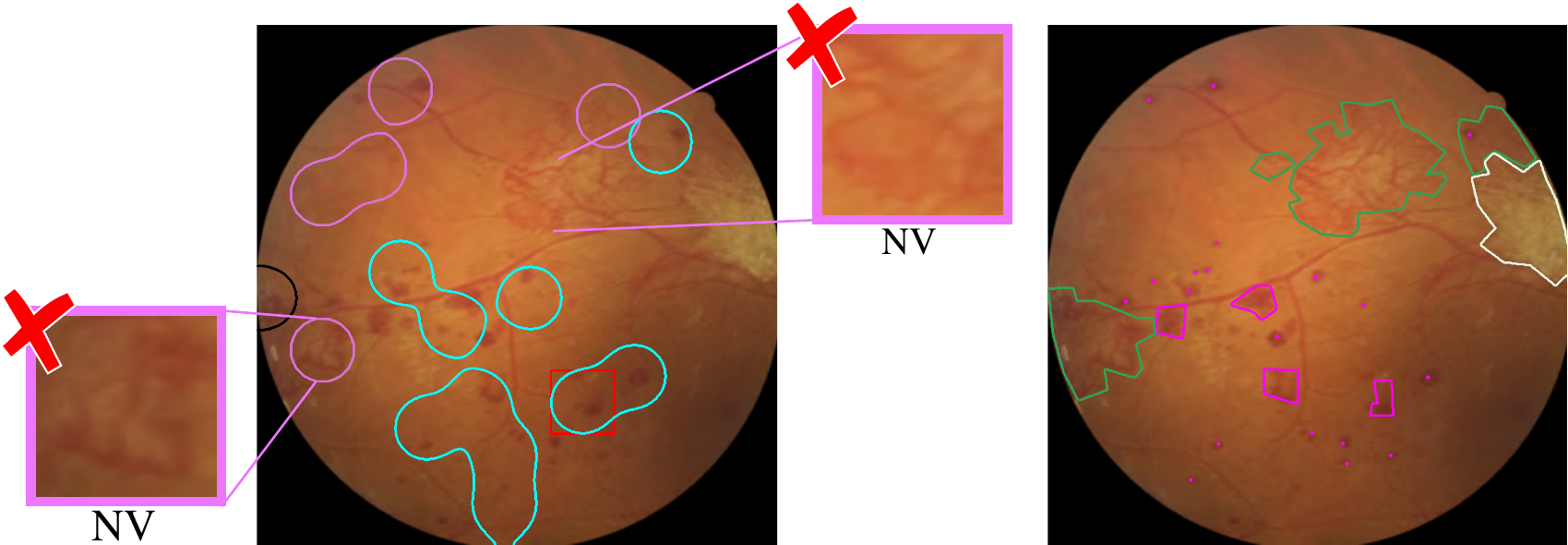}
\caption{$\hat{y}_g=3, y=4$. \label{fig:gt4_pred3_miss_nv_pfib}}
\end{subfigure}

\caption{Maps predicted by our method along with the ophthalmologist's annotated lesions in the SCREEN-DR dataset. The curves are the contours of the explanation maps (threshold=0.3), and the red square indicates the region of most relevance for diagnosis (corresponding to the maximum in the network's output activation map). Close-ups of relevant regions are shown, with \cmark~indicating that the region is correctly predicted, and \xmark~indicating otherwise. Below each close-up the name of the corresponding ground truth lesion (if existent) is shown. Explanation map: {\color{black}$\blacksquare$}~R1, {\color{orchid}$\blacksquare$}~R2, {\color{cyan}$\blacksquare$}~R3, {\color{springgreen}$\blacksquare$}~R4; ground truth map: {\color{black}$\blacksquare$}~MA, {\color{fuchsia}$\blacksquare$}~HEM, {\color{gold}$\blacksquare$}~EX, {\color{c1}$\blacksquare$}~CWS, {\color{c2}$\blacksquare$}~IrMA, {\color{c3}$\blacksquare$}~NV, {\color{c4}$\blacksquare$}~PHEM, {\color{c5}$\blacksquare$}~PFIB.\label{fig:maps_screendr}}
\end{figure*}

\renewcommand{\arraystretch}{1.15}
\begin{table}[tb]
\centering
\small{
\caption{Assessment of the produced attention maps (threshold=0.3) in comparison with specialist annotated lesions. These metrics are described in section~\ref{sec:exp_att_maps}.}
\label{tab:maps_eval}
\begin{tabular}{|l|l|l|l|l|l|}
\hline
$O_{obj\_g}$  & $O_{obj}$ &$O_{max}$ & $O_{class}$ & $O_{gt}$ & $O_{any}$                                                              \\ \hline
0.506 & 0.677 & 0.712 & 0.784 &  0.526 & 0.729
                                                                                \\ \hline
\end{tabular}
}
\end{table}


Several examples of \model's attention maps \textit{vs} the specialist's annotation on SCREEN-DR dataset images are shown in Fig.~\ref{fig:maps_screendr}. In each predicted map, the region with maximum activation is surrounded by a square. Figs.~\ref{fig:gt1_pred1}, \ref{fig:gt2_pred2}, \ref{fig:gt3_pred3_cws}~and~\ref{fig:gt3_pred3_irma} illustrate examples where the maximum value of the explanation map properly explains the predicted grade. The remaining images show cases of incorrect predictions and/or incorrect explanations.

\begin{figure}[tb]
\centering
\includegraphics[width=0.3\textwidth]{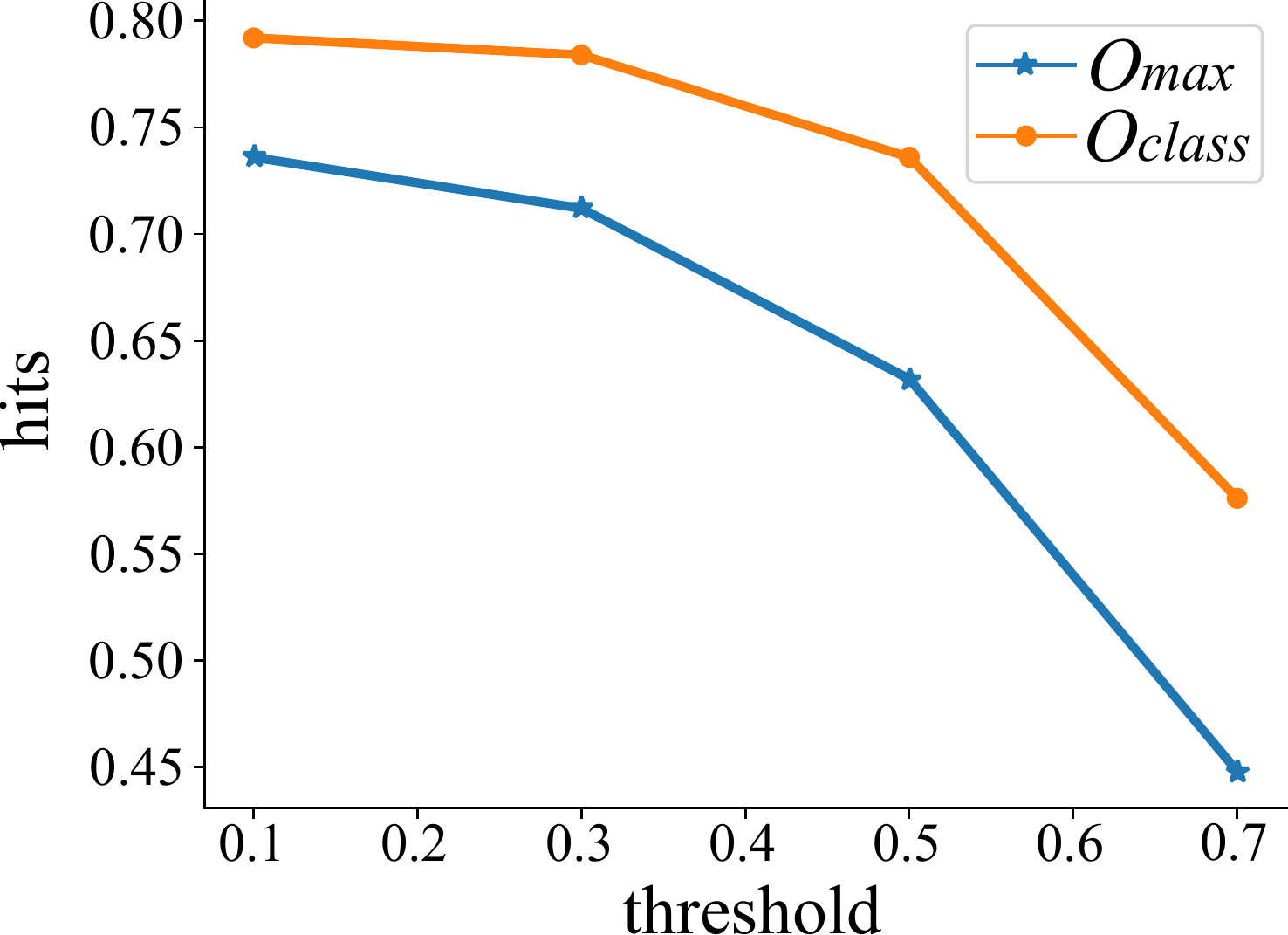} 
\caption{$O_{max}$ and $O_{class}$ for different threshold values (0.1, 0.3, 0.5 and 0.7) applied to the attention maps (Eq.~\ref{eq:gaussian_maps}).
\label{fig:hits_ths}}
\end{figure}

Table~\ref{tab:maps_eval} shows the explanation-wise quantitative performance in terms of the number and type of highlighted lesions, $O_{obj\_g}$, $O_{obj}$, $O_{max}$, $O_{class}$, $O_{gt}$ and $O_{any}$ (refer to Section~\ref{sec:exp_att_maps}). Fig.~\ref{fig:hits_ths} shows the values of $O_{max}$ and $O_{class}$ obtained through the application of different thresholds to the attention maps. The higher the threshold the smaller the objects on the map, resulting in higher specificity and lower sensitivity.

The threshold 0.3 seems to be a good compromise since it allows to retain a large percentage of the lesions while producing regions not too coarse.

\section{Discussion}

\subsection{DR grading}

\begin{figure}[t]
\centering
\begin{subfigure}{0.23\textwidth}
\includegraphics[width=\textwidth]{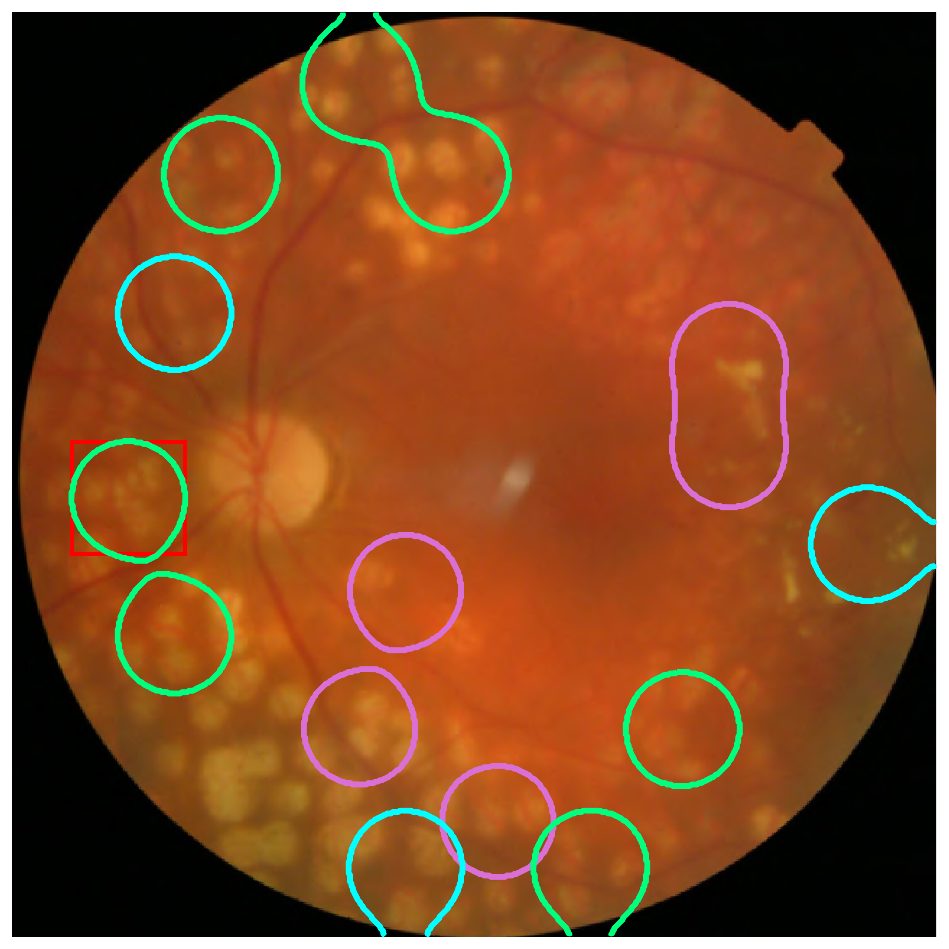}
\caption{$\hat{y}_g=4, y=4$ (Messidor-2). \label{fig:mess_p4_gt4}}
\end{subfigure}
\hfill
\begin{subfigure}{0.23\textwidth}
\includegraphics[width=\textwidth]{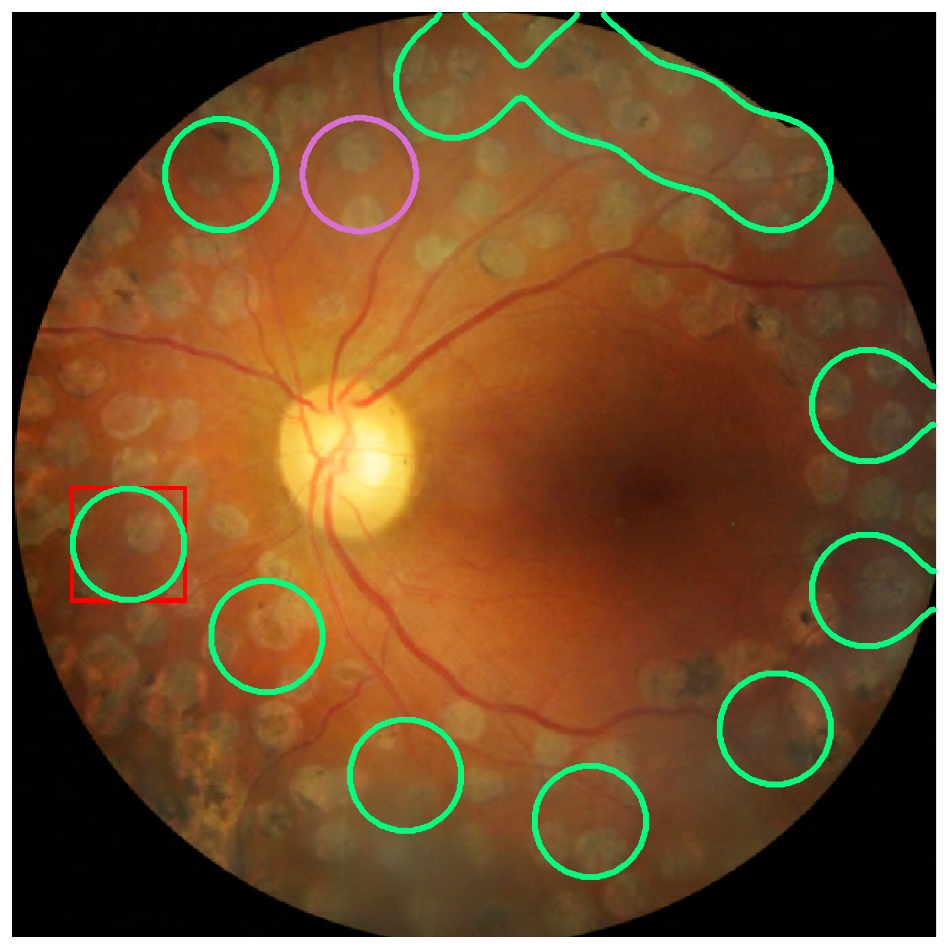}
\caption{$\hat{y}_g=4, y=4$ (\kaggle{} test). \label{fig:kaggle_p4_gt4}} 
\end{subfigure}

\begin{subfigure}{0.23\textwidth}
\includegraphics[width=\textwidth]{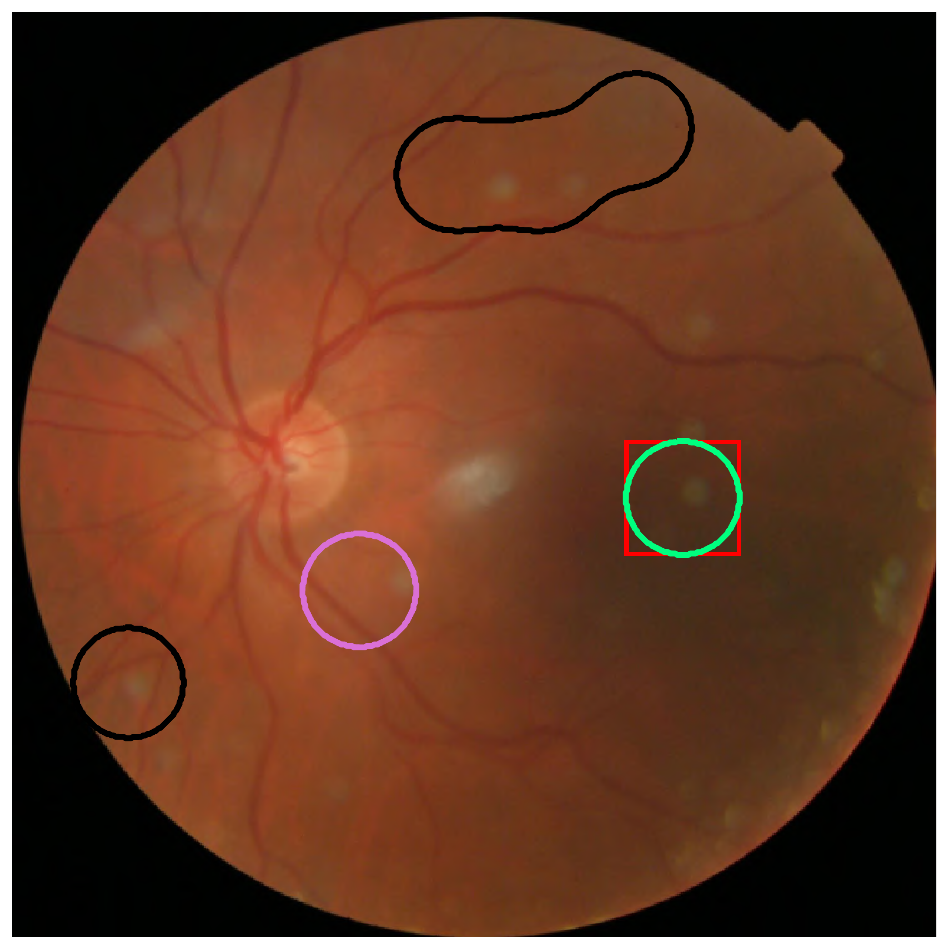} 
\caption{$\hat{y}_g=4, y=~0$ (Messidor-2). \label{fig:mess_p4_gt0}}
\end{subfigure}
\hfill
\begin{subfigure}{0.23\textwidth}
\includegraphics[width=\textwidth]{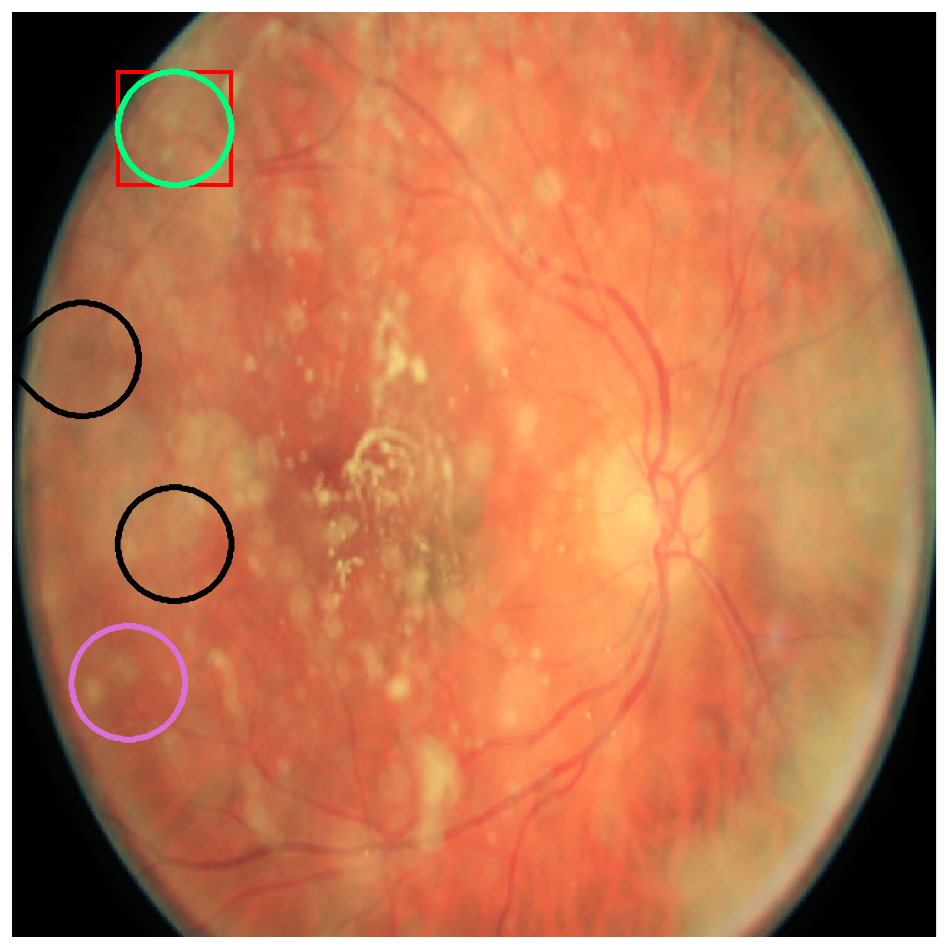} 
\caption{$\hat{y}_g=4, y=0$ (\kaggle{} test).\label{fig:kaggle_p4_gt0}} 
\end{subfigure}
\caption{Examples of images from Messidor-2 and \kaggle{} test sets that are predicted by \model{} as R4. The curves are the contours of the attention maps (threshold=0.3). {\color{black}$\blacksquare$}~R1, {\color{orchid}$\blacksquare$}~R2, {\color{cyan}$\blacksquare$}~R3, {\color{springgreen}$\blacksquare$}~R4. \label{fig:r4_laser}}
\end{figure}

\model{}'s performance does not degrade between the \kaggle{} test set, the most similar image- and grading-wise to the training data, and other datasets, showing the generalization capability of the model.
Furthermore, considering the misclassified cases, the 2$\times$ higher over-diagnosis percentage of \model{} in comparison to the expected human performance moderately promotes patient follow-up, challenging specialists to reassess their opinion.
Despite the high performance achieved, \model{} tends to fail to distinguish between R3 and R4 grades. As shown in Figs.~\ref{fig:mess_p4_gt4}~and~\ref{fig:kaggle_p4_gt4}, \model{} learned to associate photo-coagulation treatment and laser marks to R4, following the training set labels. Indeed, these images may not have signs of R4, such as neovessels and pre-retinal hemorrhages, and thus should either be marked as photocoagulated and removed from the dataset or be included with the grade corresponding to the visible lesions. This issue is further exacerbated by the high class imbalance of the data (Fig.~\ref{fig:datasets_classes}), since the data augmentation and class balancing scheme can contribute to the regulation of the weights update but not the limited variability of the samples. 
Because of this, when inferring images with different annotation criteria or with the presence of artifacts, \model{} misclassifies images as can be observed in Fig.~\ref{fig:mess_p4_gt0}~and~\ref{fig:kaggle_p4_gt0}.

\begin{figure}[tb]
\centering

\begin{subfigure}{0.23\textwidth}
\includegraphics[width=\textwidth]{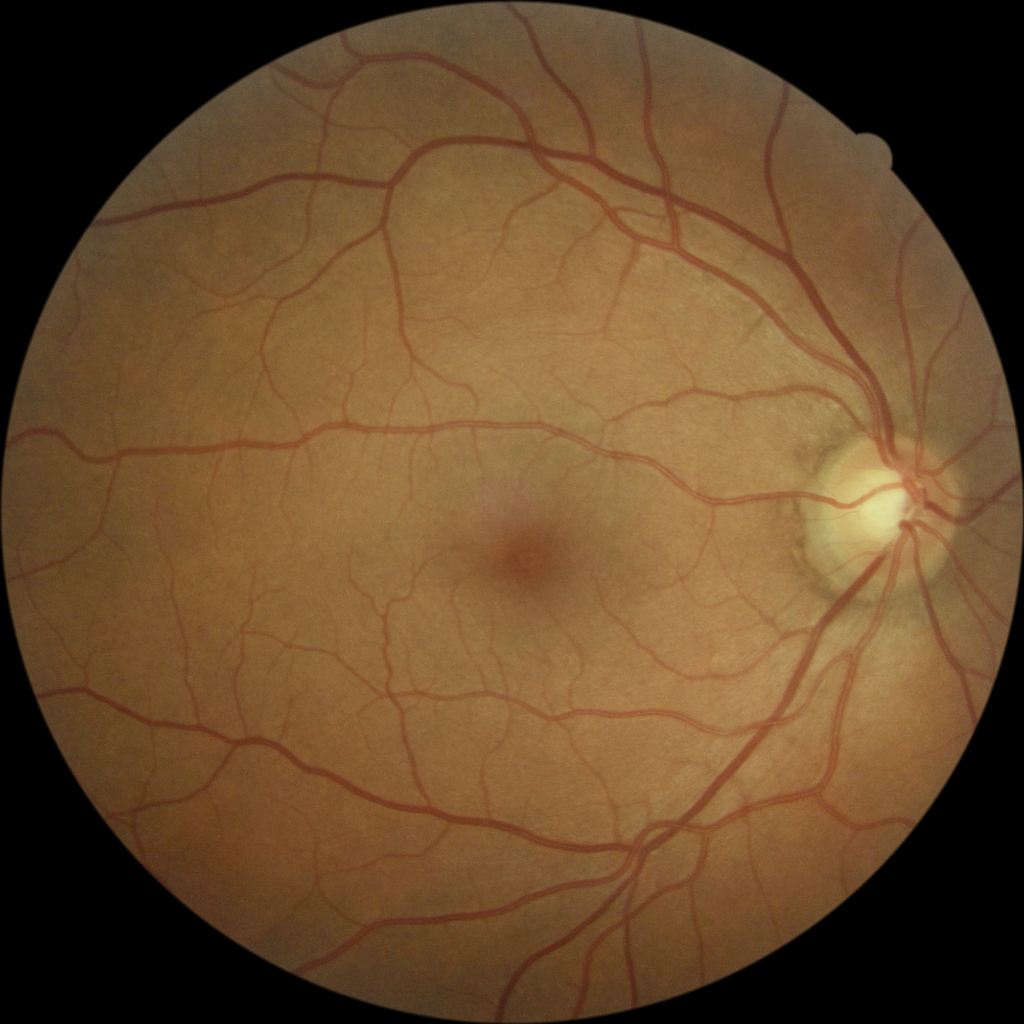} 
\caption{$\hat{y}_g=0, y=1$.\label{fig:pred0_gt1}} 
\end{subfigure}
\hfill
\begin{subfigure}{0.23\textwidth}
\includegraphics[width=\textwidth]{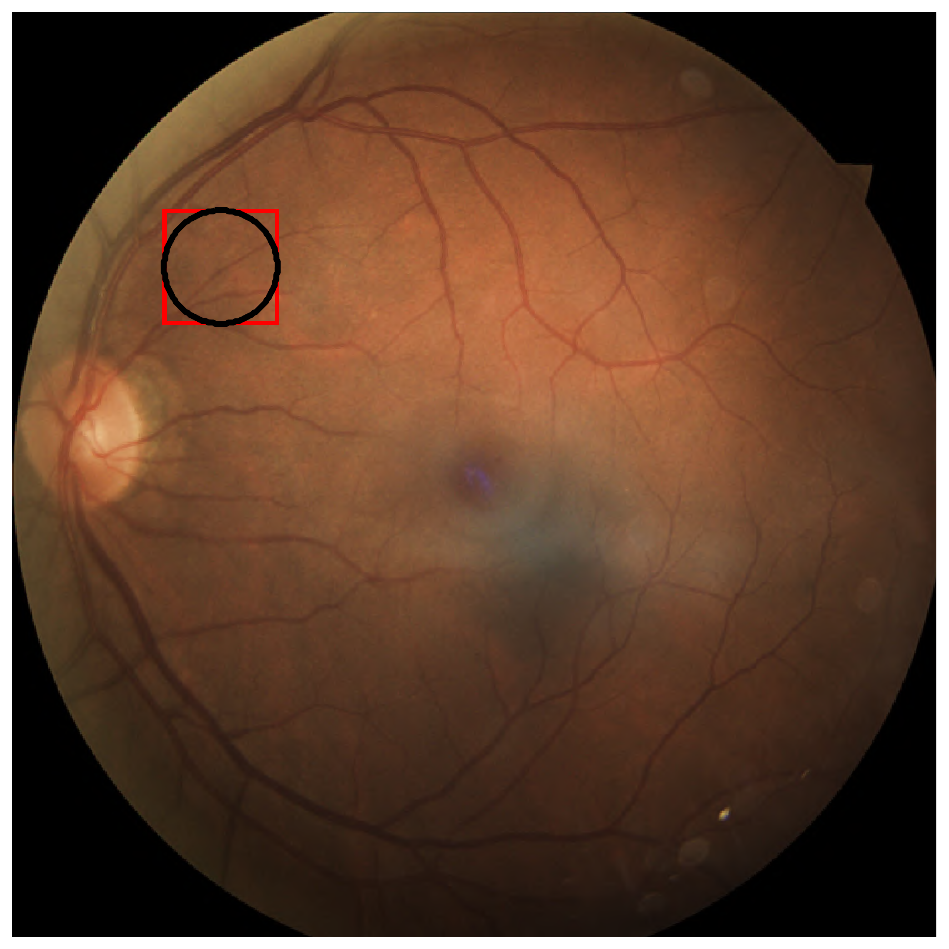} 
\caption{$\hat{y}_g=1, y=0$. \label{fig:pred1_gt0}} 
\end{subfigure}

\begin{subfigure}{0.23\textwidth}
\includegraphics[width=\textwidth]{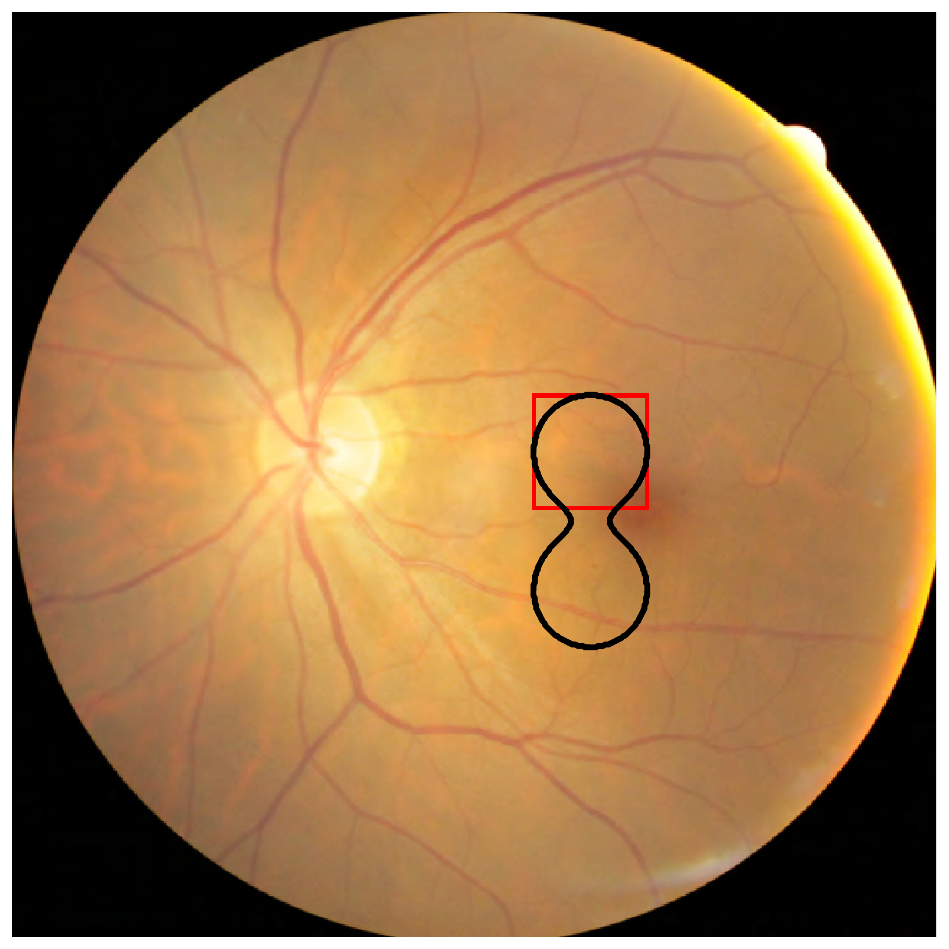} 
\caption{$\hat{y}_g=1, y=2$. \label{fig:pred1_gt2}} 
\end{subfigure}
\hfill
\begin{subfigure}{0.23\textwidth}
\includegraphics[width=\textwidth]{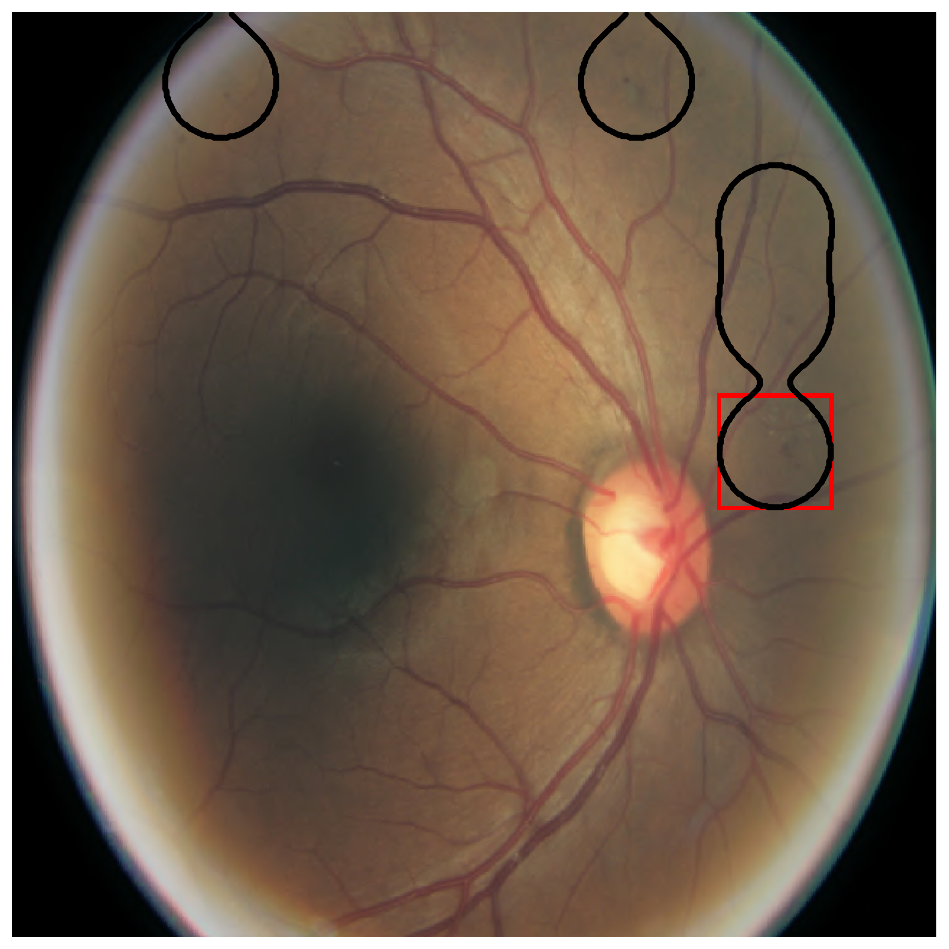} 
\caption{$\hat{y}_g=1, y=2$. \label{fig:pred1_gt2b}} 
\end{subfigure}

\caption{Examples of R0-R2 \kaggle{} dataset images (not used for training) along with the \model{} predictions. The curves are the contours of the attention maps (threshold=0.3). {\color{black}$\blacksquare$}~R1, {\color{orchid}$\blacksquare$}~R2, {\color{cyan}$\blacksquare$}~R3, {\color{springgreen}$\blacksquare$}~R4. \label{fig:r0_r1_r2}}
\end{figure}

In other scenarios, misclassifications occur due to failed lesion detection or similarities between different lesion types. Namely, failures such as predicting a R1 image as R0 can be due to missing a single MA. Indeed, the small size and subtlety of these lesions make their detection particularly challenging (Fig.~\ref{fig:pred0_gt1}, where no apparent lesion is seen) and easily confoundable with acquisition or anatomical noise (Fig.~\ref{fig:pred1_gt0}). 
Confusions between R1 and R2 are likely related with the difficulty of distinguishing between the red lesions MAs and HEMs (Fig.~\ref{fig:pred1_gt2}~and~\ref{fig:pred1_gt2b}), a task that also challenges medical experts since they can be very similar in size and color. Furthermore, different grading standards also account for these classification errors. Namely, for the DMR dataset, the conversion of SDR~=~R1 $\cup$~R2 does not holds necessarily true in all cases. For instance, when an image is R3 by the international scale due to the presence of $>20$ HEMs, it would fit in the SDR category of the Davis scale. However, through the conversion, this image would fall in the PPDR category, which can justify the model's confusion between R1 and R2 that is verified in Fig.~\ref{fig:conf_dmr}.

\begin{figure}[t]
\centering
\begin{subfigure}{0.23\textwidth}
\includegraphics[width=\textwidth]{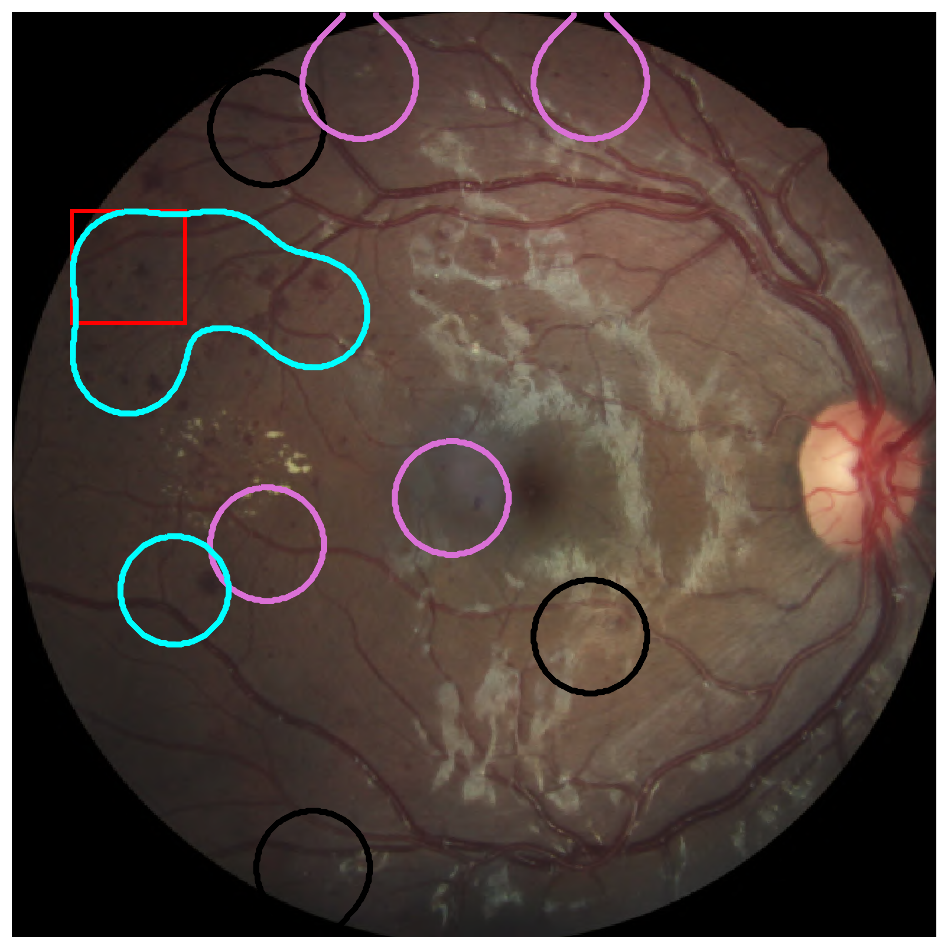} 
\caption{$\hat{y}_g=3, y=3$. \label{fig:pred3_gt3_good}}
\end{subfigure}
\hfill
\begin{subfigure}{0.23\textwidth}
\includegraphics[width=\textwidth]{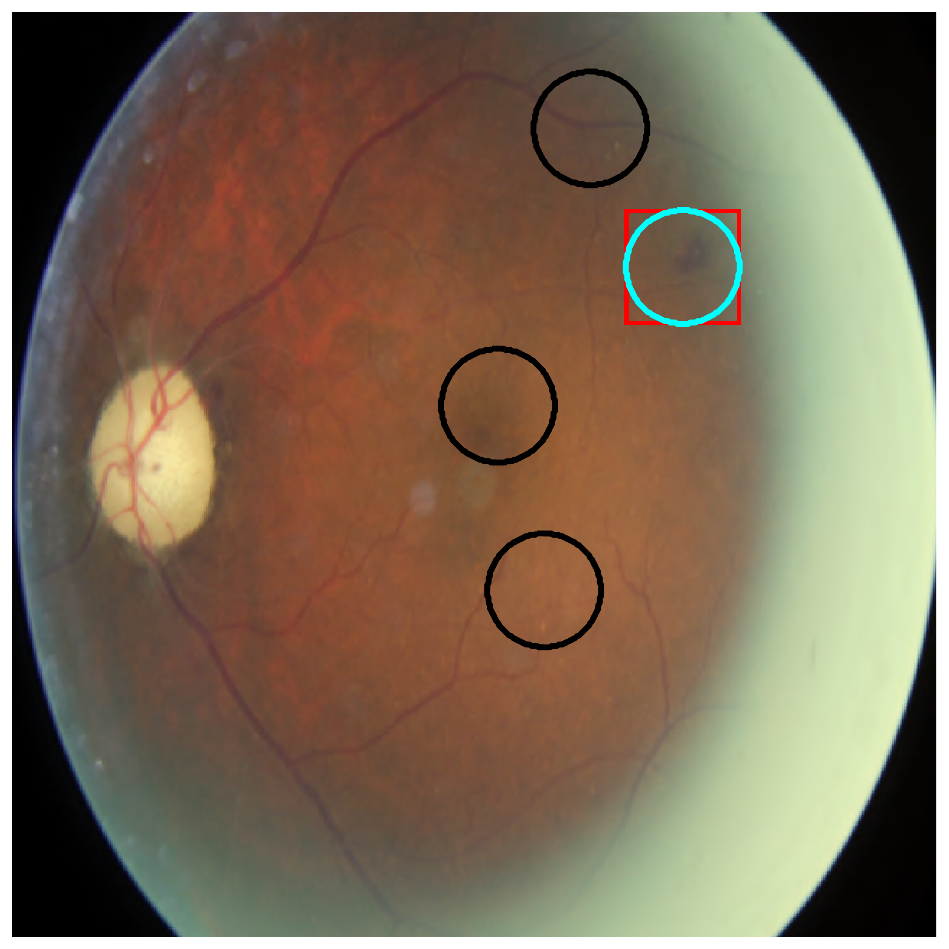} 
\caption{$\hat{y}_g=3, y=2$.\label{fig:pred3_gt2_bad}} 
\end{subfigure}
\caption{Examples of R2 and R3 \kaggle{} dataset images (not used for training) along with \model{} predictions. The curves are the contours of the attention maps (threshold=0.3). {\color{black}$\blacksquare$}~R1, {\color{orchid}$\blacksquare$}~R2, {\color{cyan}$\blacksquare$}~R3, {\color{springgreen}$\blacksquare$}~R4. \label{fig:r2_r3_assump}}
\end{figure}

Classification errors for R2 and R3 grades are mostly due to the grading rules. Indeed, in cases where there are no other signs of R3 (Table~\ref{tab:dr_scale}), grading depends on HEM counting. However, this task is non-trivial for specialists since HEM may appear in clusters, hindering lesion counting and thus introducing noise on the data. Furthermore, the MIL assumption fails for these cases, as there is no novel lesion to find, leading to prediction errors (Fig.~\ref{fig:pred3_gt2_bad}). Instead, \model{} may have associated lesion size to the grade. Alternatively, despite relying on a single lesion to perform grade inference, the construction of the lesion map $L$ depends on the integration of features with different receptive fields and thus the lesion with the maximum response may indirectly encode lesion counting, serving as a proxy of the overall relevant lesion density (Fig.~\ref{fig:pred3_gt3_good}).

\begin{figure}[tb]
\centering
\begin{subfigure}{0.23\textwidth}
\includegraphics[width=\textwidth]{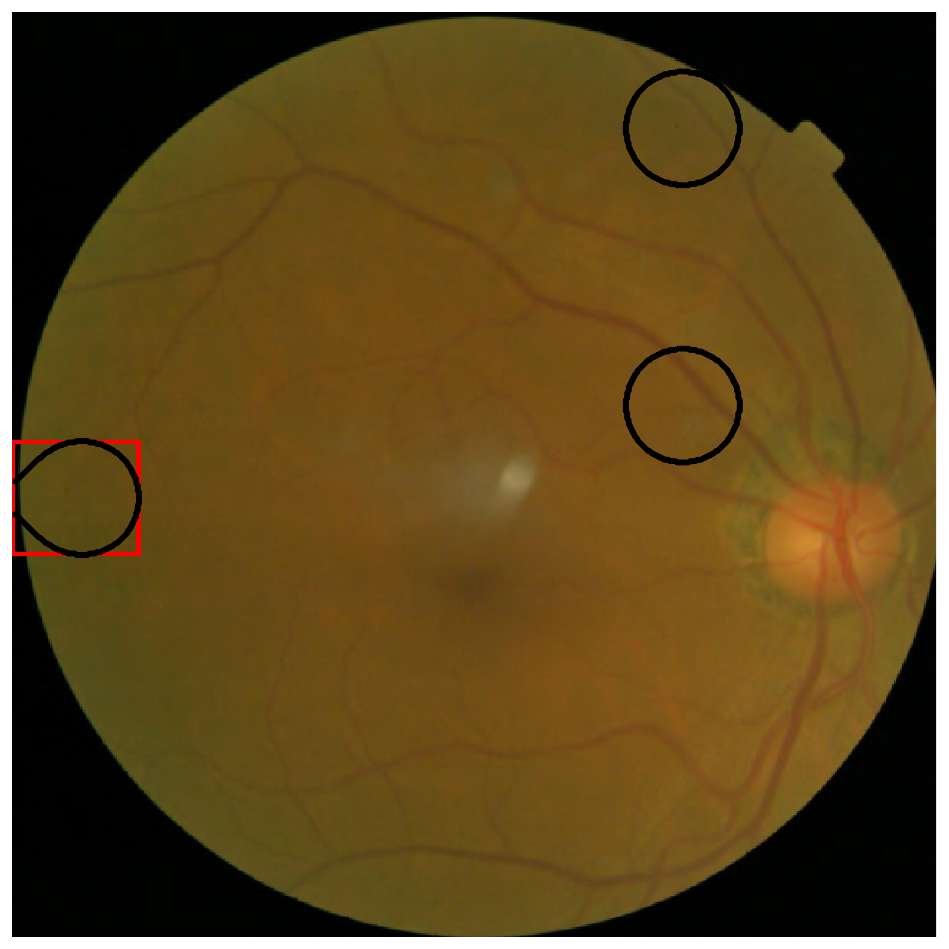} 
\caption{$\hat{y}_g=1, y=0$.}
\end{subfigure}
\hfill
\begin{subfigure}{0.23\textwidth}
\includegraphics[width=\textwidth]{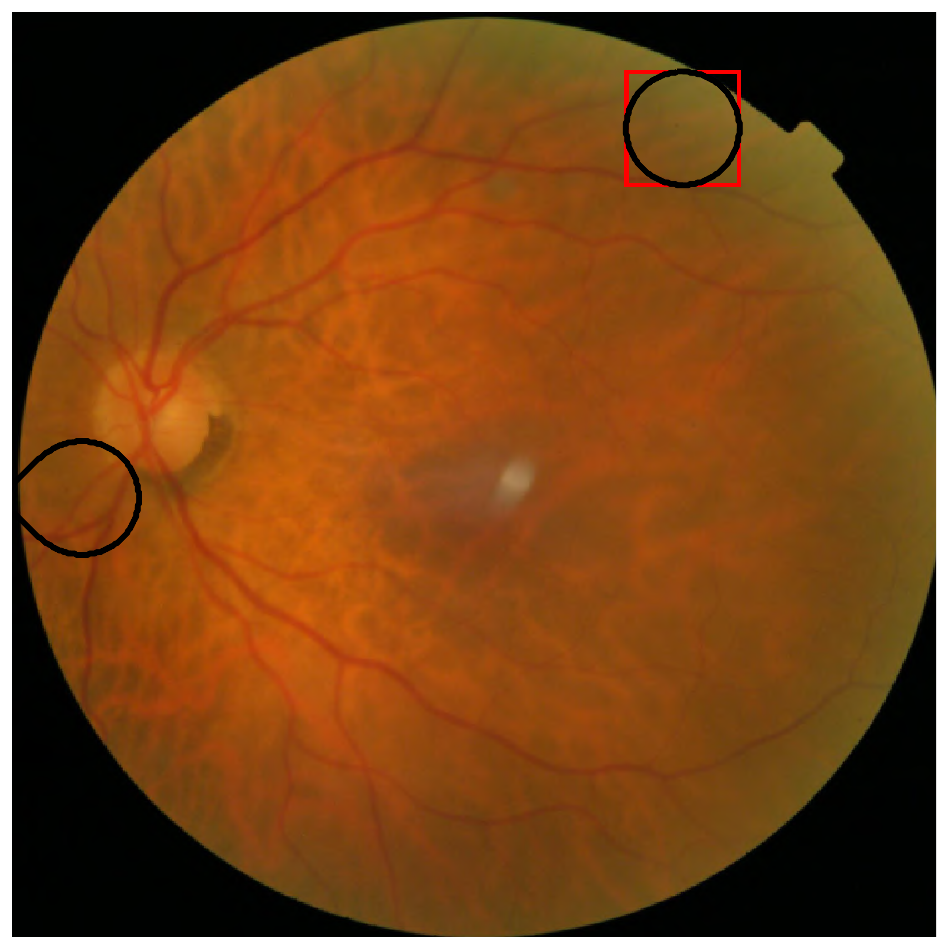}
\caption{$\hat{y}_g=1, y=0$.}
\end{subfigure}
\caption{Examples of healthy Messidor-2 images mispredicted as R1. These images show camera artifacts (present in the same location across several different images) which are detected as R1 typical lesions. The curves are the contours of the  attention maps (threshold=0.3). {\color{black}$\blacksquare$}~R1, {\color{orchid}$\blacksquare$}~R2, {\color{cyan}$\blacksquare$}~R3, {\color{springgreen}$\blacksquare$}~R4. \label{fig:messidor_marks}}
\end{figure}


\subsubsection{Comparison with the state-of-the-art}

Overall, \model{} achieves similar performance to other state-of-the-art methods and shows high generalization capability. Indeed, our model has a $\qwk{}$ similar to \cite{Takahashi2017}'s method on the DMR dataset (value computed by data provided by the authors) while being trained on a different dataset and considering a different grading scale (international instead of Davis).
Note that class-wise performance comparison is limited because confusion matrices are not commonly available in the literature. The proposed system achieves a performance similar to \cite{DelaTorre2017}, while having only 2$\%$ of the parameters and not requiring highly dedicated loss functions. Indeed, \model{}'s architecture already promotes the distance minimization between the ground-truth and the prediction. This allows the use of the common categorical crossentropy as the main term of the loss function and thus avoids the need for the direct optimization of the $\qwk{}$, which requires higher computational power. On the other hand, \model{}'s performance is inferior to the 0.84 $\qwk{}$ reported by \cite{Krause2018} on EyePACS-2 dataset (1818 images) (data publicly unavailable at the time of writing).
However, their model was trained on an expensive dataset composed of over 1.6 million images, and with ground truth labels adjudicated by multiple ophthalmologists. Further, the authors apply an empirically defined cascade of thresholds over their continuous network output in order to define the grade, ensuring that a baseline sensitivity was achieved for the higher grade classes. Consequently, the system also tends to overgrade images, with a normalized over-grading of 72\%, which can lead to a high number of unnecessary follow-ups. On the other hand, \model{} presents a lower over-estimation and thus a more moderate trade-off between over and under diagnosis of DR.

\subsection{Uncertainty estimation}

Ideally, \model{} should infer higher uncertainties for misclassified cases.
Indeed, high $\qwk{}$ values occur for low uncertainty (Fig.~\ref{fig:unc_kappa_datasets}), indicating that the \model{}'s uncertainty estimation is a valid measure of the quality of the prediction. Fig.~\ref{fig:unc_matrix} further indicates that low uncertainty is associated with correct or close classification, since the uncertainties are considerably lower in the diagonal, and tend to increase with the class distance. Furthermore, all grades follow a similar cumulative histogram, as shown in Fig.~\ref{fig:unc_kappa_grade}, indicating that the tendency illustrated in Fig.~\ref{fig:unc_kappa_datasets} does not result from the different proportions of each grade among the analysed image subsets.
Low uncertainty of misclassifications tends to be due to the presence of lesions that greatly resemble another type, characteristic of a different grade (e.g., large MAs which are similar to HEMs), or to the erroneous identification of acquisition errors as lesions (Fig.~\ref{fig:messidor_marks}).

The lower $\qwk{}$ for higher uncertainty cases suggests that \model{} can be used cooperatively in the clinical practice by asking ophthalmologists to review cases that are most likely wrong. Indeed, case selection via uncertainty may significantly reduce the workload during the screening pipeline. For instance, discarding the $25\%$ most uncertain images from \kaggle{} test set allows to achieve a 0.8 $\qwk{}$, a performance similar to human experts~\citep{Krause2018}, thus allowing \model{} to be used as an independent reader capable of identifying dubious cases.

High uncertainty tends to occur in very obscured and blurred images (Fig.~\ref{fig:unc_14_p1},~\ref{fig:unc_13_p0},~\ref{fig:unc_1_p2},~\ref{fig:unc_09_p2},~\ref{fig:unc_07_p0}). In these cases, it is not possible to make a diagnosis - at least not fully -, and thus predictions are associated with high uncertainty estimations. This will be further discussed in section~\ref{sec:disc_low_qual}.
Fig.~\ref{fig:unc_1_p4} shows an interesting case of a R4 image correctly predicted by \model{} but with high uncertainty (0.999). This image shows a very large PFIB (PDR sign), occupying almost half of the FOV. However, as referred, \model{} did not correctly learn these R4 signs (even less of this size) and thus it is not expected to be certain in making a diagnosis based on such image. This large white structure may have been interpreted as partially obscuring the image and thus the high uncertainty. 
Further, high uncertainties can also result from the resemblance of certain structures in the images with DR signs, such as CWSs (Fig.~\ref{fig:unc_045_p3}), and with photocoagulation marks (Fig.~\ref{fig:unc_058_p4}). 
Contrastingly, \model{} correctly predicts, with low uncertainty, the good quality and diagnosable images from Fig.~\ref{fig:unc_029_p1}-\ref{fig:unc_005_p4}, apart from Fig.~\ref{fig:unc_005_p3}. We believe this case is an example of the noise present in \kaggle{} dataset annotations, since the image is labeled as R2, but shows several CWSs (R3 sign), leading \model{} to predict R3 with very low uncertainty (0.05).


\subsubsection{Low quality images} 
\label{sec:disc_low_qual}
Poor image quality can lead to the occlusion or distortion of lesions, which hinders a meaningful and complete clinical decision~\citep{Niemeijer2006}. Lack of focus, sharpness or illumination often characterize bad quality images~\citep{PiresDias2014}.
In theory, and as suggested by Fig.~\ref{fig:unc}, the algorithm should output higher uncertainties for images of poor quality, since in the these cases it can not be as sure of the diagnosis as in a good quality image. 
Indeed, Fig.~\ref{fig:unc_14_p1}-\subref{fig:unc_07_p0} mainly show examples of largely obscured images which are associated with very high uncertainty estimations. Usually, when the image is partially available for diagnosis, if \model{} can find lesions in the viable image region it predicts a disease grade with high uncertainty (HEMs from R2, in Fig.~\ref{fig:unc_05_p2}), whereas if it cannot find any signs it predicts it as healthy, again with high uncertainty (Fig.~\ref{fig:unc_07_p0}).

However, since image quality is not trivial to define within the context of clinical diagnosis, image quality assessment tends to be a subjective process~\citep{Wanderley2019}. One can consider more than two levels of quality, such as good, partial, bad, where a partial quality image still has a region suitable for diagnosis. However, most datasets are labeled in a binary manner (good or bad quality), forcing partially obscured images to considered bad or good.  

Further, as image quality is not the only factor influencing the uncertainty of the model prediction, a full correspondence between image quality and the uncertainty values is not likely. However, a tendency of bad quality images being associated, overall, with higher uncertainties is expected.

\begin{figure}[tb]
\centering
\begin{subfigure}{0.23\textwidth}
\includegraphics[width=\textwidth]{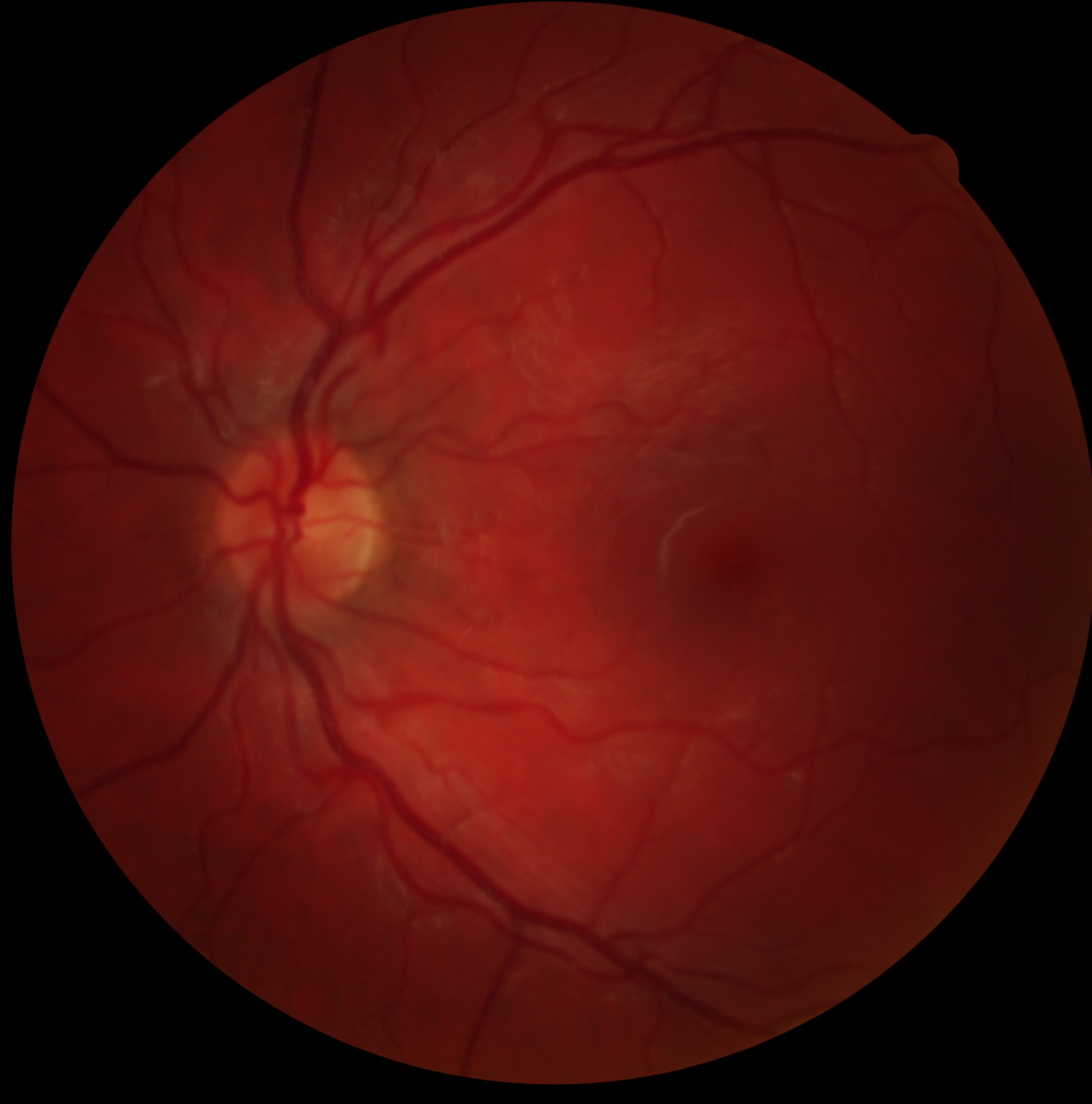} 
\caption{Bad quality image, $u$=0.415.}
\end{subfigure}
\hfill
\begin{subfigure}{0.23\textwidth}
\includegraphics[width=\textwidth]{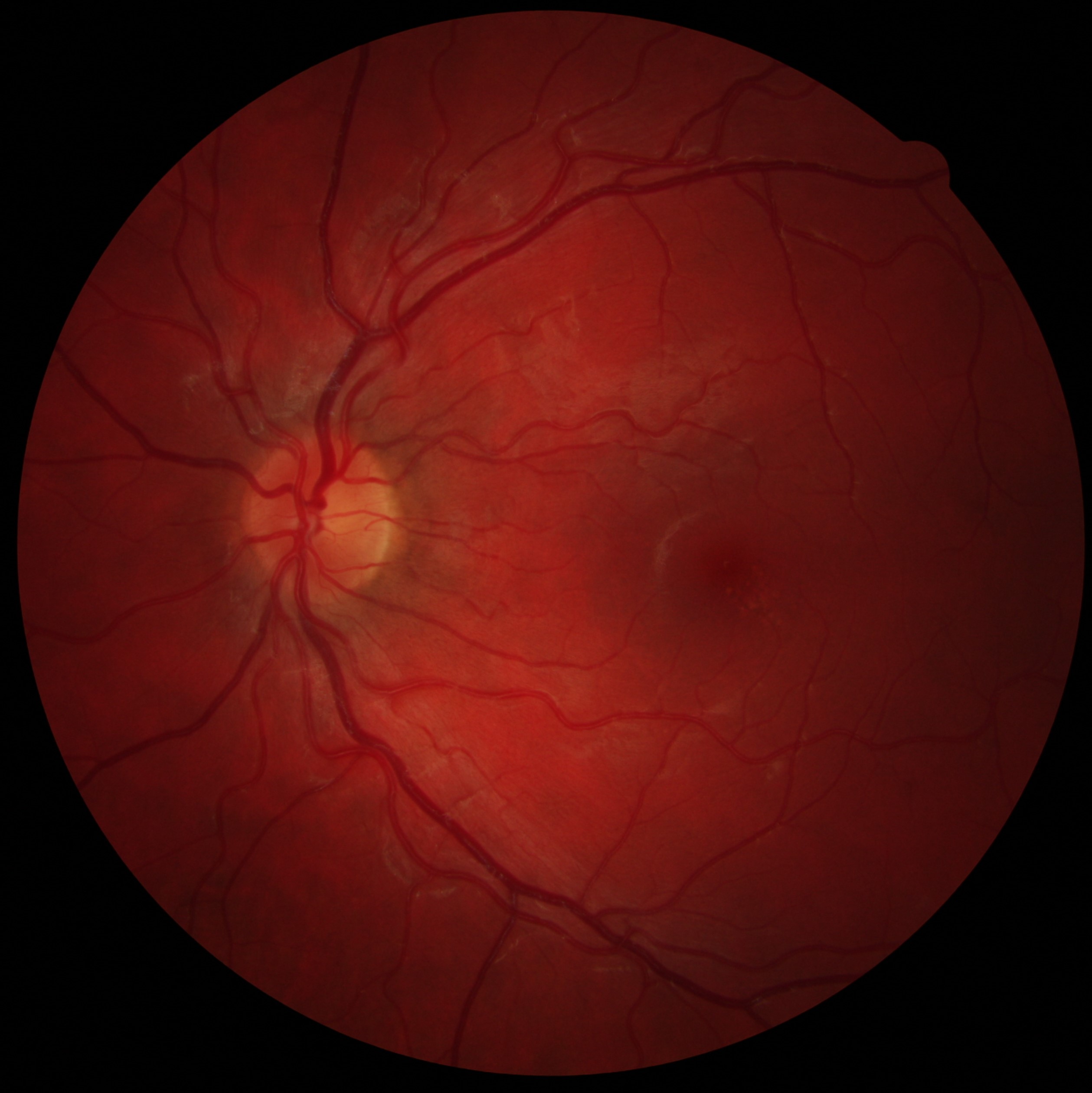} 
\caption{Good quality image, $u$=0.295.}
\end{subfigure}
\caption{Example of a pair of good/bad quality images from the HRF dataset. \label{fig:hrf_qual}}
\end{figure}

\model{} produced significantly higher uncertainties for bad than for good quality images in each dataset (Fig.~\ref{fig:bad_good_qual}), which suggests it can potentially be used in the DR diagnosis pipeline without the need for an \textit{a priori} image quality detection step.
The HRF datasets is particularly interesting since it contains pairs of good/bad quality images, but has the inconvenient of having only 18 pairs of images.
Some of the HRF bad quality images differ only slightly from the good quality counterparts (Fig.~\ref{fig:hrf_qual}), which leads to similar uncertainty estimation between the two sets. 
Overall, the bad quality images seem to still be suitable for diagnosis, which may justify the lower uncertainties for the bad quality subset comparing with what is verified on the other datasets.
DRIMDB is the more dichotomic dataset, where images are clearly bad or good, apparently not presenting a middle term. This is reflected on the estimated uncertainties, with the bad quality set showing an average uncertainty of 0.75$\pm$0.22 and the good set of 0.30$\pm$0.12.
Note that both HRF and DR1 contemplate only blur in the bad quality images, and thus low contrast or bad illumination are not considered.

\hl{Regarding the sensitivity analysis with the blurring operations, we can see based on} Fig.~\ref{fig:sens_plot} \hl{that the uncertainty shows a tendency to increase with the increase of the image blur, which decreases image quality. This corroborates our previous findings that image quality is correlated with uncertainty.}
\hl{When it comes to the contrast variation experiments, we believe that the uncertainty did not show any correlation with it due to data augmentation scheme performed during} \model{}'s \hl{training, which contemplated brightness adjustments and contrast normalization, and most likely made the network robust to these changes.}
\hl{As the major factor accounting for bad quality in the tests with the bad/good quality datasets} (Section~\ref{subsec:datasets_qual} and Fig.~\ref{fig:bad_good_qual}) \hl{was blur/decreased sharpness, not image contrast/illumination conditions, the sensitivity analysis results are according to the expectations.}


\subsubsection{Unfamiliar data types}

The uncertainty estimation on the unfamiliar data types suggests that unfamiliar medical image types, i.e., which are not eye fundus images, could be detected as outliers by the analysis of the uncertainty values produced by the algorithm, which are in average higher than for the eye fundus images.

Colonoscopy images are probably the most similar to the eye fundus, showing a similar circular FOV and  color (Fig.~\ref{fig:unfamiliar_data}), and sometimes even red and bright structures which may resemble retinal lesions and anatomical structures. Considering that the \model{} was trained with eye fundus images, images that most resemble these can lead to more uncertainty since they are somehow familiar but do not present the usual and expected structures for diagnosis, similarly to what happens in bad quality eye fundus images. 
H$\&$E stained microscopy images of breast tissue are the closest to the eye fundus in terms of uncertainty estimation, and are visually far from retinal images. This may be due to the presence of several small rounded structures, darker than the background (cells' nuclei) which may resemble retinal lesions. Considering that our MIL approach learns to identify the local structures responsible for a given DR grade, it makes sense that resemblance at a local level (as happens in microscopy images) leads to lower uncertainties than resemblance at the global image level (colonoscopy images).


\subsection{Explainability}

\model{}'s MIL-based architecture leads to the explanation maps to pinpoint regions relevant for diagnosis, and thus highlighting every single lesion is not expected. Because of this, the global lesion-wise is not reliable enough to fully characterize an image. 
The detection performance is further hindered by the fact that specialists may not annotate all abnormalities (Fig.~\ref{fig:maps_screendr}) since this can be a large time consuming task. 
On the other hand, \model{} is able to correctly identify plausible regions for the reasoning behind the grade of a given image.
The analysis of the produced attention maps show the ability of the system to detect lesions as small as MAs (Fig.~\ref{fig:gt1_pred1}).
As expected, when image quality is poor, image diagnosis is hindered and lesions are often missed (Fig.~\ref{fig:gt1_pred0_badqual}).
Confusion between MAs and HEMs occurs for some images (Fig.~\ref{fig:gt2_pred2}), which is justifiable by the high resemblance of small HEMs and MAs.
The detection of R3 is particularly interesting due to the variety of configurations that are associated with this grade (Table~\ref{tab:dr_scale}), which \model{} seems to be able to identify: a high number of HEMs (Fig.~\ref{fig:gt2_pred3_hems}), some CWSs (Fig.~\ref{fig:gt3_pred3_cws}) or IrMAs (Fig.~\ref{fig:gt3_pred3_irma}). Regarding HEMs, one can see that the model detects this lesion type both as a R2 or a R3 sign (Fig.~\ref{fig:gt2_pred3_hems}), which is understandable since a low number of HEMs is a sign for R2 and a high number a sign of R3. Despite this image being annotated as R2, the high number of HEMs may justify the R3 prediction.
However, \model{} fails to detect NVs and other R4 lesions such as PHEMs (Fig.~\ref{fig:gt4_pre4_badmap}) and PFIB (Fig.~\ref{fig:gt4_pred3_miss_nv_pfib}). Most of the times the reason for a R4 prediction is related with the presence of photocoagulation marks (Fig.~\ref{fig:gt4_pre4_badmap}) which corroborates the idea that these were learn by the algorithm as being R4 signs. Nevertheless, R4 lesions are sometimes highlighted in the maps as signs of disease. For instance, NVs and PHEMs are detected as R2 signs in these two images, since they slightly resemble HEMs.

Interestingly, the inherent explainability of \model{} allows for the \textit{a posteriori} identification of systematic errors during image acquisition. For instance, several Messidor-2 images have been wrongly classified as R1 due to the presence of dust particles in the camera lens (Fig.\ref{fig:messidor_marks}). However, the clear explanation provided would easily allow the clinician to correct the acquisition error and thus improve the overall quality of the screening pipeline.


\section{Conclusions}
We propose \model{}, a novel deep learning-based approach for DR grading that provides an uncertainty and explanation associated with each prediction. 
\model{} was trained on the \kaggle{} training set, and achieved state-of-the-art performance in several DR-labeled datasets.
The classification of the R4 grade was the least accomplished task, suggesting the training did not correctly capture the PDR signs. This is most likely due to the type of annotation in the \kaggle{} dataset, in which images that underwent photocoagulation treatment and show laser marks appear to be labeled as PDR even when not presenting the typical signs of this grade.  

The analysis of the predictions' uncertainties showed the tendency of lower $\qwk{}$ for higher uncertainty cases, thus suggesting that the estimated uncertainty is a viable measure of the quality of the prediction. This means that at test time, \textit{i.e.} where the performance of the system is not known, the uncertainty can be a measure of how much a specialist can trust the algorithm's decision, allowing to rank the predictions and select which images would benefit most from a second evaluation. 
Further, the explanation map produced by our algorithm should be easily interpretable by the retinal specialists, and thus constitutes a valuable tool for mitigating the black-box behaviour associated with deep NNs. 

Comparing with the state-of-the-art, \model{} has the advantage of producing both an uncertainty estimation associated with the prediction and an explanation map highlighting the most relevant regions for the classification, without compromising the DR grading performance. We believe that coupling an uncertainty estimation and explanation with the DR grade will ease the application of \model{} in a clinical setting.

\section*{Acknowledgments}

Teresa Ara\'{u}jo is funded by the FCT grant contract SFRH/BD/122365/2016. Guilherme Aresta is funded by the FCT grant contract SFRH/BD/120435/2016. 
This work is financed by the ERDF – European Regional Development Fund through the Operational Programme for
Competitiveness and Internationalisation - COMPETE 2020 Programme, and by National Funds through the FCT - Funda\c{c}\~{a}o para a Ci\^{e}ncia e a Tecnologia within project CMUP-ERI/TIC/0028/2014.


\bibliographystyle{model2-names}\biboptions{authoryear}
\bibliography{references.bib}

\end{document}